\documentclass[]{jfm}

\usepackage{graphicx}
\usepackage{newtxtext}
\usepackage{newtxmath}
\usepackage{natbib}
\usepackage{hyperref}
\hypersetup{
    colorlinks = true,
    urlcolor   = blue,
    citecolor  = black,
}

\newcommand{\RomanNumeralCaps}[1]

\usepackage{epstopdf, epsfig}
\usepackage{amsmath}
\usepackage{psfrag}
\usepackage{subfigure}
\usepackage{mathrsfs}
\usepackage{color}
\usepackage{overpic}
\usepackage{soul}
\usepackage[normalem]{ulem}
\usepackage{comment}

\newcommand{\x}{\bar{x}}

\def\pipe#1{\left.{#1}\right\vert}

\shorttitle{Nonlinear compressible G\"ortler vortices and streaks}
\shortauthor{D. Xu, P. Ricco and E. Marensi}

\title{Excitation and stability of nonlinear compressible G\"ortler vortices and streaks induced by free-stream vortical disturbances}

\author{Dongdong Xu, Pierre Ricco\corresp{\email{p.ricco@sheffield.ac.uk}}, Elena Marensi}

\affiliation{School of Mechanical, Aerospace and Civil Engineering, The University of Sheffield, Sheffield, S1 3JD, United Kingdom}

\begin{document}

\maketitle

\begin{abstract}
We study the generation, nonlinear development and secondary instability of unsteady G\"ortler vortices and streaks in compressible boundary layers exposed to free-stream vortical disturbances and evolving over concave, flat and convex walls. The formation and evolution of the disturbances are governed by the compressible nonlinear boundary-region equations, supplemented by initial and boundary conditions that characterise the impact of the free-stream disturbances on the boundary layer. Computations are performed for parameters typical of flows over high-pressure turbine blades, where the G\"ortler number, a measure of the curvature effects, and the disturbance Reynolds number, a measure of the nonlinear effects, are order-one quantities. At moderate intensities of the free-stream disturbances, increasing the G\"ortler number renders the boundary layer more unstable, while increasing the Mach number or the frequency stabilises the flow. As the free-stream disturbances become more intense, {\color{black}vortices over concave surfaces no longer develop into the characteristic mushroom-shaped structures,} while the flow over convex surfaces is destabilised. An occurrence map identifies G\"ortler vortices or streaks for different levels of free-stream disturbances and G\"ortler numbers. Our calculations capture well the experimental measurements of the enhanced skin friction and wall-heat transfer over turbine-blade pressure surfaces. The time-averaged wall-heat transfer modulations, termed hot fingers, are elongated in the streamwise direction and their spanwise wavelength is half of the characteristic wavelength of the free-stream disturbances. Nonlinearly saturated disturbances are unstable to secondary high-frequency modes, whose growth rate increases with the G\"ortler number. A new varicose even mode is reported, which may promote transition to turbulence at the stem of nonlinear streaks.
\end{abstract}

\section{Introduction}

G\"ortler instability originates in boundary layers over concave walls from an inviscid imbalance between pressure and centrifugal forces. The resulting boundary-layer disturbances are steady or low-frequency streamwise-elongated structures -- known as G\"ortler vortices -- which play a primary role in driving the laminar-to-turbulence transition in a wide range of industrial and technological applications. In high-speed flows, G\"ortler vortices are a major concern for the design of hypersonic vehicles, atmospheric re-entry capsules and jet engines, where the intensified wall-shear stresses and wall-heat transfer caused by these vortices pose a severe risk for surface thermal protection \citep{schneider1999flight,sun2017review}. G\"ortler vortices are also critical for the design of nozzles in high-speed wind tunnels because they rapidly promote transition to turbulence, which radiates aerodynamic noise that often prevents accurate measurements in the test section and, more seriously, renders the test condition drastically different from that of flight \citep{beckwith1973control,schneider-2008,schneider-2015}.

Of particular interest in our study is the influence of compressible G\"ortler vortices  on the efficiency of turbomachinery, such as high-pressure turbines, characterised by highly curved blade profiles and high levels of ambient disturbances. Despite the ubiquity of G\"ortler vortices in turbomachinery flows, we note that the literature on G\"ortler vortices does not often mention turbomachinery applications. At the same time, most studies on turbine blades recognise the presence of disturbed transitional flows, but only a few have paid attention to G\"ortler vortices. A clear conceptual link between studies on G\"ortler vortices and turbomachinery flows is therefore missing,  although effort and progress to connect the two have been made by \citet*{wu2011excitation} and \citet*{xu_zhang_wu_2017}. Furthermore, one of the key challenges in understanding transitional boundary layers populated by G\"ortler vortices is their extreme sensitivity to external disturbances, such as free-stream turbulence, whose intensity in turbomachinery flows can reach 20\%. The strong influence of external disturbances on G\"ortler instability needs to be accounted for via a receptivity formalism \citep{wu2011excitation,xu_zhang_wu_2017,marensi2017growth}.

In this work, we develop a rigorous mathematical and numerical framework to investigate the generation, nonlinear evolution and secondary stability of compressible G\"ortler vortices excited by free-stream vortical disturbances (FVD) for a range of parameters that are relevant to high-pressure turbine blades. We also study nonlinear compressible streaks evolving over flat surfaces, often called Klebanoff modes \citep*{ricco2007response,marensi2017nonlinear}, and elongated streaky structures appearing over convex surfaces. Receptivity to external vortical disturbances is central in our analysis as it allows linking our work to studies on turbomachinery flows. In \S1.1, we summarise theoretical studies of compressible G\"ortler vortices, including linear stability theory, initial-value theory and initial-boundary-value receptivity theory. Comprehensive reviews of incompressible G\"ortler instability were given by \citet{Hall1990}, \citet{floryan1991gortler} and \citet{Saric1994}. A recent review on theoretical, numerical and experimental studies of compressible G\"ortler vortices can be found in \citet*{xu2024review}. Flows over the pressure side of turbine blades are discussed in \S1.2. Further details on the scope of our study are given in \S1.3.

\subsection{Theoretical studies of compressible G\"ortler vortices}
Early studies on incompressible and compressible G\"ortler vortices neglected the spatial evolution of boundary layers and resorted to a local eigenmode approach by adopting the parallel mean-flow assumption. However, due to the growing nature of free-stream boundary-layer flows, in general, G\"ortler instability has to be formulated as an initial-value problem, as first rigorously demonstrated in the incompressible case by \citet{hall1983thelinear}. \citet{hall1983thelinear} realised that the non-parallel-flow terms cannot be neglected or included in an approximate manner in the study of G\"ortler instability in the case of order-one G\"ortler number and characteristic wavelength comparable to the boundary-layer thickness. The non-parallel-flow terms in the equations of motion are of leading order because the streamwise length scale of G\"ortler vortices is comparable to that of the base flow. \citet{hall1983thelinear} also showed that the asymptotic limit of large Reynolds number renders the Navier-Stokes equations parabolic along the streamwise direction, i.e.  the streamwise diffusion and the streamwise pressure gradient of the perturbations are negligible because they are asymptotically small. The parabolised equations are nowadays called the boundary-region equations \citep*{leib1999effect}, although this terminology was not used by \citet{hall1983thelinear}. The spanwise diffusion is retained because the spanwise wavelength of the disturbance is comparable to the boundary-layer thickness.  It should be noted that the initial-boundary-value formulation of \citet{leib1999effect} is the only theory that takes the external-disturbance receptivity into account. The  eigenvalue-problem formulation becomes tenable only when the G\"ortler number is asymptotically large \citep{Hall1982TaylorGortlerVI}.

\citet{hall1989growth} and \citet{hall1989gortler} studied compressible G\"ortler vortices with a wavelength smaller than the boundary-layer thickness under the assumptions of order-one and large Mach numbers, respectively. They concluded that compressibility has a stabilising effect on G\"ortler instability. A major difference between G\"ortler vortices in incompressible and compressible flows is the presence of the temperature adjustment layer  in the hypersonic limit of large Mach number \citep{hall1989gortler}. This layer is located at the edge of the boundary layer, where the temperature of the base flow changes rapidly to its free-stream value. In the limits of large Mach number  and large G\"ortler number, \citet{hall1989gortler} analysed G\"ortler vortices trapped in the adjustment layer by using an eigenvalue approach. The adjustment-layer mode grew the most and therefore the adjustment layer was deemed to be the most dangerous site for secondary instability \citep{fu1991effects}. \citet{dando1993compressible} and \citet{ren2014competition} studied the competition between the adjustment-layer mode and the conventional wall-layer mode and showed that the former becomes dominant in the hypersonic regime, but it is overtaken by the wall-layer mode for sufficiently large G\"ortler numbers.

The nonlinear interaction of disturbances in a boundary layer generates harmonics and a mean-flow distortion. Nonlinearity saturates the G\"ortler vortices when they acquire a significant amplitude. \citet{fu1991nonlinear} first studied the nonlinear development of G\"ortler vortices in the large Mach-number limit. \citet{bogolepov2001asymptotic} investigated the nonlinear evolution of long-wavelength G\"ortler vortices in hypersonic boundary layers and showed the effects of wall temperature.  The eigensolutions of the linear stability problem were used by \citet{ren2015secondary} to initiate the downstream computation of the nonlinear parabolised stability equations (this mathematical framework differs from the boundary-region approach, as amply discussed in \cite{xu2024review}).  It should be noted that the use of eigenfunctions as initial conditions is a common $ad$ $hoc$ practice and is only justified when the G\"ortler number is large. Mushroom-shaped structures of the streamwise velocity, common in flows dominated by G\"ortler vortices, were found to be replaced by bell-shaped structures during the initial flow evolution. \cite{ren2015secondary} ascribed this result to the dominance of the adjustment-layer mode.

\citet{viaro2018neutral,viaro2019neutral, viaro2019compressible} extended the receptivity theory of incompressible G\"ortler vortices by \citet{wu2011excitation} to the compressible regime and studied the neutral curves of G\"ortler instability excited by weak FVD. They tackled the receptivity problem by solving the linear compressible boundary-region equations complemented by initial and boundary conditions that synthesise the influence of physically realizable FVD. As opposed to the parabolised stability equations, where the streamwise diffusion and streamwise pressure-gradient terms are modelled by an $ad$ $hoc$ numerical procedure, the boundary-region equations are parabolic to leading-order accuracy as they are the rigorous asymptotic limit of the Navier-Stokes equations for low-frequency and long-wavelength perturbations, to which the boundary layer is most receptive.

\cite{marensi2017nonlinear} solved the nonlinear boundary-region equations to extend the work of \citet{ricco2007response} on linear compressible streaks to take into account nonlinear effects. \citet{sescu2020streaks} focused on the nonlinear evolution of steady G\"ortler vortices excited by FVD and computed the wall-shear stress and the wall-heat transfer for Mach numbers varying from 0.8 to 6.

\subsection{Flows over high-pressure turbine blades}
High-pressure turbine blades are subject to extreme inlet conditions, including high levels of temperature, pressure and unsteadiness of the oncoming turbulence, rendering these flows extremely difficult to measure experimentally and to simulate numerically \citep{mayle19911991,zhao2020bypass}. Additional difficulties arise from the strong blade curvature and the effects of wall temperature and pressure gradients. Due to these complexities, most experiments and simulations have been conducted in incompressible flow conditions \citep{radomsky2002detailed,varty2016experimental,morata2012effects,kanani2019large,dhurovic2021free,lengani2022receptivity}.
\cite*{arts1990aero} carried out unique experimental measurements in a compressible wind tunnel and reported data of quantities at the wall. Further boundary-layer measurements, such as those by \citet{radomsky2002detailed}, are still needed for realistic turbomachinery flow conditions. In a few studies, compressible-flow simulations have been performed \citep{bhaskaran2010large,wheeler2016direct,zhao2020bypass}, but a systematic parameter study has not been carried out due to computational limitations.

According to \cite*{gourdain2012comparison}, streamwise vortices are excited in boundary layers over the pressure and suction surfaces of turbine blades. These vortices impact the wall-shear stress and the wall-heat transfer, but their prediction is challenging due to the multitude of factors mentioned earlier. In particular, the influence of the blade curvature on the excitation and evolution of the induced vortices remains obscure. Previous studies have suggested that centrifugal forces could trigger vortices on the pressure surface, as evidenced by the detection of typical G\"ortler-vortex structures, such as mushrooms and wall `hot fingers' (elongated regions of high wall-heat transfer), as reported by \citet{gourdain2012comparison} and \citet{baughn1995experimental}, respectively. However, recent direct numerical simulations have revealed that the concave curvature of the blade is not the sole cause of these vortices, as they also appear in the leading-edge region of both suction and pressure surfaces \citep{wheeler2016direct, zhao2020bypass}. Furthermore, the effect of curvature was not detected in simulations and experiments with elevated free-stream turbulence levels as G\"ortler vortices with the typical mushroom-shaped structure were not observed \citep{wheeler2016direct,zhao2020bypass,arts1990aero}. \citet{dhurovic2021free} numerically identiﬁed the appearance of longitudinal vortical structures on the pressure side of low-pressure turbine blades, but ruled out the possibility that these structures were produced by G\"ortler instability. In their incompressible receptivity study, \cite{xu_zhang_wu_2017} found that, under high-intensity FVD, G\"ortler vortices took on the character of streaks, also known as Klebanoff modes, disturbances typically observed in boundary-layer flows over flat plates \citep{ricco2007response,marensi2017nonlinear}.

Despite these research endeavours, a full characterisation of the nature of these structures -- G\"ortler vortices or streaks -- in the compressible regime is unavailable. Most importantly, previous incompressible studies, such as those mentioned earlier, can neither predict the temperature field in the boundary layer nor capture typical compressible-flow structures, such as the hot fingers. Understanding the formation of these structures is crucial as it informs the design of cooling techniques to protect the blade surface \citep{wright2014review}.

\subsection{Scope of our study}
Our objective is to study the receptivity, nonlinear evolution and secondary instability of FVD-induced G\"ortler vortices and streaks in compressible boundary layers. A direct application of our investigation is the dynamics of boundary layers that are typically observed over the pressure and suction surfaces of high-pressure turbines. Our study is based on the earlier investigations of \citet{marensi2017nonlinear} and \citet{viaro2019compressible} and it can be viewed as an extension of the former to include centrifugal effects and a generalisation of the latter to the nonlinear case (the reader is refereed to table 2 of \cite{xu2024review} for an overview of boundary-region receptivity studies). The present work is also an extension of \citet{xu_zhang_wu_2017} to the compressible regime. The flow parameters are chosen as representative of common turbomachinery flows, in particular with reference to the unique compressible experiments of \cite{arts1990aero}.

We focus on unsteady disturbances because they are likely to be present in boundary layers exposed to high free-stream turbulence environments, such as those over turbine blades \citep{schultz2003effects}. A systematic investigation of the effects of Mach number, wall curvature and FVD intensity on the nonlinear development of G\"ortler vortices has been carried out, thus uncovering the intricate interplay between these factors in realistic turbomachinery conditions. The unexplained absence of G\"ortler vortices in flows over turbine blades is elucidated by studying the competition between wall curvature and FVD intensity, thus providing a novel link between G\"ortler vortices and turbomachinery flow systems. Comparisons with experimental measurements are also presented, showing the key role of the mean-flow distortion in the nonlinear generation of hot fingers over pressure surfaces. Finally, a secondary-instability analysis of the nonlinearly saturated disturbances has revealed the occurrence of a new varicose mode, never reported in previous studies,  which may promote transition to turbulence at the stem of streaks.

A limitation of our fundamental analysis is the absence of a pressure gradient, which may impact the flows on both surfaces of a turbine blade and, in particular, induce boundary-layer separation over the suction surface \citep*{NAGARAJAN_LELE_FERZIGER_2007}. Furthermore, leading-edge bluntness, also absent in the present work, can influence the receptivity of the base flow and the evolution of boundary-layer disturbances through the induced streamwise pressure gradient and by distorting the flow around the stagnation point \citep{xu2020gortler,NAGARAJAN_LELE_FERZIGER_2007}. Inclusion of these effects in our future work is discussed in the concluding remarks (\S\ref{sec:conclusions}).

\section{Mathematical framework}
\label{sec:pro-form}

\begin{figure}
\centering
\includegraphics[width=1\textwidth]{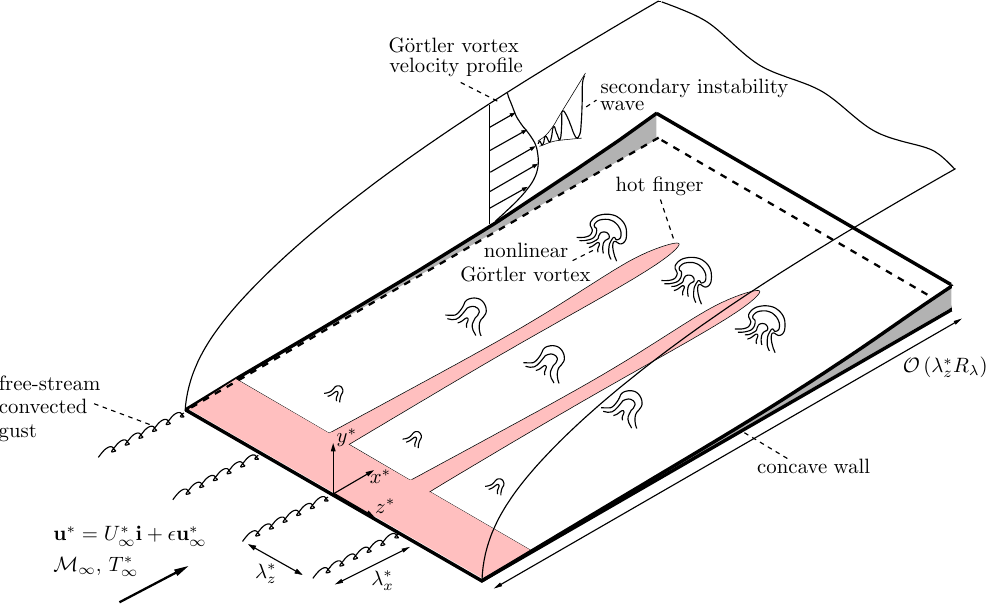}
\caption{Schematic of the physical domain for the concave-wall case. The sketches of the G\"ortler vortices and the hot fingers are simply illustrative and do not represent their actual relative positions. The dynamics between the G\"ortler vortices and the hot fingers is discussed in \S\ref{sec:cf}.}
\label{fig:domain}
\end{figure}

We consider compressible boundary layers flowing over concave, flat and convex surfaces. The radius of curvature of the surface is denoted by $r^*_0.$ Hereafter, the superscript $^*$ indicates dimensional quantities. Figure \ref{fig:domain} shows a schematic of the flow domain in the concave-wall case. The oncoming base flow is uniform with free-stream velocity $U_\infty^*$ and temperature $T_\infty^*,$ superimposed on which are unsteady free-stream disturbances. Although free-stream turbulence is of broadband nature, as in \cite{marensi2017nonlinear} we consider the simplified case of FVD consisting of a pair of vortical modes with the same frequency (and hence streamwise wavenumber), but opposite spanwise wavenumbers $\pm k_z^*$. As streamwise-elongated vortices in a boundary layer typically exhibit a well-defined spanwise spacing $\Lambda^*$, it is reasonable to study vortices that are excited by a pair of dominant oblique FVD components.

The flow is described in an orthogonal curvilinear coordinate system, $\boldsymbol{x}^*=\left\{x^*,y^*,z^*\right\}$, that defines the streamwise, wall-normal and spanwise directions. The conversion from the Cartesian to the curvilinear coordinate system is achieved through the  Lam\'e coefficients $\left\{h_x,h_y,h_z\right\}=\left\{(r_0^*-y^*)/r_0^*,1,1\right\}$ \citep{wu2011excitation, viaro2019compressible}. Lengths are normalised using the length scale $\Lambda^*=1/k_z^*$, while $U_\infty^*$ and $T_\infty^*$ are the velocity and temperature scales. The fluid properties, such as the density $\rho^*$, the dynamic viscosity $\mu^*$ and the thermal conductivity $\upkappa^*$, are scaled by their respective constant free-stream values, $\rho_\infty^*$, $\mu^*_\infty$ and $\upkappa^*_\infty$. The time $t^*$ and the pressure $p^*$ are non-dimensionalised by $\Lambda^*/U_\infty^*$ and $\rho^*_\infty U^{*2}_\infty$, respectively. The free-stream disturbance $\boldsymbol{u}_\infty$ is expressed as
\begin{eqnarray}
\boldsymbol{u}-{\rm \bf{i}}= \epsilon \boldsymbol{u}_\infty(x-t, y,z) = \epsilon \left(\boldsymbol{\hat u}_+^\infty {\rm e}^{{\rm i} k_z z} +\boldsymbol{\hat u}_-^\infty {\rm e}^{-{\rm i} k_z z}\right) {\rm e}^{{\rm i} k_x (x-t) + {\rm i} k_y y}+{\rm c.c.},
\label{eq:gust}
\end{eqnarray}
where $\epsilon\ll 1$ is a measure of the disturbance intensity, $\rm \bf{i}$ is the unit vector along the streamwise direction and $\rm c.c.$ indicates the complex conjugate. The gust disturbance \eqref{eq:gust} is passively advected by the free-stream base flow, i.e.  the phase velocity is $U_\infty^*$ because the disturbance is of small amplitude and specified at small $x$ distances, where viscous effects play a secondary role, and at large $y$ distances, where the displacement effect induced by the boundary layer is negligible.
The vector $\boldsymbol{\hat u}_{\pm}^\infty=\left\{{\hat u}_{x, \pm}^\infty,{\hat u}_{y, \pm}^\infty,{\hat u}_{z, \pm}^\infty\right\}=O(1)$ satisfies the solenoidal condition
\begin{equation}
k_x \hat{u}_{x,\pm}^{\infty} + k_y \hat{u}_{y,\pm}^{\infty} \pm k_z \hat{u}_{z,\pm}^{\infty}=0.
\label{eqn:fsvd-continuity}
\end{equation}
The Reynolds number $R_\Lambda$ is defined as
\begin{equation}
R_\Lambda=\dfrac{\rho_\infty^* U_\infty^* \Lambda^*} {\mu_\infty^*}
\end{equation}
\noindent
and is taken to be asymptotically large, i.e.  $R_\Lambda$$\gg$1. The scaled wavenumbers $\kappa_y=k_y/\sqrt{k_xR_\Lambda}$$=$$\mathcal{O}(1)$ and $\kappa_z=k_z/\sqrt{k_xR_\Lambda}$$=$$\mathcal{O}(1)$ are also defined. To account for centrifugal effects, a G\"ortler number is introduced,
\begin{eqnarray}
  \mathcal{G}=\dfrac{R_\Lambda^{1/2}\Lambda^* }{k_x^{3/2} r_0^*}=\mathcal{O}(1).
\label{eq:gortler}
\end{eqnarray}
In the present study only unsteady disturbances ($k_x \neq 0$) are considered and therefore the G\"ortler number is well defined. The G\"ortler number $G_\Lambda$ defined in \citet{viaro2019compressible} is related to $\mathcal{G}$ by $\mathcal{G}=(\kappa_z/k_z)^3G_\Lambda.$ Note that $\mathcal{G}=\mathcal{O}(1)$ only if $\kappa_z=\mathcal{O}(1),$ which is the case in the present analysis. As a measure of nonlinear effects, we introduce the disturbance Reynolds number $r_t=\epsilon R_{\Lambda}= \mathcal{O}(1)$,  as in \citet{leib1999effect} and \citet{ricco2011evolution}. The oncoming flow is isentropic and air is assumed to be a perfect gas. The free-stream Mach number is defined as $\mathcal{M}_\infty=U_\infty^*/a_\infty^*=\mathcal{O}(1)$, where $a_\infty^*=(\gamma R^* T_\infty^*)^{1/2}$ is the speed of sound in the free stream, $R^*=287.06$ $\rm J$ $\rm kg^{-1}$ $\rm K^{-1}$ is the ideal gas constant for air and $\gamma=1.4$ is the ratio of the specific heat capacities.

We focus on low-frequency, long-streamwise-wavelength free-stream disturbances ($k_x\ll1$) because boundary layers are most receptive to these perturbations. Experimental evidence has shown that low-frequency disturbances are those that amplify the most inside wall-bounded shear layers \citep{matsubara-alfredsson-2001}. The plate is thin and the Mach number is moderate so that shocks are assumed to be weak and distant from the boundary layer. The effects of shocks on the free-stream perturbations and the boundary layer are therefore neglected. The reader is referred to \citet{qin2016response} for the response of a flat-plate hypersonic boundary layer to free-stream acoustic, vortical and entropy disturbances downstream of a shock.

The flow domain is divided into four asymptotic regions, described in \citet{viaro2019compressible}. The region of interest is region III, where the spanwise and wall-normal viscous effects are comparable and the streamwise coordinate is scaled with the streamwise wavenumber of the free-stream disturbance, i.e.  $\bar x=k_x x=\mathcal{O}(1)$. The distinguished relationship $k_x$$=$$\mathcal{O}\left(R_\Lambda^{-1}\right)$ emerges from the asymptotic balance and the slow time variable $\bar t=k_x t=\mathcal{O}(1)$ is defined. The streamwise velocity is larger than the wall-normal and spanwise velocities by a factor $\mathcal{O}(R_\Lambda)$ and larger than the pressure by a factor $\mathcal{O}\left(R_\Lambda^2\right)$. The velocity, pressure and temperature variables are rescaled as
\begin{eqnarray}
\left\{u^*,v^*,w^*\right\}/U_\infty^*=
\left\{\Tilde u,\sqrt{k_x/R_\Lambda} \Tilde  v,k_x \Tilde w\right\}, \;\; p^*/\left(\rho^*_\infty U_\infty^{*2}\right)=k_x R_\Lambda^{-1} \Tilde p,\;\;
 T^*/T_\infty^*=\Tilde T.
\label{eqn:u-p}
\end{eqnarray}
By substituting expression \eqref{eqn:u-p} into the compressible Navier-Stokes equations written in curvilinear coordinates and by performing the change of variable $(x, t) \to (\bar{x}, \bar{t})$, we obtain the following leading-order nonlinear boundary-region equations:

\begin{equation}
\label{eqn:nubr-ta}
\dfrac{\p \Tilde \rho}{\p \bar t}
+\dfrac{\p \Tilde \rho \Tilde u}{\p \bar x}
+\dfrac{\kappa_z}{k_z}\dfrac{\p \Tilde \rho \Tilde v}{\p y}
+\dfrac{\p \Tilde \rho \Tilde w}{\p z}=0,
\end{equation}
\begin{equation}
\Tilde \rho \dfrac{\p \Tilde u}{\p \bar t}
+\Tilde \rho \Tilde u\dfrac{\p \Tilde u}{\p \bar x}
+\Tilde \rho \Tilde v\dfrac{\kappa_z}{k_z}\dfrac{\p \Tilde u}{\p y}
+\Tilde \rho \Tilde w\dfrac{\p \Tilde u}{\p z}=
\dfrac{\kappa_z^2}{k_z^2}
\left[
\dfrac{\p }{\p y}
\left(
\Tilde \mu\dfrac{\p \Tilde u}{\p y}
\right)
+\dfrac{\p }{\p z}
\left(
\Tilde \mu\dfrac{\p \Tilde u}{\p z}
\right)
\right],
\end{equation}
\begin{equation}
\begin{split}
&
\Tilde \rho\dfrac{\p \Tilde v}{\p \bar t}
+\Tilde \rho \Tilde u\dfrac{\p \Tilde v}{\p \bar x}
+\Tilde \rho \Tilde v\dfrac{\kappa_z}{k_z}\dfrac{\p \Tilde v}{\p y}
+\Tilde \rho \Tilde w\dfrac{\p \Tilde v}{\p z}+\mathcal{G}\Tilde u^2
=
\\ &
\dfrac{\kappa_z}{k_z}
\left\{ -\dfrac{\p \Tilde p}{\p y}
+
\dfrac{\p }{\p y}
\left[
\dfrac{2}{3}\Tilde \mu
\left(
\dfrac{2\kappa_z}{k_z}\dfrac{\p \Tilde v}{\p y}
-\dfrac{\p \Tilde w}{\p z}
\right)
\right]
+
\dfrac{\p }{\p z}
\left[
\Tilde \mu
\left(
\dfrac{\kappa_z}{k_z}\dfrac{\p \Tilde v}{\p z}
+\dfrac{\p \Tilde w}{\p y}
\right)
\right]
-
\dfrac{\p }{\p y}
\left(
\dfrac{2}{3}\Tilde \mu
\dfrac{\p \Tilde u}{\p \bar x}
\right)
+\dfrac{\p }{\p \bar x}
\left(
\Tilde \mu\dfrac{\p \Tilde u}{\p y}
\right)
\right\},
\end{split}
\end{equation}
\begin{equation}
\begin{split}
&
\Tilde \rho\dfrac{\p \Tilde w}{\p \bar t}
+\Tilde \rho \Tilde u\dfrac{\p \Tilde w}{\p \bar x}
+\Tilde \rho \Tilde v\dfrac{\kappa_z}{k_z}\dfrac{\p \Tilde w}{\p y}
+\Tilde \rho \Tilde w\dfrac{\p \Tilde w}{\p z}
=
\\ &
\dfrac{\kappa_z^2}{k_z^2}
\left\{ -\dfrac{\p \Tilde p}{\p z}
+
\dfrac{\p }{\p z}
\left[
\dfrac{2}{3}\Tilde \mu
\left(
2\dfrac{\p \Tilde w}{\p z}
-\dfrac{\kappa_z}{k_z}\dfrac{\p \Tilde v}{\p y}
\right)
\right]
+
\dfrac{\p }{\p y}
\left[
\Tilde \mu
\left(
\dfrac{\kappa_z}{k_z}\dfrac{\p \Tilde v}{\p z}
+\dfrac{\p \Tilde w}{\p y}
\right)
\right]
-
\dfrac{\p }{\p z}
\left(
\dfrac{2}{3}\Tilde \mu
\dfrac{\p \Tilde u}{\p \bar x}
\right)
+
\dfrac{\p }{\p \bar x}
\left(
\Tilde \mu\dfrac{\p \Tilde u}{\p z}
\right)
\right\},
\end{split}
\end{equation}
\begin{equation}
\begin{split}
&
\Tilde \rho\dfrac{\p \Tilde T}{\p \bar t}
+\Tilde \rho \Tilde u\dfrac{\p \Tilde T}{\p \bar x}
+\Tilde \rho \Tilde v\dfrac{\kappa_z}{k_z}\dfrac{\p \Tilde T}{\p y}
+\Tilde \rho \Tilde w\dfrac{\p \Tilde T}{\p z}
=
\\ &
\dfrac{\kappa_z^2}{k_z^2}
\left\{
\dfrac{1}{Pr}
\left[
\dfrac{\p }{\p y}
\left(
\Tilde \mu\dfrac{\p \Tilde T}{\p y}
\right)
+\dfrac{\p }{\p z}
\left(
\Tilde \mu\dfrac{\p \Tilde T}{\p z}
\right)
\right]
+
(\gamma-1)\mathcal{M}_\infty^2\Tilde \mu
\left[
\left(
\dfrac{\p \Tilde u}{\p y}
\right)^2
+
\left(
\dfrac{\p \Tilde u}{\p z}
\right)^2
\right]
\right\}.
\label{eqn:nubr-te}
\end{split}
\end{equation}

The flow is decomposed as the sum of the compressible Blasius flow and the perturbation flow induced by the FVD, namely
\begin{equation}
\left\{\Tilde u,\Tilde  v,\Tilde w,\Tilde p,\Tilde T \right\}
=\left\{U,V,0,\dfrac{1}{\gamma \mathcal{M}_\infty},T \right\}
+r_t \left\{ \bar u,\bar v,\bar w,\bar p,\bar \tau \right\}\left(\bar x, \eta,z,\bar t \right),
\label{eqn:field-all}
\end{equation}
where $\left\{U,V\right\}$=$\left\{F'(\eta), T(\eta_c F'-F)/\sqrt{2\bar x}\right\}$, $T$=$T(\eta)$,
\begin{eqnarray}
\eta=\sqrt{\dfrac{R_\Lambda}{2 x}}\int_0^y\rho(\bar x,\check y) {\rm d}{\check y}, \quad \eta_c=\frac{1}{T}\int_0^{\eta} T(\check \eta){\rm d}{\check \eta},
\end{eqnarray}
and $\rho=T^{-1}$. The prime denotes differentiation with respect to the similarity variable $\eta$. The compressible Blasius functions $F(\eta)$ and $T(\eta)$ are solutions to the boundary-value problem,
\begin{equation}
\left.\begin{array}{c}
(\mu F''/T)'+FF'' =0, \\
( \mu T'/T )'+PrFT'+ \mu(\gamma -1)Pr \mathcal{M}^2_\infty (F'')^2/T=0, \\
F=F'=0, \quad T=T_w, \quad {\rm at } \quad \eta=0, \\
F'\rightarrow1, \quad T'=0,\quad {\rm as } \quad \eta \rightarrow\infty,
\end{array}\right\}
\label{eqn:com-blasius}
\end{equation}
where the Prandtl number $Pr$ is assumed constant, $Pr=0.707$, the dynamic viscosity is $\mu(T)=T^{\omega}$ with $\omega=0.76$ \citep{stewartson1964theory} and the thermal conductivity is $\upkappa=\mu$.  Curvature effects are negligible at leading order in system \eqref{eqn:com-blasius} because of the assumptions $R_\Lambda \gg 1$ and $r_0 \gg 1$ \citep{hall1983thelinear}.

The density is decomposed as $\Tilde \rho=T^{-1}+r_t \bar \rho$, where, using the equation of state for a perfect gas, $\bar \rho=-\bar \tau/T^2-r_t \bar \rho \bar \tau/T+\mathcal{O}\left(k_xR_\Lambda^{-1}\right)$.
The viscosity is expressed as $\Tilde \mu = \left(T+ r_t \tau\right)^{\omega}$ and expanded using the binomial formula as in equation (2.21) of \cite{marensi2017nonlinear}.

The boundary-layer disturbance consists of all temporal and spanwise harmonics
\begin{eqnarray}
\left\{\bar u ,\bar v,\bar w,\bar p, \bar \tau\right\}
=
\sum^{\infty}_{m,n=-\infty}
\Big{\{}\hat u_{m,n}(\bar x ,\eta ), \sqrt{2\bar x} \hat v_{m,n}(\bar x,\eta ), k_z^{-1} \hat w _{m,n}(\bar x,\eta ),
  \qquad
  \qquad
  \qquad
  \nonumber
  \\
\hat p_{m,n}(\bar x,\eta ),\hat \tau_{m,n}(\bar x,\eta ) \Big{\}}
\, {\rm e}^{\textrm{i}m \bar t+\textrm{i}n k_z z}.
\quad
\label{eqn:fourier}
\end{eqnarray}
As the physical quantities are real, the Fourier coefficients are Hermitian, $\hat q_{-m,-n}=(\hat q_{m,n})_{\rm cc},$ where $\hat q$ stands for any of $\{\hat u,\hat v,\hat w,\hat p,\hat \tau\}.$ Inserting expressions \eqref{eqn:field-all} and \eqref{eqn:fourier} into the nonlinear boundary-region equations \eqref{eqn:nubr-ta}-\eqref{eqn:nubr-te} yields the governing equations for the disturbance Fourier coefficients.

\noindent The continuity equation
\begin{eqnarray}
&&\dfrac{\eta_c}{2\bar x} \dfrac{T'}{T}\hat u_{m,n}
+\dfrac{\p \hat u_{m,n}}{\p \bar x}
-\dfrac{\eta_c}{2\bar x} \dfrac{\p \hat u_{m,n}}{\p \eta}
-\dfrac{T'}{T^2}\hat v_{m,n}
+\dfrac{1}{T}\dfrac {\p \hat v_{m,n}}{\p \eta}
+{\rm i}n\hat w_{m,n} \nonumber \\
&&-\left(\dfrac{{\rm i}m}{T}
+\dfrac{1}{2 \bar x}\dfrac{FT'}{T^2}\right)\hat \tau_{m,n}
- \dfrac{F'}{T}\dfrac{\p \hat \tau_{m,n}}{\p \bar x}
+\dfrac{1}{2\bar x}\dfrac{F}{T}\dfrac{\p \hat \tau_{m,n}}{ \p \eta}
=r_t\mathcal{\hat C}_{mn}.
\label{eqn:nubr-con}
\end{eqnarray}
The $x$-momentum equation
\begin{eqnarray}
\label{eq:LUBR-X}
&&
\left({\rm i}m-\dfrac{\eta_c}{2\bar x}F''+n^2\kappa_z^2\mu T\right)\hat u_{m,n}
+F'\dfrac{\p \hat u_{m,n}}{\p \bar x}
-\dfrac{1}{2\bar x} \left(F+\dfrac{\mu'T'}{T}-\dfrac{\mu T'}{T^2}\right)\dfrac{\p \hat u_{m,n}}{\p \eta} \nonumber\\
&&-\dfrac{1}{2\bar x}\dfrac{\mu}{T}\dfrac{\p^2 \hat u_{m,n}}{\p \eta^2}
+\dfrac{F''}{T}\hat v_{m,n}
+\dfrac{1}{2\bar x T}\left(FF''-\mu''F''T'
+\dfrac{\mu'F''T'}{T}-\mu'F'''\right)\hat\tau_{m,n}
\nonumber\\
&&-\dfrac{1}{2 \bar x}\dfrac{\mu'F''}{T}\dfrac{\p \hat \tau_{m,n}}{\p \eta}=r_t\mathcal{\hat X}_{mn}.
\end{eqnarray}
The $y$-momentum equation
\begin{eqnarray}
\label{eq:LUBR-Y}
  &&
  \dfrac{1}{4 \x^2} \left[ \eta_c \left( F T^\prime
      - F^\prime T \right)
    - \eta_c^2 F^{\prime \prime} T
    + F T
  \right] \hat{u}_{m,n}
  + \dfrac{\mu^\prime T^\prime}{3 \x}
  \dfrac{\p \hat{u}_{m,n}}{\p \x}
  - \dfrac{\mu}{6 \x}
  \dfrac{\p^2 \hat{u}_{m,n}}{\p \x \p{\eta}}\nonumber \\
  &&
  +\dfrac{\eta_c \mu}{12 \x^2} \dfrac{\p^2 \hat{u}_{m,n}}{\p \eta^2}+
  \nonumber
   \dfrac{1}{12 \x^2} \left( \eta_c \mu^\prime T^\prime
    + \mu
    -  \dfrac{\eta_c \mu T^\prime}{T}
  \right) \dfrac{\p \hat{u}_{m,n}}{\p \eta}\\
  &&
  + \left[ \dfrac{1}{2 \x}
    \left(
      F^\prime
      + \eta_c F^{\prime \prime}
      - \dfrac{ F T^\prime}{ T}
    \right)
    +{\rm i} m
    + n^2 \kappa_z^2 \mu T  \right] \hat{v}_{m,n}
  \nonumber \\
  &&
  +F^\prime  \dfrac{\p \hat{v}_{m,n}}{\p \x}
  + \dfrac{1}{\x}
  \left[
    \dfrac{2}{3 T}
    \left(
      \dfrac{\mu T^\prime}{T}
      - \mu^\prime T^\prime
    \right)
    - \dfrac{F}{2}
  \right]
  \dfrac{\p \hat{v}_{m,n}}{\p \eta}
  - \dfrac{2}{3 \x} \dfrac{\mu}{T}
  \dfrac{\p^2 \hat{v}_{m,n}}{\p \eta^2}
  + {\rm i}n\dfrac{\mu^\prime T^\prime}{3 \x} \hat{w}_{m,n} \nonumber\\
&&
- {\rm i}n\dfrac{\mu}{6 \x} \dfrac{\p \hat{w}_{m,n}}{\p \eta}
  + \dfrac{1}{2 \x}\dfrac{\p \hat{p}_{m,n}}{\p \eta}
  \nonumber \\[0.2cm]
  &&
  +\bigg[
  \dfrac{1}{3 \x^2 T}
  \bigg(
  \mu^{\prime \prime} F {T^{\prime}}^2
  -  \dfrac{\mu^{\prime} F {T^{\prime}}^2}{T}
  + \mu^{\prime} F T^{\prime \prime}
  + \mu^\prime F^\prime T^\prime
  \bigg)
  -\dfrac{1}{4 \x^2}
  \bigg(
  F^\prime F
  - \eta_c {F^{\prime}}^2
  - \eta_c F F^{\prime \prime}
  \nonumber \\
  &&
  +\dfrac{F^2 T^\prime}{T}
  + \mu^{\prime} F^{\prime \prime}
  + \eta_c \mu^{\prime \prime} F^{\prime \prime}   T^\prime
  - \dfrac{\eta_c  \mu^\prime F^{\prime \prime} T^\prime}{T}
  + \eta_c F^{\prime \prime \prime} \mu^\prime
  \bigg)
  \bigg] \hat{\tau}_{m,n}
  + \dfrac{\mu^\prime}{\x^2}
  \left(
    \dfrac{F T^\prime }{3  T}
    - \dfrac{\eta_c F^{\prime \prime}}{4}
  \right)
  \dfrac{\p \hat{\tau}_{m,n}}{\p \eta}
  \nonumber \\
  &&  - \dfrac{\mu^\prime F^{\prime \prime}}{2 \x}
  \dfrac{\p \hat{\tau}_{m,n}}{\p \x}
  + \dfrac{\mathcal{G}}{\sqrt{2 \x}} \left(
      2 F^\prime \hat{u}_{m,n}
      -\dfrac{{F^{\prime}}^2}{T} \hat{\tau}_{m,n} \right)\nonumber \\
  &&=r_t \left[\mathcal{\hat Y}_{mn}  -\dfrac{\mathcal{G}}{\sqrt{2 \x}} \left(2F'T\widehat{\bar \rho \bar u}+\widehat{\bar u \bar u}+r_tT\widehat{\bar \rho \bar u\bar u}\right)
-F'^2\widehat{\bar \rho \bar \tau}\right].
\end{eqnarray}
The $z$-momentum equation
\begin{eqnarray}
\label{eq:LUBR-zMOM}
  &&  \dfrac{{\rm i}n\kappa_z^2 \eta_c \mu^\prime T T^\prime}{2 \x} \hat{u}_{m,n}
  -  \dfrac{{\rm i}n\kappa_z^2 \mu T}{3}  \dfrac{\p \hat{u}_{m,n}}{\p \x}
  +  \dfrac{{\rm i}n\kappa_z^2 \eta_c \mu T}{6 \x}
  \dfrac{\p \hat{u}_{m,n}}{\p \eta}
  - {\rm i}n\kappa_z^2 \mu^\prime T^\prime \hat{v}_{m,n}
  -  \dfrac{ {\rm i}n\kappa_z^2 \mu}{3} \dfrac{\p \hat{v}_{m,n}}{\p \eta}
  \nonumber\\
  && +\left(
    \dfrac{4}{3} n^2 \kappa_z^2 \mu T
    +{\rm i} m
  \right)
  \hat {w}_{m,n}
  + F^\prime \dfrac{\p \hat{w}_{m,n}}{\p \x}
  + \dfrac{1}{2 \x}
  \left(
     \dfrac{\mu T^\prime}{T^2}
    - F
    - \dfrac{\mu^\prime T^\prime}{T}
  \right)
  \dfrac{\p \hat{w}_{m,n}}{\p \eta}
  - \dfrac{1}{2 \x} \dfrac{\mu}{T}
  \dfrac{\p^2 \hat{w}_{m,n}}{\p \eta^2}
  \nonumber\\ 
 & & +{\rm i}n \kappa_z^2  T \hat{p}_{m,n}
  - \dfrac{ {\rm i}n \kappa_z^2}{3 \x} \mu^\prime F T^\prime \hat{\tau}_{m,n} =r_t\mathcal{\hat Z}_{mn}.
\end{eqnarray}
The energy equation
\begin{eqnarray}
  &&-  \dfrac{\eta_c}{2 \x} T^\prime \hat{u}_{m,n}
  + \dfrac{T^\prime}{T} \hat {v}_{m,n}
  + \left[
    \dfrac{F T^\prime }{2 \x T}
    + {\rm i} m
    +  \dfrac{n^2\kappa_z^2 \mu T}{Pr}
    - \dfrac{1}{2 \x Pr}
    \left(
      \dfrac{\mu^\prime T^\prime}{T}
    \right)'
  \right]
    \hat{\tau}_{m,n}
  + F^\prime \dfrac{\p \hat{\tau}_{m,n}}{\p \x}
  \nonumber\\[0.2cm]
 & & +\dfrac{1}{2 \x}
  \left(
    \dfrac{\mu T^\prime}{Pr T^2}
    - F
    - \dfrac{2 \mu^\prime T^\prime}{Pr T}
  \right)
  \dfrac{\p \hat{\tau}_{m,n}}{\p \eta}
  - \dfrac{1}{ 2 \x Pr}
  \dfrac{\mu}{T} \dfrac{\p^2 \hat{\tau}_{m,n}}{\p \eta^2}
  \nonumber
 \\
  && - \mathcal{M}_\infty^2 \dfrac{\gamma - 1}{\x T}
  \left(
    \mu F^{\prime \prime}
    \dfrac{\p \hat{u}_{m,n}}{\p \eta}
    +\dfrac{ \mu^\prime {F^{\prime \prime}}^2}{2} \hat{\tau}_{m,n}
  \right) = r_t \mathcal{\hat E}_{mn}.
   \label{eqn:nubr-e}
\end{eqnarray}
where  $\mu'={\rm d}\mu /{\rm d} T$ and the nonlinear terms $\mathcal{\hat C}_{mn},\mathcal{\hat X}_{ mn}, \mathcal{\hat Y}_{ mn}, \mathcal{\hat Z}_{mn},\mathcal{\hat E}_{mn}$ are given in equations (A1)-(A5) of \citet{marensi2017nonlinear}. The nonlinear terms collected on the right-hand sides of equations of \eqref{eqn:nubr-con}-\eqref{eqn:nubr-e} vanish as $r_t\rightarrow 0$ and the linearised boundary-region equations of \citet{viaro2019compressible} are recovered.

In the boundary layer, the velocity and temperature fluctuations induced near the leading edge are of small amplitude, and thus evolve linearly in this region. Curvature effects near the leading edge are also negligible and therefore the initial conditions for the forced modes $(m, n)= (1, \pm 1)$ are the same as those in the linear flat-plate case \citep{ricco2007response}. The initial conditions are given in Appendix \ref{app:in}. Matching the boundary-region solution with the outer solution gives the outer boundary conditions
\begin{equation}
\left\{\hat{u}_{m,n},\hat{v}_{m,n},\hat{w}_{m,n},\hat{p}_{m,n},\hat{\tau}_{m,n}\right\}
\rightarrow \left\{0,
\dfrac{\kappa_z}{\sqrt{2\bar{x}}}v^{\dagger}_{m,n}, \kappa_z^2
w^{\dagger}_{m,n}, \dfrac{\epsilon}{k_x} p^{\dagger}_{m,n}, 0
\right\}  \ \ \text{as} \ \ \eta \rightarrow \infty,\label{eqn:BC}
\end{equation}
{where $v^{\dagger}_{m,n}, w^{\dagger}_{m,n}, p^{\dagger}_{m,n}$ are given by equations (2.76) in \citet{marensi2017nonlinear}. The initial-boundary-value problem, consisting of equations \eqref{eqn:nubr-con}-\eqref{eqn:nubr-e}, \eqref{eq:BC-x0_u}-\eqref{eq:BC-x0_tau} and \eqref{eqn:BC}, governs the excitation and nonlinear evolution of G\"ortler vortices in the presence of FVD for $r_t=\mathcal{O}(1)$, $\mathcal{G}=\mathcal{O}(1)$ and $\mathcal{M}_{\infty}=\mathcal{O}(1).$

\subsection{Secondary instability}

The velocity and temperature profiles altered by nonlinearity are sensitive to high-frequency secondary disturbances as they exhibit inflection points in the transverse and spanwise directions during certain phases of the oscillations. These high-frequency secondary disturbances  amplify} and ultimately cause transition to turbulence in boundary layers over the pressure surface of turbine blades \citep{butler2001effect} and in wind-tunnel experiments \citep{ghorbanian2011experimental}. A secondary instability analysis of the boundary-layer flow perturbed by nonlinear disturbances is therefore carried out to elucidate the transition process.

The flow $q$ is decomposed into a base flow $\tilde q(y,z;\overline{x},\overline{t})$, given by \eqref{eqn:field-all}, and a secondary perturbation flow $q'_s(x,y,z,t)$, namely
\begin{equation}
    q(y,z;x,t)=\tilde q + \epsilon_s q'_s=\tilde q+ \epsilon_s\left\{\rho'_s,u'_s,v'_s,w'_s,T'_s\right\}(x,y,z,t),
\label{eq:sec-inst-exp}
\end{equation}
where $\epsilon_s \ll 1$. Substituting expression \eqref{eq:sec-inst-exp} into the full compressible Navier-Stokes equations and neglecting the $\mathcal{O}\left(\epsilon_s^2\right)$ nonlinear terms, we obtain the linearised compressible Navier-Stokes equations.
Since the base-flow $\tilde{u}$ and $\tilde{T}$ vary slowly with $\bar{x}$ and $\bar{t}$, the dependence on these two variables can be treated as parametric when the short-wavelength (of order $\delta^*$) and the high-frequency (of order $U_{\infty}/\delta^*$) instability is considered.
A solution is sought in the normal-mode form
\begin{eqnarray}
    q'_s(x,y,z,t)=q_s(y,z)\exp[{\rm i}(\alpha x-\omega t)]+{\rm c.c.},
\end{eqnarray}
where $\alpha$ is the streamwise wavenumber and $\omega$ is the frequency of the secondary disturbance. The shape function $q_s(y,z)=\left\{u_s,v_s,w_s,T_s\right\}$ is governed by a system of partial differential equations, supplemented by homogeneous boundary conditions, $\left\{u_s,v_s,w_s,T_s\right\}=0$ at $y=0$ and $\left\{u_s,v_s,w_s,T_s\right\}\rightarrow 0$ as $y\rightarrow \infty.$

For a spanwise-periodic base flow $\tilde{q}$, the solution for $q_s$ can be expressed using Floquet theory as
\begin{eqnarray}
q_s={\rm e}^{{\rm i} \gamma \beta z} \sum_{k=-\infty}^{\infty}\phi_{s,k}(y) {\rm e}^{{\rm i} k\beta z},
\end{eqnarray}
where $\beta$ is the spanwise wavenumber and $0\le \gamma \le 1/2.$
Fundamental modes ($\gamma = 0$), subharmonic modes ($\gamma = 1/2$) and detuned modes ($0 < \gamma < 1/2$) are all part of the same branch of instability modes but with varying spanwise wavelengths. The growth rate of the modes was found to be insensitive to the Floquet parameter \citep{ren2015secondary}.

\section{Numerical procedures}
\label{sec:num}
The initial-boundary-value problem, i.e.  the nonlinear boundary-region equations \eqref{eqn:nubr-con}-\eqref{eqn:nubr-e} supplemented by the initial conditions \eqref{eq:BC-x0_u}-\eqref{eq:BC-x0_tau} and the outer boundary conditions \eqref{eqn:BC}, is solved numerically. The boundary-region equations are parabolic in the streamwise direction and therefore can be solved by a marching procedure in the $\bar x-$direction. A second-order backward finite-difference scheme in the $\bar x-$direction and a second-order central finite-difference scheme in the $\eta-$direction are employed. In order to avoid the pressure decoupling phenomenon, the pressure is computed on a grid that is staggered in the $\eta-$direction with respect to the grid for the velocity components and temperature. The nonlinear terms are evaluated using the pseudo-spectral method. In order to prevent aliasing errors, i.e.  the spurious energy cascade from the unresolved high-frequency modes into the resolved low-frequency ones, the 3/2-rule is applied \citep{canuto1988spectral}. The resulting block tri-diagonal system is solved using a standard block-elimination algorithm. A second-order predictor--corrector under-relaxation scheme is used to calculate the nonlinear terms while marching downstream, as in the computation of incompressible G\"ortler vortices by \citet{xu_zhang_wu_2017}. The use of under-relaxation for capturing the generation of nonlinear streaks was deemed unnecessary by \citet{marensi2017nonlinear}. However, it is needed in our analysis to stabilise the computations, given the high growth rate and intensity exhibited by G\"ortler vortices. The wall-normal domain extends to $\eta_{max}=60$ and 2000 grid points are used in this direction. The typical step size in the marching direction is $\Delta \bar x=0.01.$ To capture the nonlinear effects, it is sufficient to use $N_t$=17 modes to discretise time and $N_z$=17 modes to discretise the spanwise direction.

The equations governing the secondary instability are discretised using a five-point finite-difference scheme with fourth-order accuracy along the wall-normal direction and Fourier spectral expansion along the spanwise direction. The code was used by \cite*{song2020secondary} to perform a secondary-instability analysis of nonlinear stationary vortices.

\section{Results}
\label{sec:results}

\subsection{Flow parameters}
\label{sec:flow-parameters}

The nonlinear boundary-layer disturbances are studied for parameters that characterise flows over high-pressure turbine blades. The flow parameters chosen as reference are given in table \ref{tab:para}. As discussed in \cite{marensi2017nonlinear}, they are inspired by typical experimental works on turbomachinery applications, such as \cite{arts1990aero} and \cite{camci-arts-1990}. In the figure captions, only the parameters that are varied in the figure are given. In all our computations, the scaled amplitudes of the free-stream velocity components are $\hat{u}_{x,\pm}^{\infty}=\hat{u}_{y,\pm}^{\infty}=1$ and $\hat{u}_{z,\pm}^{\infty}=\mp 1$. The continuity relation \eqref{eqn:fsvd-continuity} reduces to $k_x + k_y \pm 1 = 0$.
\begin{table}
\begin{center}
\def~{\hphantom{0}}
\begin{tabular}{ccccccccc}
$\mathcal{M}_\infty$ & {$T_w$}  & $R_{\Lambda}$& $\mathcal{G}$  &$k_x$ &  $\kappa_z$ & $\kappa_y$ & $Tu$& $\epsilon \cdot 10^{2}$ \\[3pt] 0.69  & 0.75 & 1124 & 35.2 & 0.0073 & 0.35 & 0.35&1\% 4\% 6\% &0.35 1.41 2.11
\end{tabular}
\caption{Reference flow parameters.}
\label{tab:para}
\end{center}
\end{table}

The adiabatic wall temperature is calculated using the relation valid for a perfect gas, $T_{ad}=1+(\gamma-1)\sqrt{Pr}\mathcal{M}_\infty^2/2$. The non-dimensional wall temperature is $T_w=0.75$ as blade cooling is often applied to avoid excessive wall-heat transfer. The axial chord length of the turbine blade is $C_{ax}^*=0.0388$ m. This length corresponds to the maximum streamwise coordinate $\bar x=0.558$ for $k_x=7.3\cdot 10^{-3},$ our chosen frequency representative of the experiments of \citet{arts1990aero} and \cite{camci-arts-1990}. The reference radius of curvature is  $r_0^*=$1.4~m and the spanwise length scale is $\Lambda^*=0.89\cdot10^{-3}$m, corresponding to a G\"ortler number $\mathcal{G}=35.2$. The FVD level varies between $Tu=1\%$ and $6\%,$ as in the experiments of \citet{arts1990aero}. For the form of perturbations assumed here, the FVD level $Tu$ is related to the FVD intensity $\epsilon$ by $Tu(\%)=100 \cdot 2\epsilon\left(\hat u_{x,+}^{\infty 2}+\hat u_{x,-}^{\infty 2}\right)^{1/2}.$

We investigate the effect of three parameters on the evolution of boundary-layer disturbances, i.e.  the G\"ortler number $\mathcal{G}$, the FVD level $Tu$ and the Mach number $\mathcal{M}_\infty$. Boundary-layer transition is also affected by the free-stream disturbance length scales (e.g. as recently shown by \citet{fransson2020effect}). The impact of $k_x$ on the evolution of the boundary-layer disturbances was studied in detail in our previous studies \citep{marensi2017nonlinear,xu_zhang_wu_2017,marensi2017growth} and similar effects are expected in the present case. Furthermore, as verified in several experimental campaigns, boundary-layer disturbances have a spanwise length that is comparable to the boundary-layer thickness and therefore we fix $\kappa_z,\kappa_y=\mathcal{O}(1)$.

The overall intensity of the disturbances is measured by the root mean square (r.m.s.) of the fluctuating quantity, defined as
\begin{eqnarray}
q_{rms,max}(\bar x)
=\max_\eta q_{rms}(\bar x, \eta)
=\max_\eta r_t\sqrt{ \sum_{m=-N_t}^{N_t}  \sum_{n=-N_z}^{N_z}\left|\hat q_{m,n}\right|^2 }, \quad m\neq 0,
\end{eqnarray}
where $q$ stands for any quantity, but we focus on the streamwise velocity and the temperature because they are the leading-order variables.

\subsection{Velocity and temperature of the nonlinear boundary-layer disturbances}
\label{sec:gortler}

\begin{figure}
\centering
   \subfigure{
      \put(0,146){$(a)$}
      \includegraphics[width=0.48\textwidth]{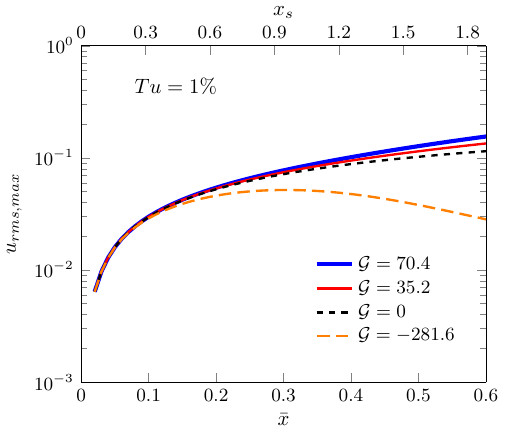}
   }
   \subfigure{
      \put(0,146){$(b)$}
      \includegraphics[width=0.48\textwidth]{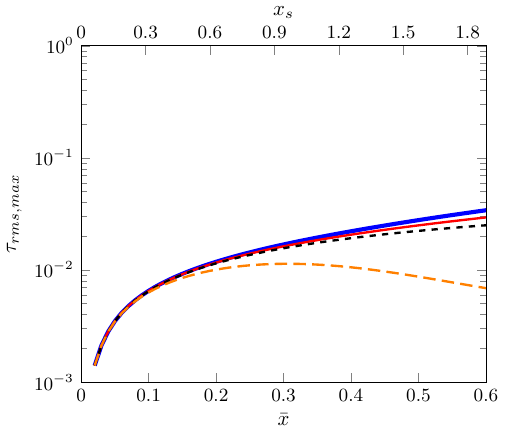}
   }
   \subfigure{
      \put(0,146){$(c)$}
      \includegraphics[width=0.48\textwidth]{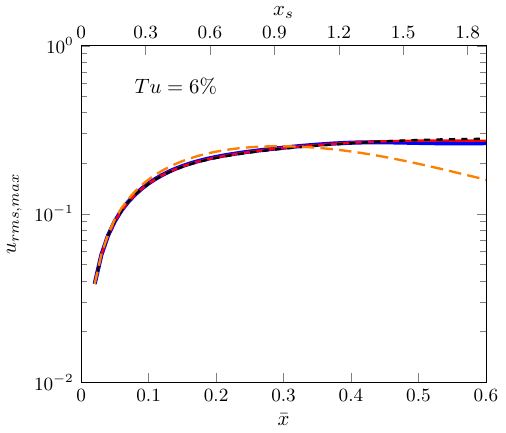}
   }
   \subfigure{
      \put(0,146){$(d)$}
      \includegraphics[width=0.48\textwidth]{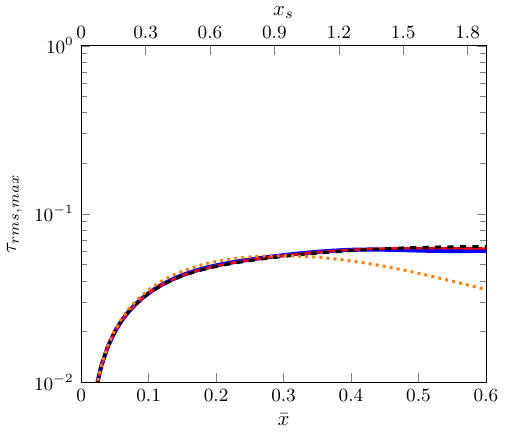}
   }
   \captionsetup{justification=raggedright}
   \caption{Effect of G\"ortler number on the downstream development of $u_{rms,max}$ and $\tau_{rms,max}$ induced by $(a,b)$ $Tu=1\%$ and $(c,d)$ $Tu=6\%$.}
   \label{fig:curvature-effect}
\end{figure}
The effect of G\"ortler number on the downstream evolution of the streamwise and temperature disturbances is studied first. The variation of G\"ortler number is achieved by adjusting the boundary-layer curvature while keeping the frequency constant. Figure \ref{fig:curvature-effect} depicts the downstream development of $u_{rms,max}$ and $\tau_{rms,max}$ for four different G\"ortler numbers, including the flat-wall case ($\mathcal{G}=0$) and a convex-wall case ($\mathcal{G}=-281.6$). Two FVD levels are tested ($Tu=1\%$ and $Tu=6\%$). The coordinate $x_s$ on top of the graphs is normalised by the axial chord length $C_{ax}^*$ (the end of the turbine blade is at $x_s=1.65$).
For $Tu=1\%,$ the concave wall destabilises the flow, whereas the convex wall has a marked stabilising effect on the growth of both the velocity and temperature disturbances. For $Tu=6\%,$ the curvature has little effect in the concave-wall case and is stabilising in the convex-wall case. The evolutions of the vortical structures for $\mathcal{G}=35.2$ and $\mathcal{G}=70.4$ are indeed almost the same as in the flat-wall  case. The convex curvature is not influential up to $\bar x=0.35$ for such a higher FVD level. For the cases considered, the boundary-layer dynamics is therefore largely independent of the curvature up to $x_s=1.2$, i.e.  for most of the extent of the turbine blade.

\begin{figure}
   \centering
   \subfigure{
      \put(0,146){$(a)$}
      \includegraphics[width=0.48\textwidth]{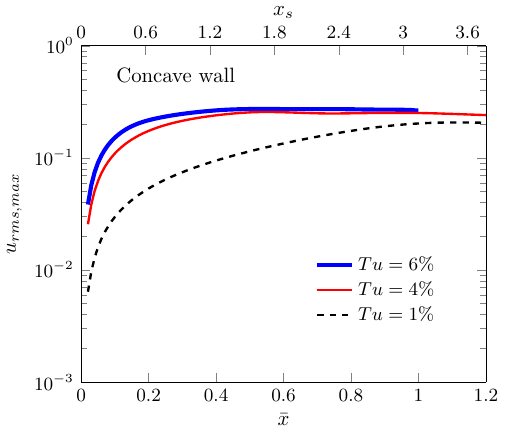}
   }
   \subfigure{
      \put(0,146){$(b)$}
      \includegraphics[width=0.48\textwidth]{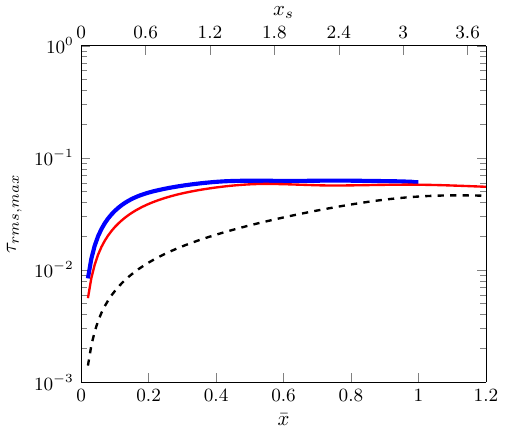}
   }
   \subfigure{
      \put(0,146){$(c)$}
      \includegraphics[width=0.48\textwidth]{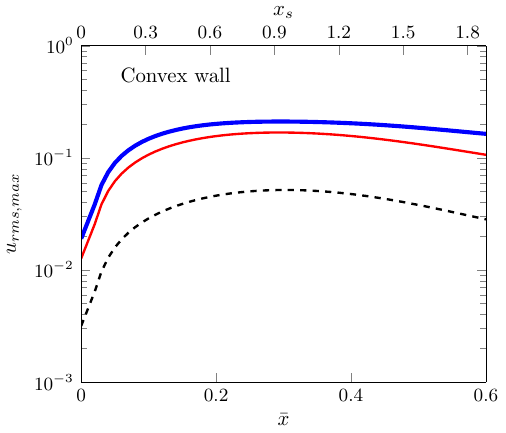}
   }
   \subfigure{
      \put(0,146){$(d)$}
      \includegraphics[width=0.48\textwidth]{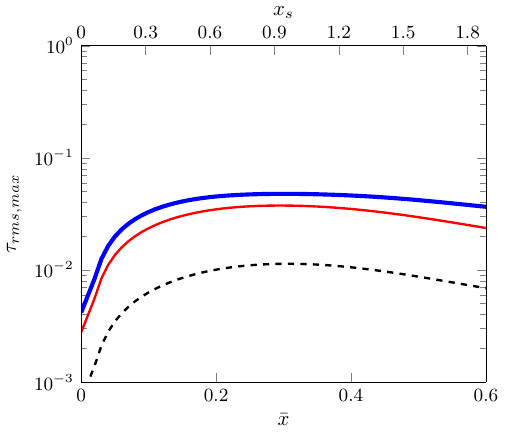}
   }
   \captionsetup{justification=raggedright}
   \caption{Effect of FVD level on the downstream development of $u_{rms,max}$ and $\tau_{rms,max}$ over $(a,b)$ concave wall ($\mathcal{G}=35.2$) and $(c,d)$ convex wall ($\mathcal{G}=-281.6$).}
   \label{fig:tur-effect}
\end{figure}
Figure \ref{fig:tur-effect}$(a,b)$ shows the effect of the FVD level on the downstream development of $u_{rms,max}$ and $\tau_{rms,max}$ for $\mathcal{G}=35.2$. For $Tu=1\%$, G\"ortler vortices undergo non-modal growth and gradually evolve to nonlinear saturation, similarly to incompressible cases \citep{xu_zhang_wu_2017,marensi2017growth}. For the high-intensity cases, $Tu=4\%$ and $Tu=6\%,$ the vortices saturate after a much shorter non-modal growth than in the $Tu=1\%$ case. The values of $u_{rms,max}$ and $\tau_{rms,max}$ saturate to almost the same level for different FVD intensities. This behaviour is different from that of compressible streaks over flat plates, where the perturbation intensity depends significantly on the FVD level \citep{marensi2017nonlinear}. As shown in figure \ref{fig:tur-effect}($c,d$), the intensity of the disturbances evolving over convex walls is enhanced by increasing the FVD level, similarly to the flat-wall case.
\begin{figure}
   \centering
   \subfigure{
      \put(0,146){$(a)$}
      \includegraphics[width=0.48\textwidth]{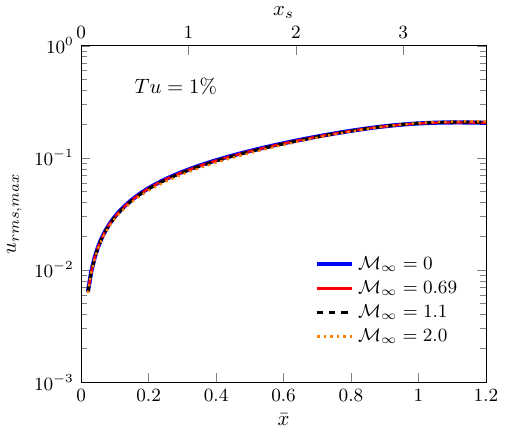}
      \label{fig:mach-effect-a}
   }
   \subfigure{
      \put(0,146){$(b)$}
      \includegraphics[width=0.48\textwidth]{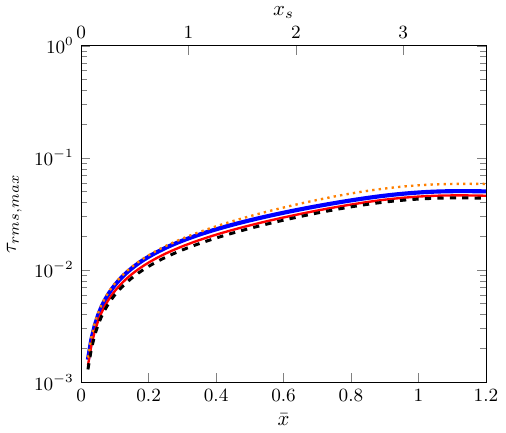}
      \label{fig:mach-effect-b}
   }
   \subfigure{
      \put(0,146){$(c)$}
      \includegraphics[width=0.48\textwidth]{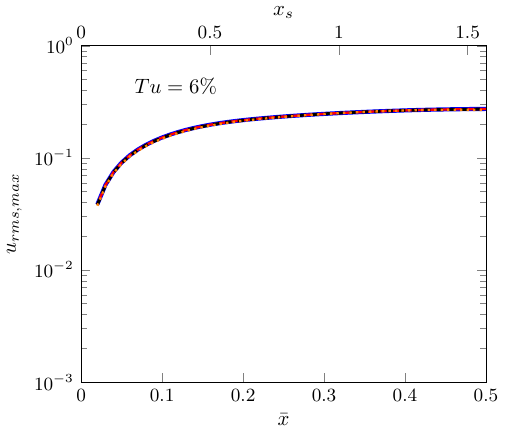}
      \label{fig:mach-effect-c}
   }
   \subfigure{
      \put(0,146){$(d)$}
      \includegraphics[width=0.48\textwidth]{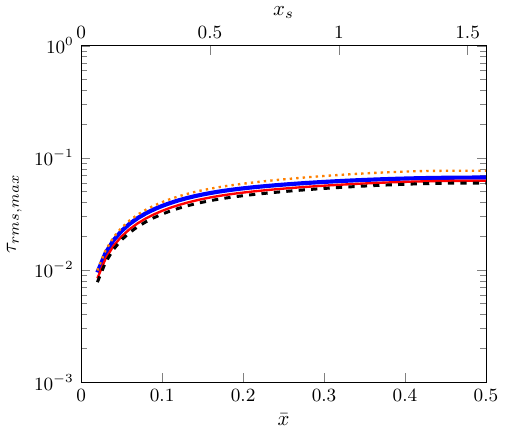}
      \label{fig:mach-effect-d}
   }
   \captionsetup{justification=raggedright}
   \caption{Effect of Mach number on the downstream development of $u_{rms,max}$ and $\tau_{rms,max}$ for $(a,b)$ $Tu=1\%$ and $(c,d)$ $Tu=6\%$. The G\"ortler number is $\mathcal{G}=$35.2.}
   \label{fig:mach-effect}
\end{figure}

The Mach-number effect on the G\"ortler vortices is studied by keeping the Reynolds number, the frequency and the radius of curvature constant. The change of Mach number with a constant Reynolds number can be achieved through an adjustment of the total pressure (hence, the density), as in the experiments of \citet*{huang2021inner}, and by use of the relation $R_\Lambda=\mathcal{M}_\infty \rho_\infty^* \Lambda^*/\left(\sqrt{\gamma R^* T_\infty^*}\mu_\infty^*\right)$, as discussed in \citet{viaro2019compressible}. Figure \ref{fig:mach-effect} shows the effect of Mach number on the evolution of G\"ortler vortices induced by low-intensity FVD ($Tu=1\%$) and high-intensity FVD ($Tu=6\%$). Figure \ref{fig:mach-effect}$(a)$ illustrates that the growth of the streamwise velocity is not influenced by the Mach number. The growth of the thermal disturbances is instead affected by the Mach number, as shown in figure \ref{fig:mach-effect}$(b)$. They are slightly stabilised as the Mach number increases within the subsonic range, unaffected in transonic conditions, and moderately enhanced in supersonic conditions.

The Mach-number effect in our cases is markedly different from that reported by \citet{viaro2019compressible} in their figure 6. \citet{viaro2019compressible} showed that, as the Mach number increases from the incompressible condition, the r.m.s. of the streamwise velocity is attenuated, while the r.m.s. of the temperature increases for a short distance from the leading edge and decreases further downstream. The difference in dynamics between our flows and those in \citet{viaro2019compressible} is due to the higher G\"ortler number and frequency of our cases. As both these quantities become larger, the boundary-layer response becomes less sensitive to a change in Mach number.

\begin{figure}
\centering
\subfigure{
   \put(82,146){$Tu=1\%$}
   \put(0,146){$(a)$}
\includegraphics[width=0.48\textwidth]{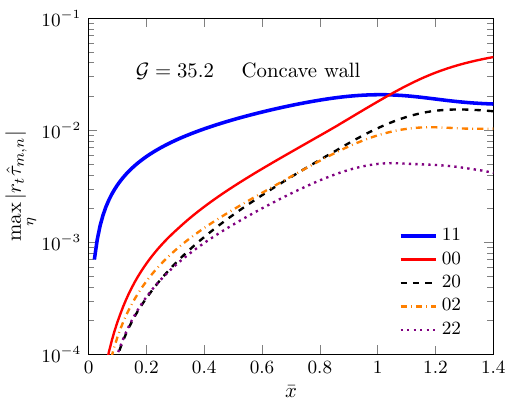}
}
\subfigure{
   \put(82,146){$Tu=6\%$}
   \put(0,146){$(b)$}
\includegraphics[width=0.48\textwidth]{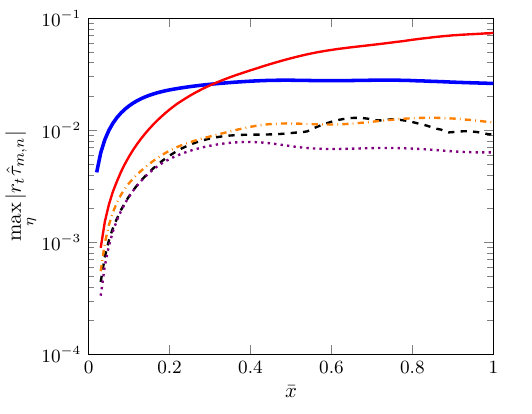}
\label{fig:fourier-modes-b}
}
\subfigure{
   \put(82,146){$Tu=1\%$}
   \put(0,146){$(c)$}
\includegraphics[width=0.48\textwidth]{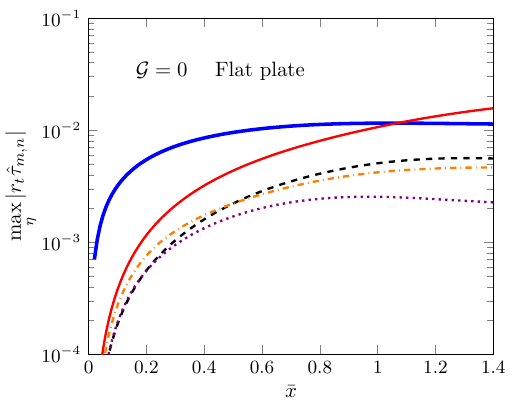}
}
\subfigure{
   \put(82,146){$Tu=6\%$}
   \put(0,146){$(d)$}
\includegraphics[width=0.48\textwidth]{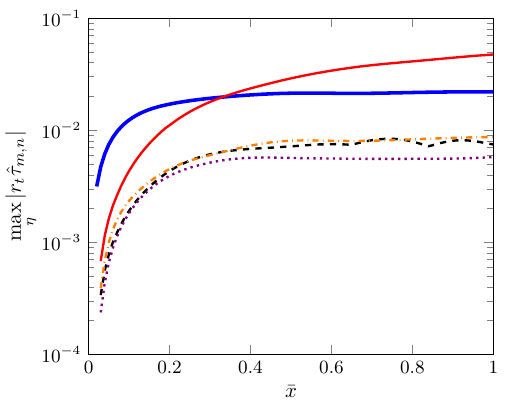}
}
\subfigure{
   \put(82,146){$Tu=1\%$}
   \put(0,146){$(e)$}
\includegraphics[width=0.48\textwidth]{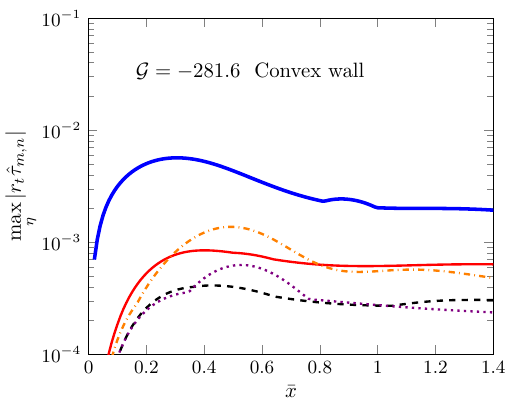}
}
\subfigure{
   \put(82,146){$Tu=6\%$}
   \put(0,146){$(f)$}
\includegraphics[width=0.48\textwidth]{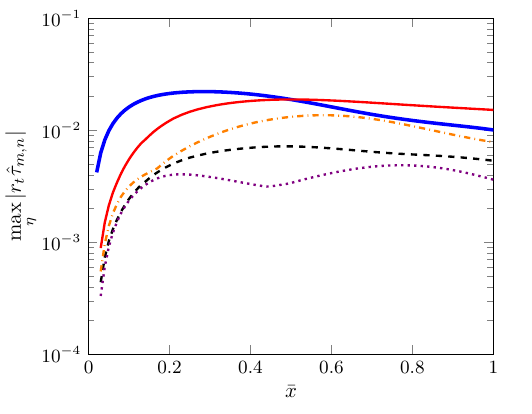}
}
\captionsetup{justification=raggedright}
\caption{Development of the fundamental mode {$(m,n)=(1,1)$} and the harmonic components {$(m,n)=(0,0), (2,0), (0,2), (2,2)$} of temperature disturbance for different G\"ortler numbers: {\it (a,b)} $\mathcal{G}=35.2$, {\it (c,d)} $\mathcal{G}=0$, {\it (e,f)} $\mathcal{G}=-281.6$, and FVD levels: {\it (a,c,e)} $Tu=1\%$, {\it (b,d,f)} $Tu=6\%$. Only modes with $n\ge0$ are shown as modes $(m, \pm n)$ have the same amplitude for the free-stream disturbance of the assumed form.}
\label{fig:fourier-modes}
\end{figure}
Figure \ref{fig:fourier-modes} shows the development of the maximum amplitudes of the fundamental and the harmonic temperature Fourier modes for $\mathcal{G}=35.2,$ $0$ and $-281.6$. The G\"ortler number plays a different role at low $(Tu=1\%)$ and high $(Tu=6\%)$ FVD levels.
In all cases, the fundamental modes (1,$\pm1$) are initially dominant over all the other modes. For the case with $\mathcal{G}=35.2,$ shown in figures \ref{fig:fourier-modes}$(a,b)$, the mean-flow distortion given by the mode (0,0) grows significantly downstream, acquiring a magnitude larger than that of the fundamental modes (1,$\pm1$). The cross-over streamwise location moves closer to the leading edge as the FVD level increases. The amplitude of the other harmonics remains smaller than that of the fundamental modes (1,$\pm1$) at any location. In the flat-wall case for $Tu=1\%$, shown in figure \ref{fig:fourier-modes}$(c)$, the cross-over of modes (1,$\pm1$) and (0,0) also occurs and all the modes keep growing downstream up to saturation, but their amplitude is lower than that in the concave case. As shown in figure \ref{fig:fourier-modes}$(e)$, for the convex-wall case and $Tu=1\%$, the fundamental modes (1,$\pm1$) are dominant over all the other harmonics and the overtake of the mean-flow distortion does not occur within the streamwise distance studied. Differently from the flat-wall case, all the modes grow  and eventually decay in the convex-wall case. Figures \ref{fig:fourier-modes}$(d,f)$ show that, in the flat-wall and convex-wall cases for $Tu=6\%$, the mode (0,0) surpasses the fundamental modes (1,$\pm1$). For $Tu=6\%,$ the cross-over location moves closer to the leading edge as the G\"ortler number increases.

\begin{figure}
   \centering
   \subfigure{
      \put(0,146){$(a)$}
      \put(82,146){$\mathcal{G}=35.2$}
      \includegraphics[width=0.48\textwidth]{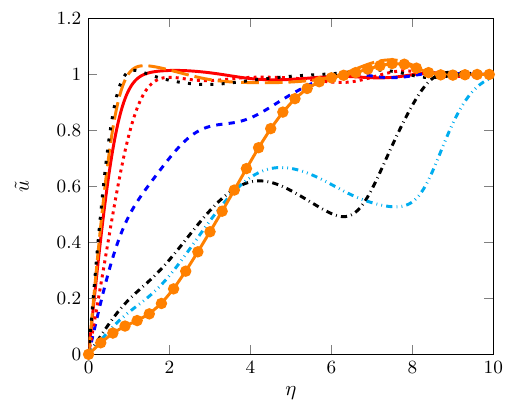}
   }
   \subfigure{
      \put(0,146){$(b)$}
      \put(82,146){$\mathcal{G}=35.2$}
      \includegraphics[width=0.48\textwidth]{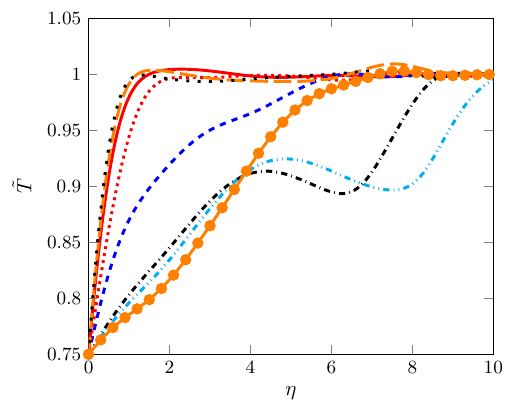}
   }
   \subfigure{
      \put(0,146){$(c)$}
      \put(82,146){$\mathcal{G}=0$}
      \includegraphics[width=0.48\textwidth]{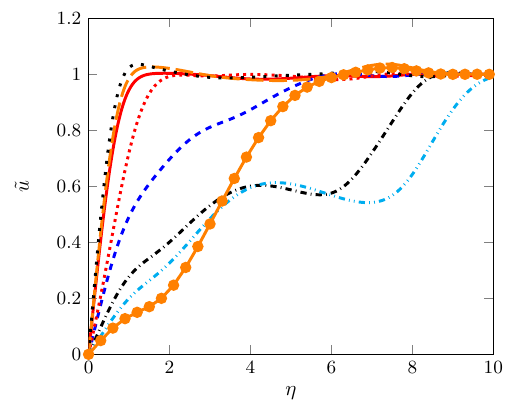}
   }
   \subfigure{
      \put(0,146){$(d)$}
      \put(82,146){$\mathcal{G}=0$}
      \includegraphics[width=0.48\textwidth]{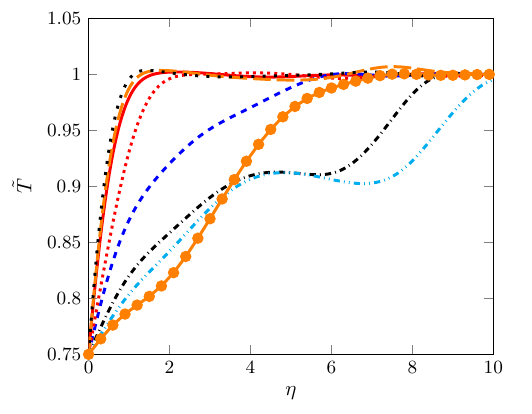}
   }
   \subfigure{
      \put(0,146){$(e)$}
      \put(82,146){$\mathcal{G}=-281.6$}
      \includegraphics[width=0.48\textwidth]{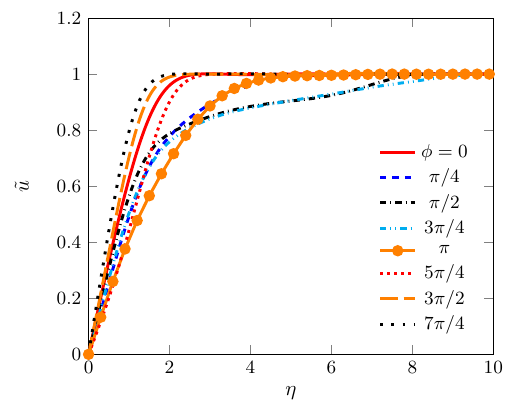}
   }
   \subfigure{
      \put(0,146){$(f)$}
      \put(82,146){$\mathcal{G}=-281.6$}
      \includegraphics[width=0.48\textwidth]{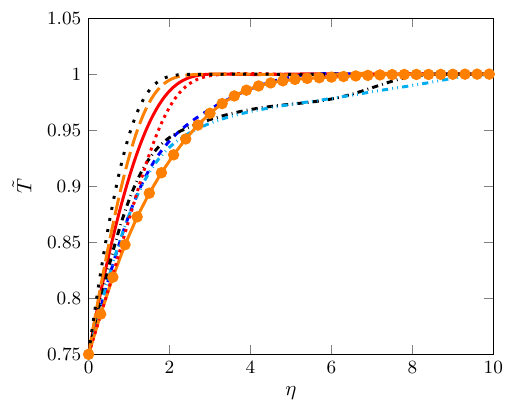}
   }
   \captionsetup{justification=raggedright}
   \caption{Profiles of instantaneous $(a,c,e)$ streamwise velocity and $(b,d,f)$ temperature at $\bar x=0.54$, $z=0$ for $Tu=6\%$ and different G\"ortler numbers.}
   \label{fig:vel-tem-pro}
\end{figure}

Of particular interest are the streamwise velocity and temperature profiles of the perturbed boundary-layer flow. Figure \ref{fig:vel-tem-pro} shows the instantaneous profiles at $z=0$ and different phases $\phi=k_x t,$ for three different G\"ortler numbers. For $\mathcal{G}=35.2,$ the profiles exhibit great variation with the phase, becoming highly inflectional at certain phases ($\phi=\pi/2$ and $3\pi/4$). This behaviour suggests that the flow may be inviscidly unstable. The variation becomes slightly weaker for the flat-wall  case and subsides in the convex-wall case, for which the profiles are much less inflectional.

\begin{figure}
\centering
   \subfigure{
   \includegraphics[width=0.315\textwidth]{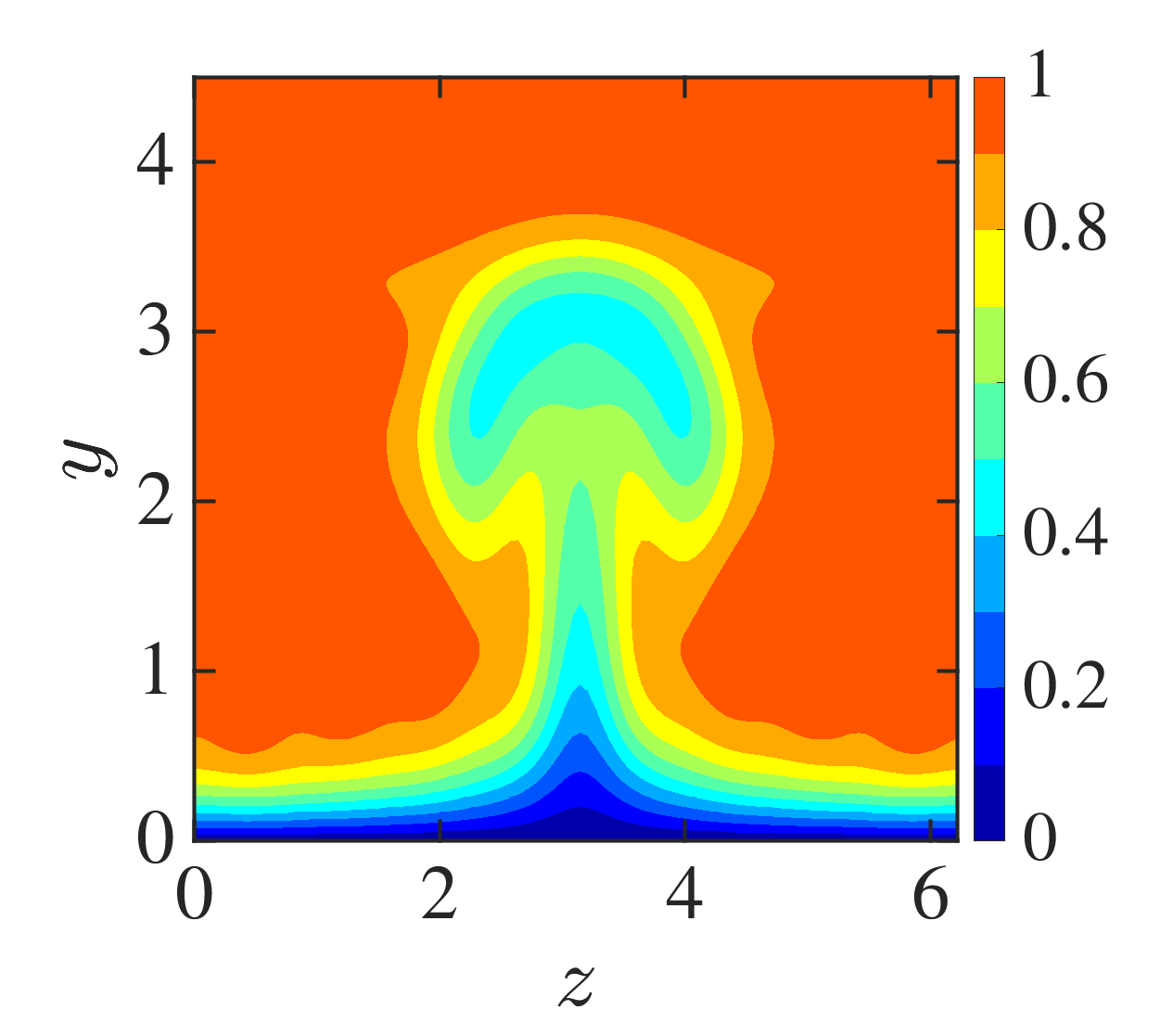}
   \put(-120,104){$(a)$}
   \put(-78,104){$\mathcal{G}=35.2$}
   }
   \subfigure{
   \includegraphics[width=0.315\textwidth]{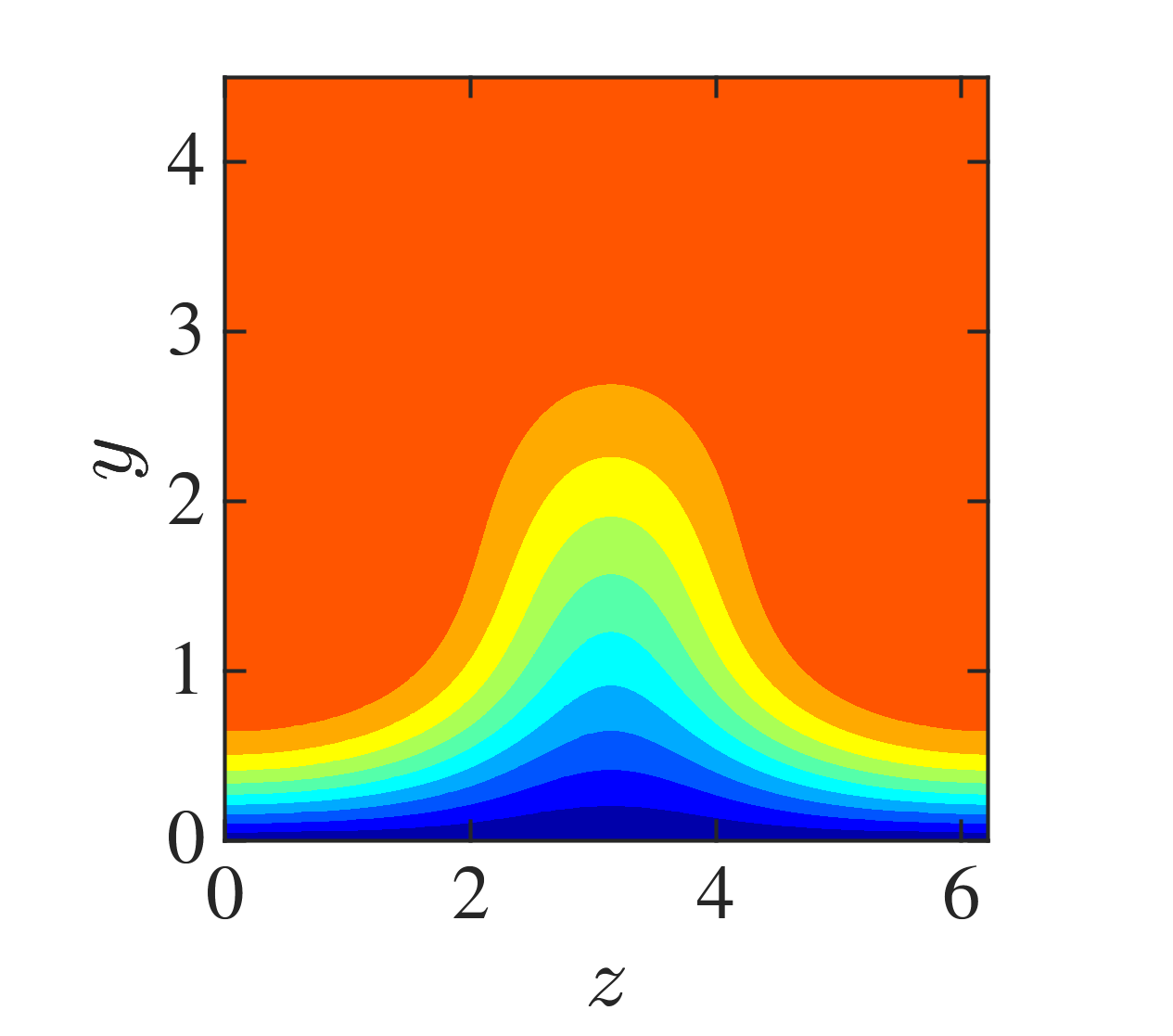}
   \put(-120,104){$(b)$}
   \put(-73,104){$\mathcal{G}=0$}
   }
   \subfigure{
   \includegraphics[width=0.315\textwidth]{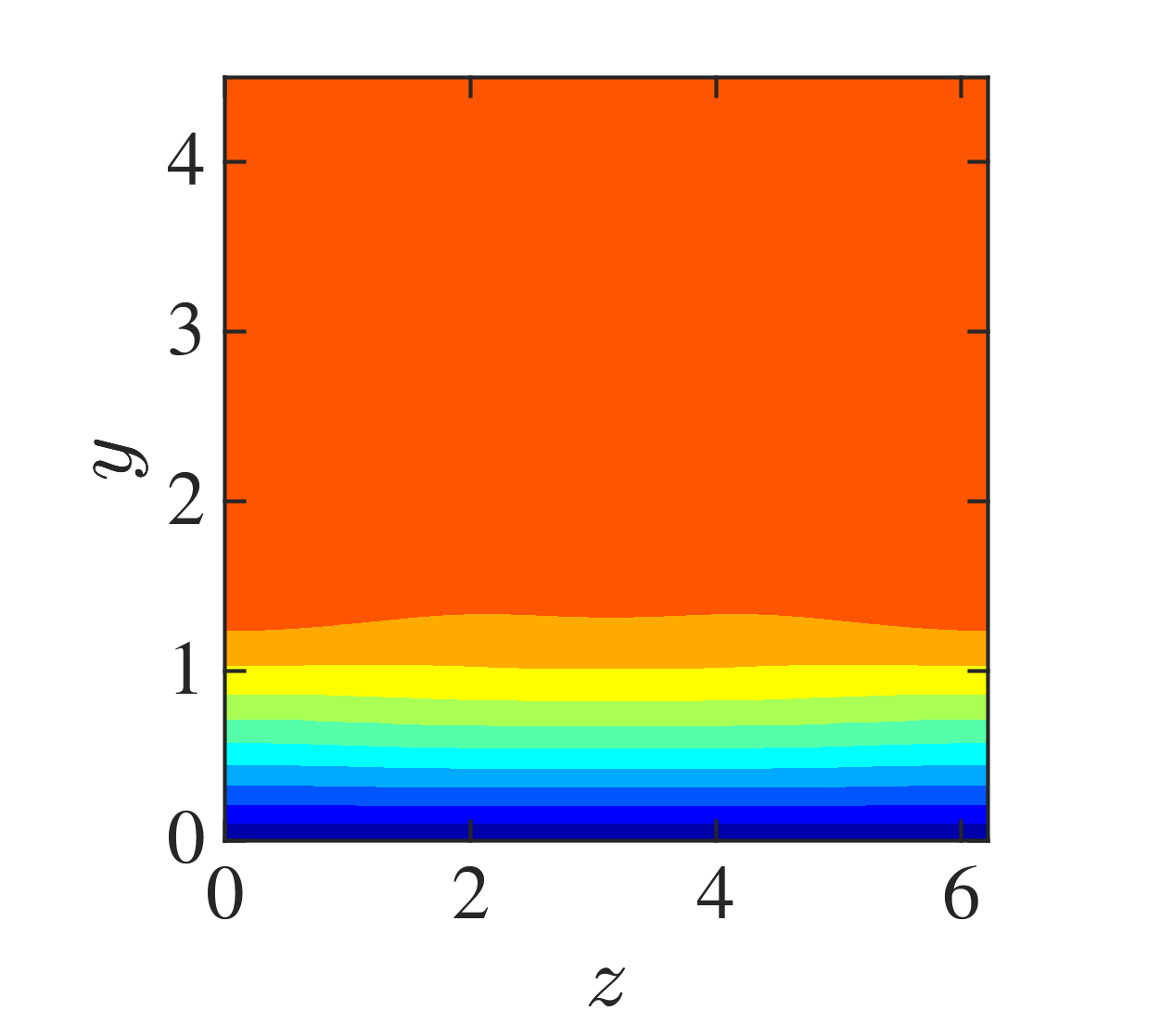}
   \put(-120,104){$(c)$}
   \put(-80,104){$\mathcal{G}=-281.6$}
   }
   \subfigure{
   \includegraphics[width=0.315\textwidth]{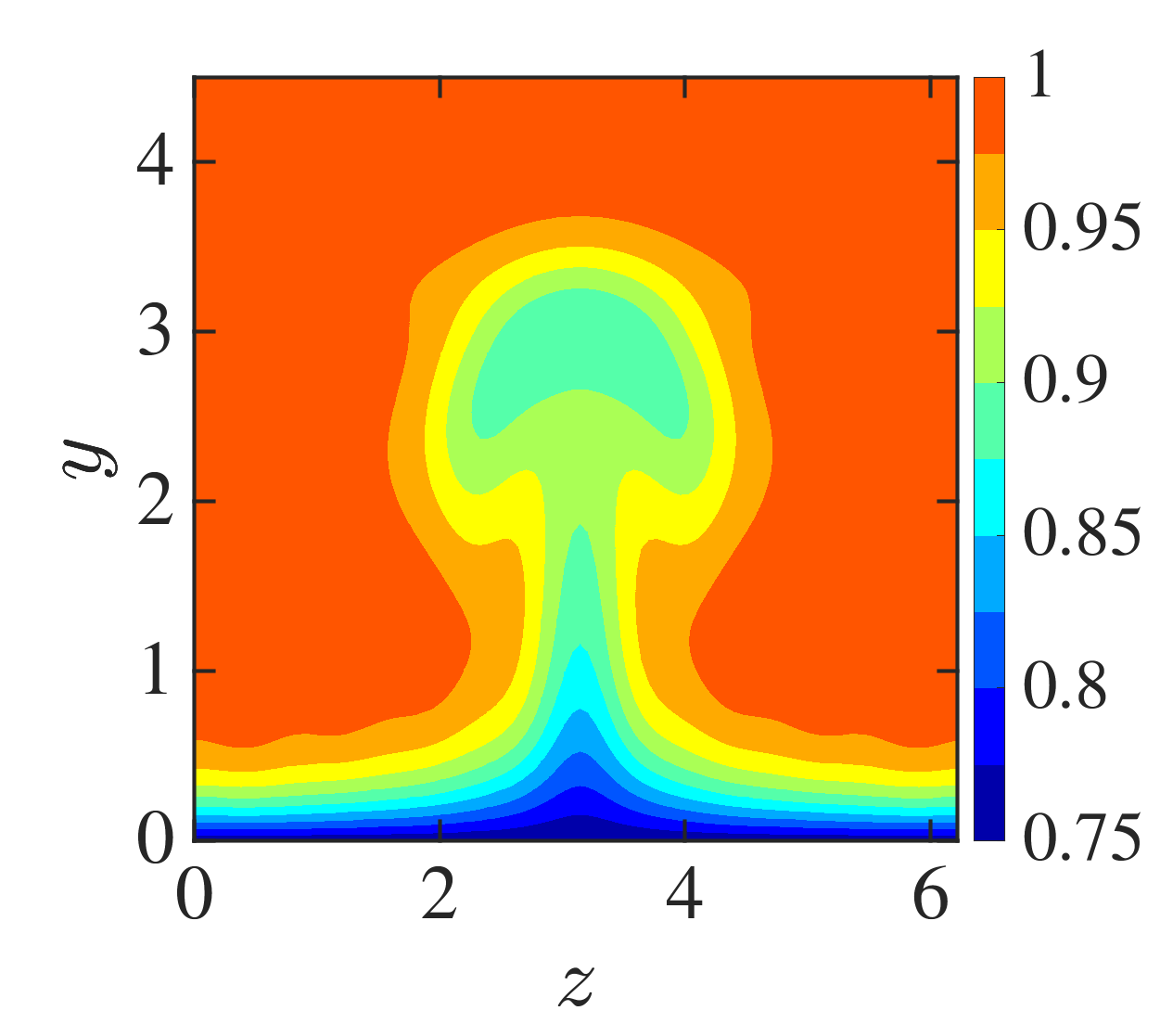}
   \put(-120,104){$(d)$}
   \put(-78,104){$\mathcal{G}=35.2$}
   }
   \subfigure{
   \includegraphics[width=0.315\textwidth]{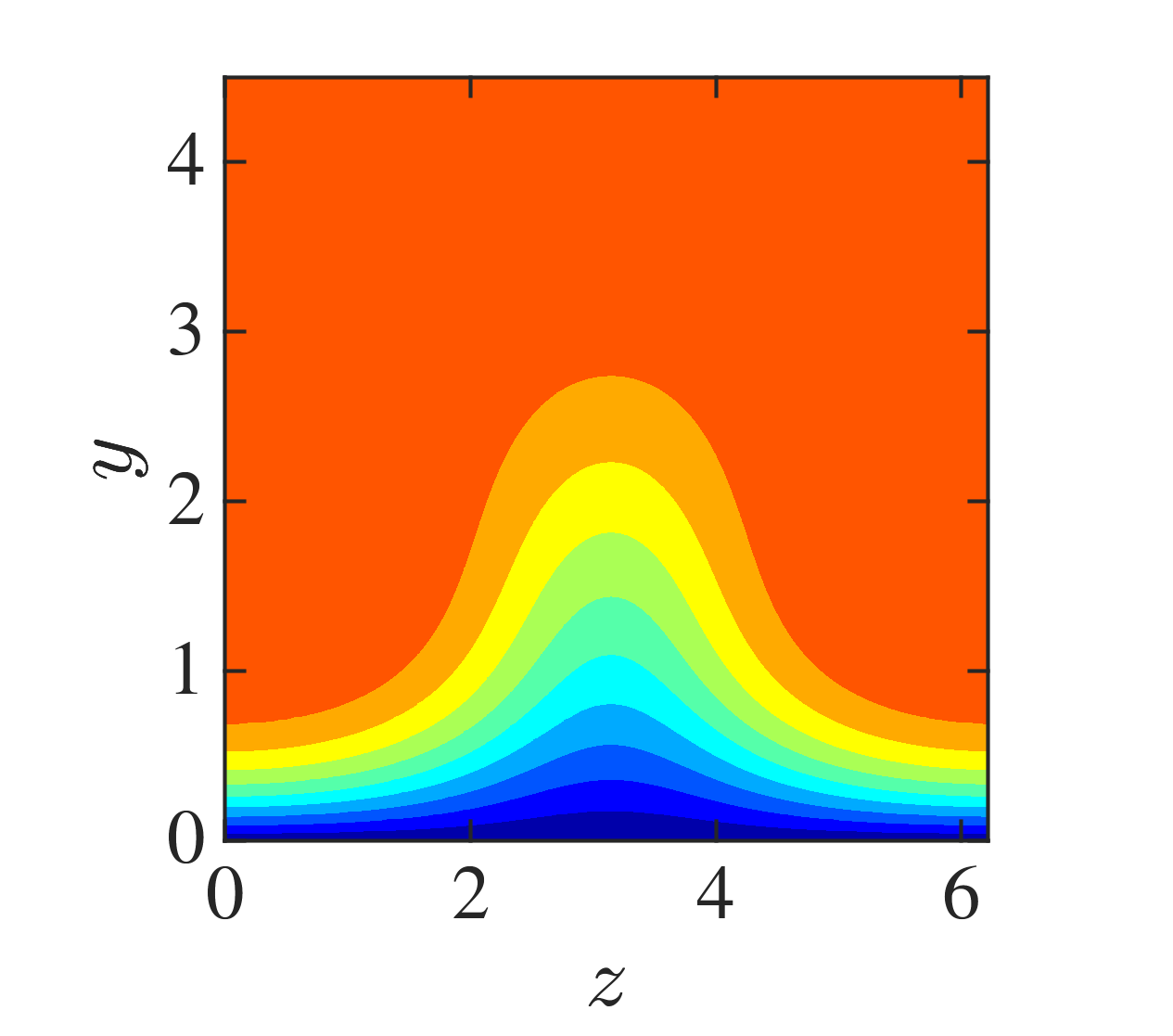}
   \put(-120,104){$(e)$}
   \put(-73,104){$\mathcal{G}=0$}
   }
   \subfigure{
   \includegraphics[width=0.315\textwidth]{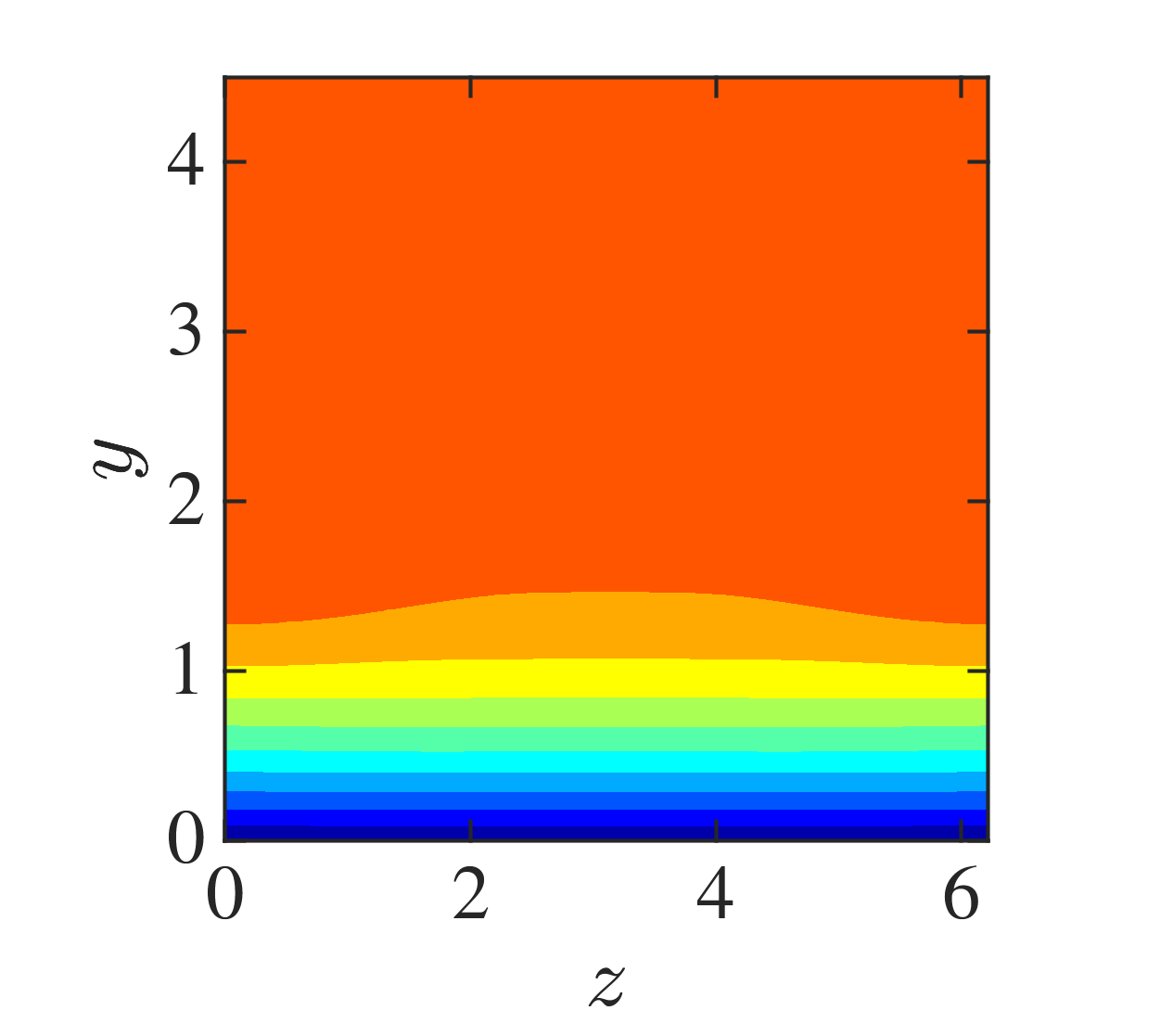}
   \put(-120,104){$(f)$}
   \put(-80,104){$\mathcal{G}=-281.6$}
   }
\captionsetup{justification=raggedright}
\vspace{-5mm}
\caption{Contours of the instantaneous $(a-c)$ streamwise velocity and $(d-f)$ temperature in the $y-z$ plane for $Tu=1\%$ at $\bar x=1.5$. The increment of the contour values is $0.1$ for the velocity and $0.05$ for the temperature. The coordinate $y$ is related to the similarity variable $\eta$ via $y =\sqrt{2 x/R_\Lambda}\int_0^\eta T(\eta){\rm d} \eta$.}
\label{fig:con-low}
\end{figure}
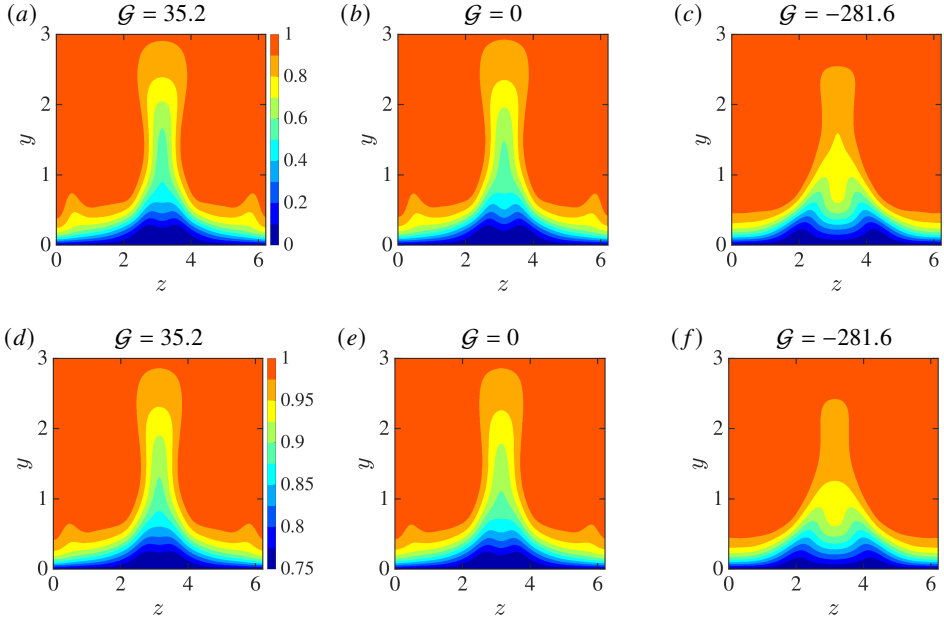
\begin{figure}
\centering
   \subfigure{
\includegraphics[width=0.315\textwidth]{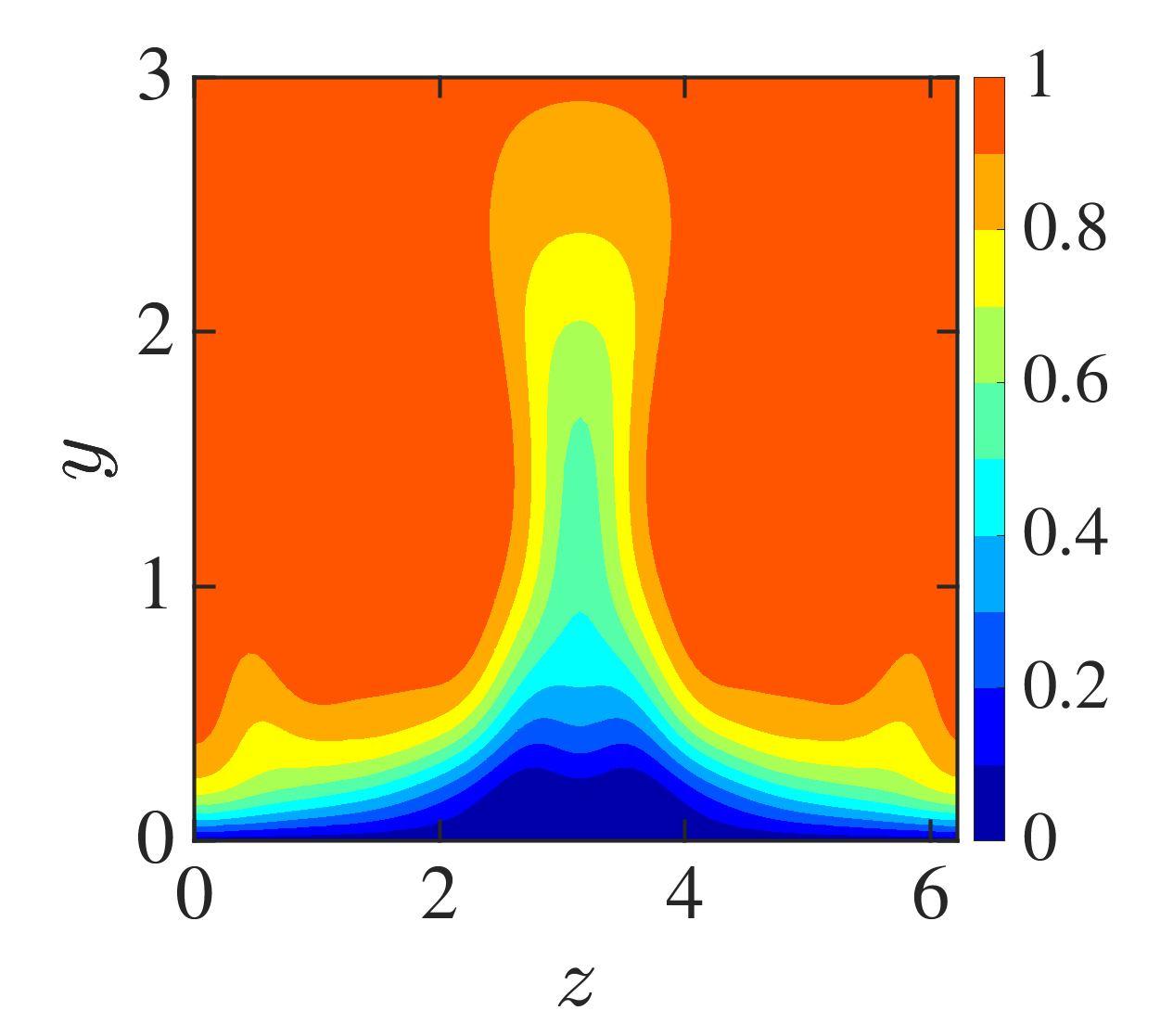}
  \put(-120,104){$(a)$}
   \put(-78,104){$\mathcal{G}=35.2$}
}
  \subfigure{
  \includegraphics[width=0.315\textwidth]{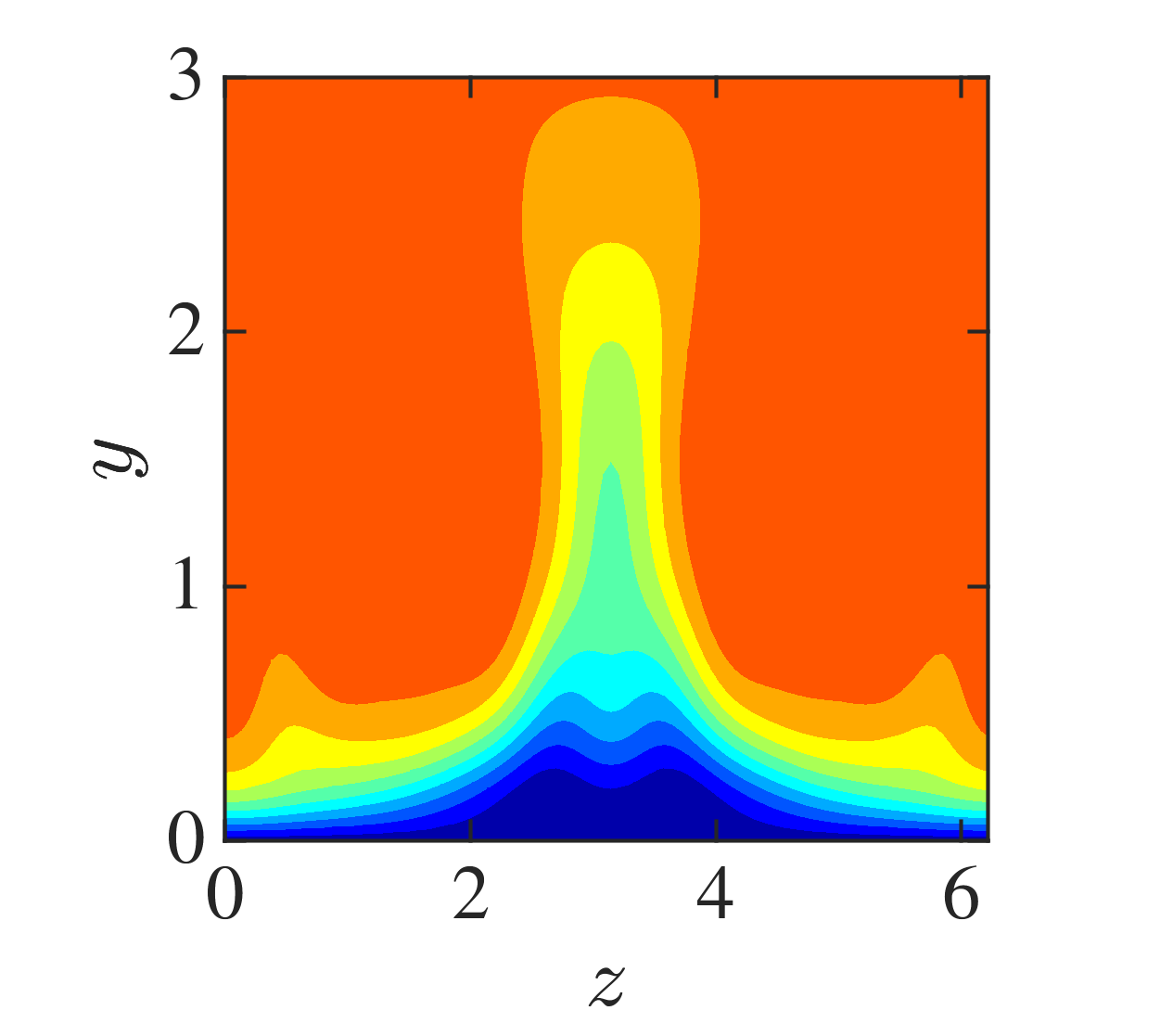}
  \put(-120,104){$(b)$}
     \put(-73,104){$\mathcal{G}=0$}
   }
     \subfigure{
  \includegraphics[width=0.315\textwidth]{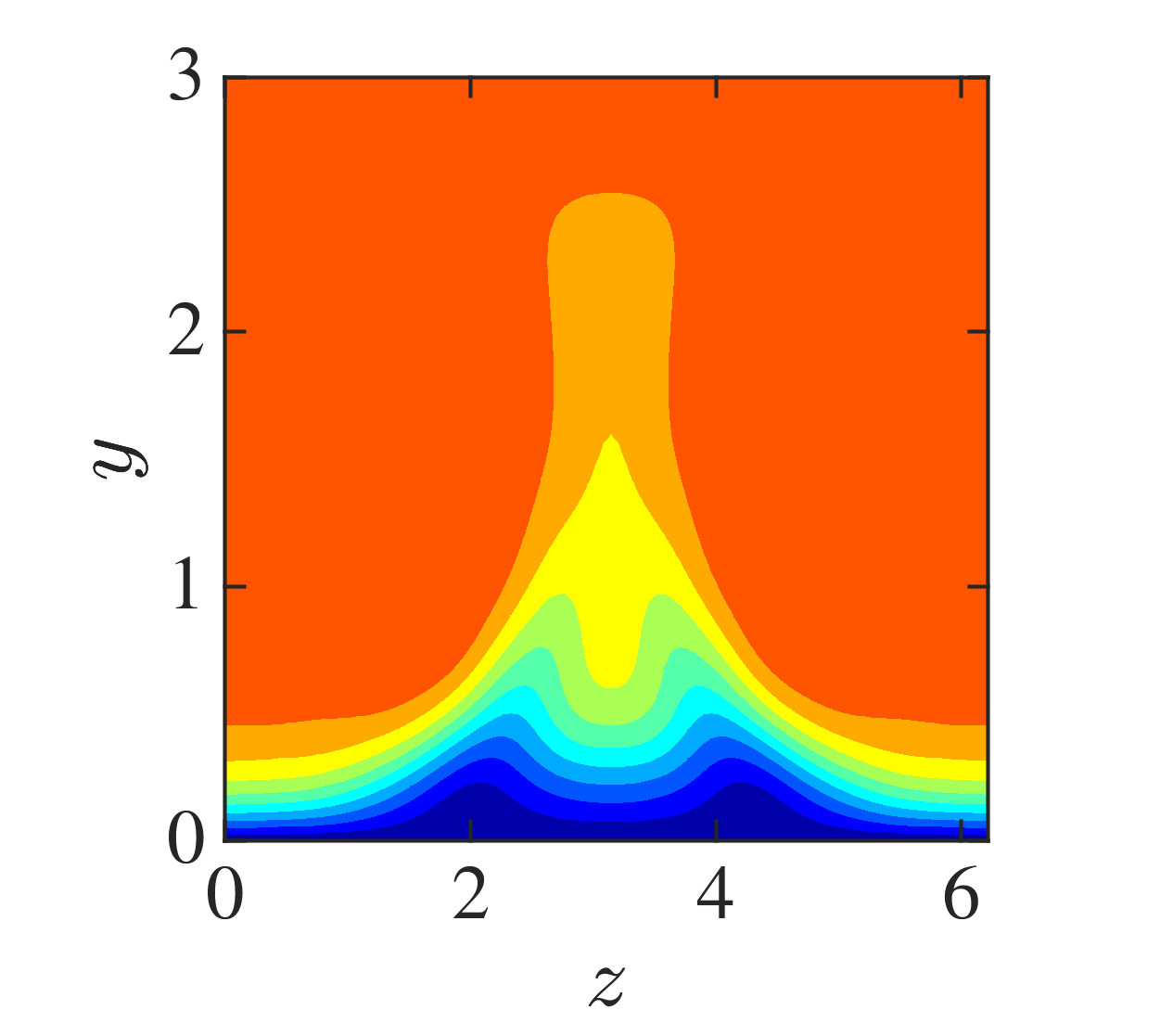}
  \put(-120,104){$(c)$}
    \put(-80,104){$\mathcal{G}=-281.6$}
   }
   \subfigure{
\includegraphics[width=0.315\textwidth]{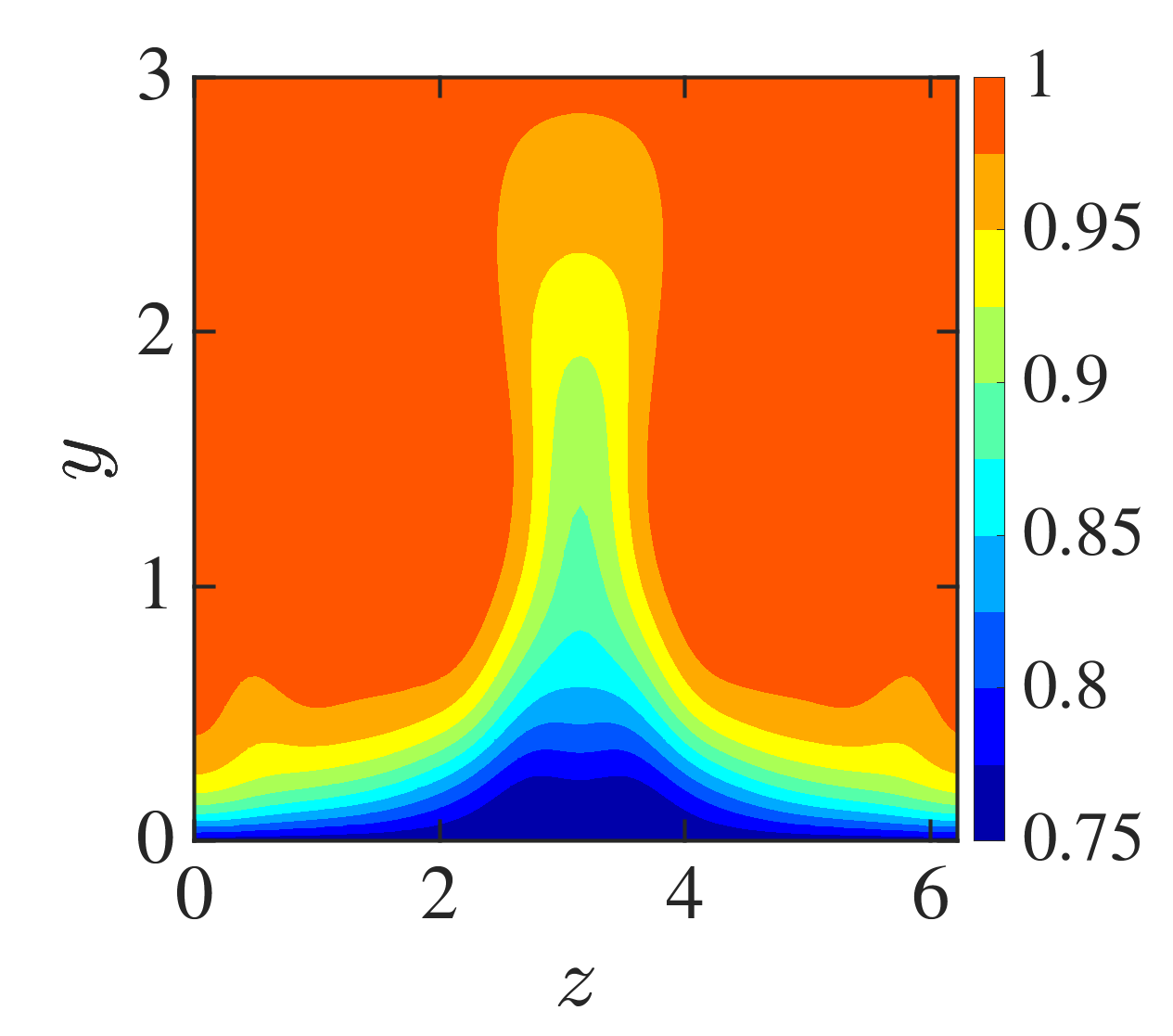}
   \put(-120,104){$(d)$}
   \put(-78,104){$\mathcal{G}=35.2$}
   }
  \subfigure{
 \includegraphics[width=0.315\textwidth]{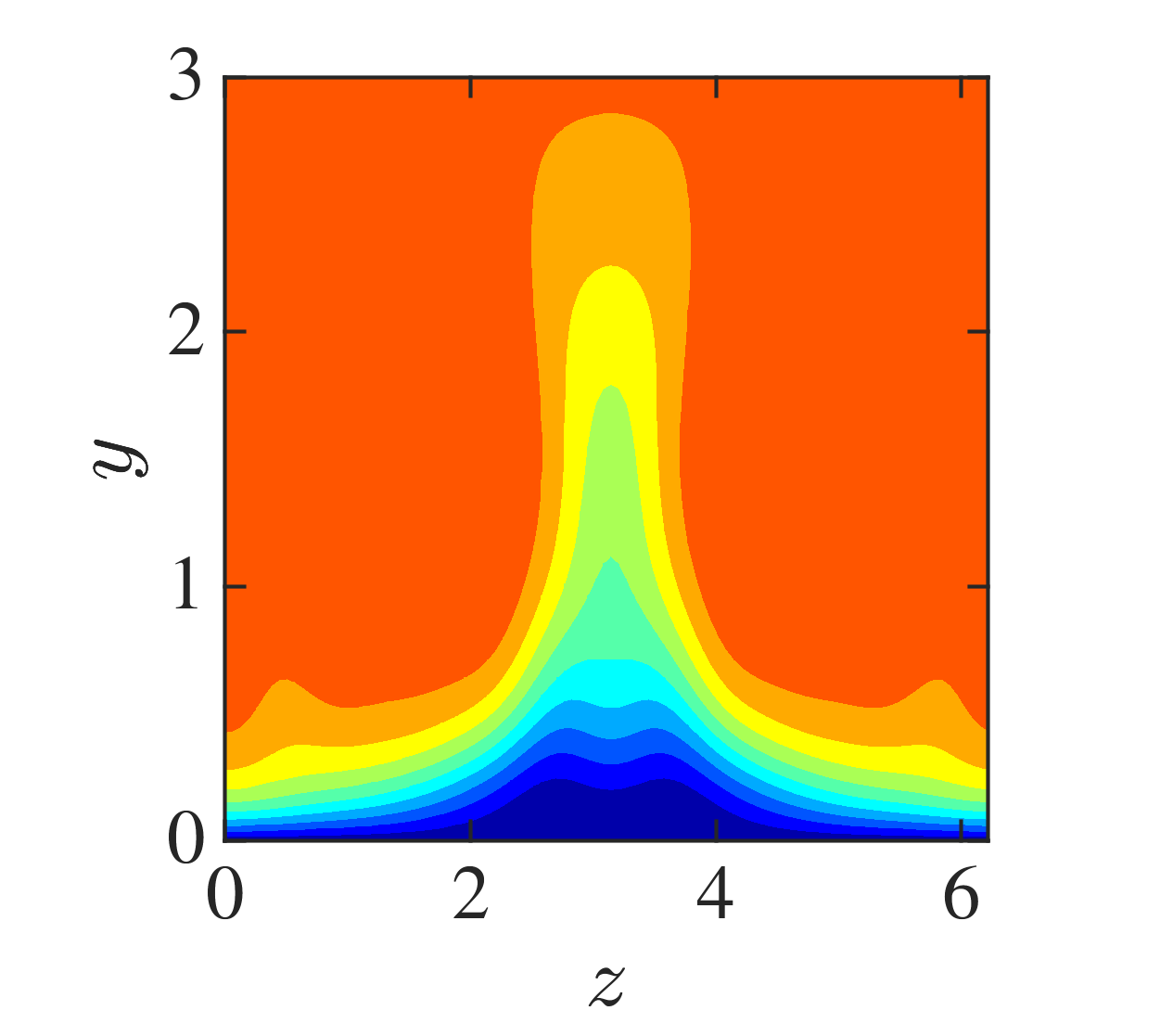}
  \put(-120,104){$(e)$}
  \put(-73,104){$\mathcal{G}=0$}
  }
   \subfigure{
 \includegraphics[width=0.315\textwidth]{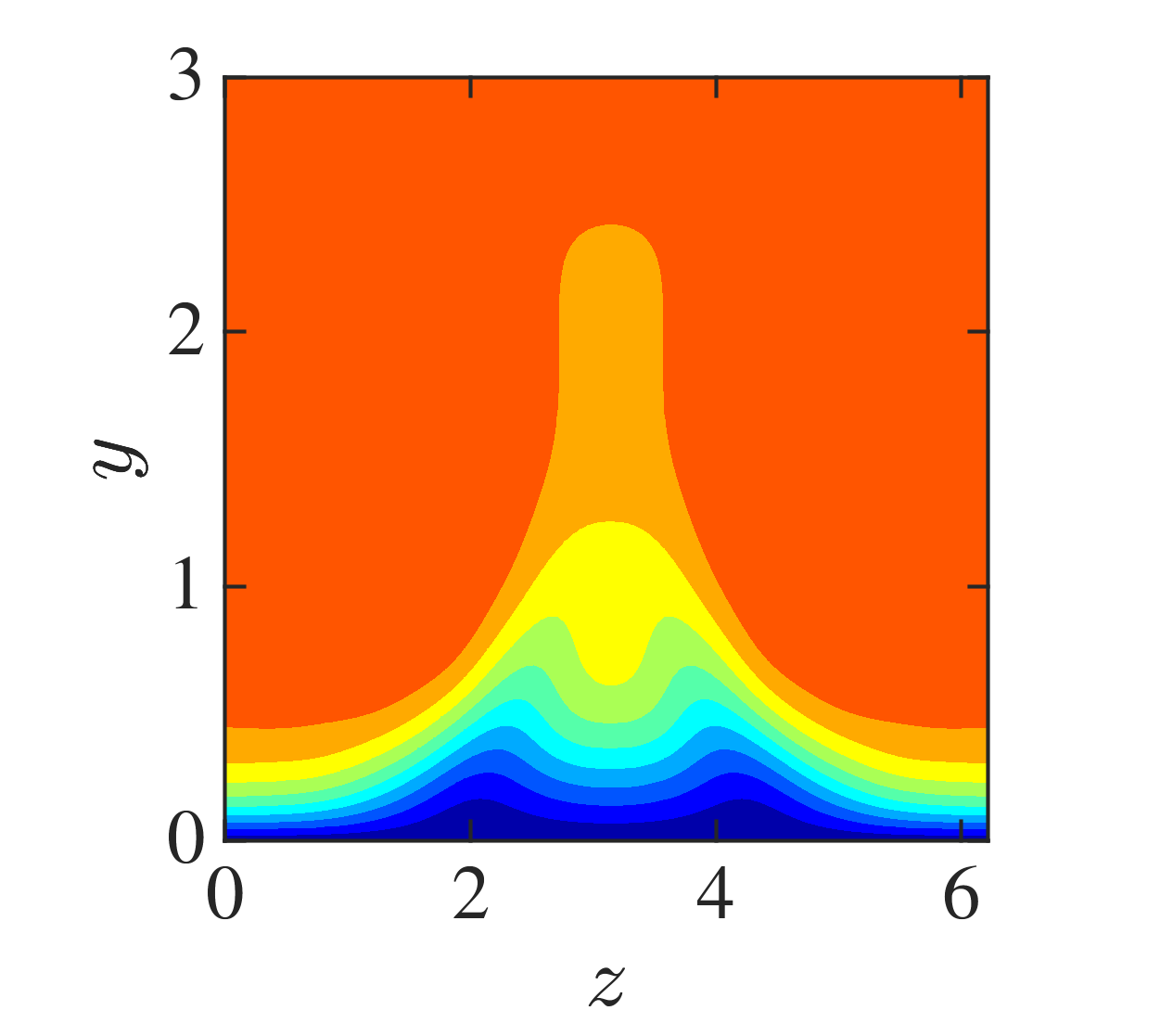}
  \put(-120,104){$(f)$}
  \put(-80,104){$\mathcal{G}=-281.6$}
  }
\captionsetup{justification=raggedright}
\vspace{-5mm}
\caption{Contours of the instantaneous $(a-c)$ streamwise velocity and $(d-f)$ temperature in the $y-z$ plane for $Tu=6\%$ at $\bar x=0.36$. The increment of the contour values is $0.1$ for the velocity and $0.05$ for the temperature.}
\label{fig:con-high}
\end{figure}

Contours of the instantaneous $\tilde u$ and $\tilde T$ in $y-z$ planes are displayed in figure \ref{fig:con-low} for a moderate FVD level ($Tu$=1\%) and in figure \ref{fig:con-high} for a high FVD level ($Tu$=6\%). The contours are shown at phases where the disturbances obtain maximum amplitude and at sufficiently downstream locations where they have saturated. Figure \ref{fig:con-low} shows that, for the moderate FVD level $Tu=1\%,$ the velocity and temperature disturbances exhibit the typical mushroom shape in the concave case, while the bell shape, characteristic of streaky structures, appears in the flat-wall case. The flow remains largely undisturbed when the wall is convex.

We note that the mushroom shapes could be observed over a concave plate in a wind tunnel because of the long extent of the test section. However, the downstream locations of figures \ref{fig:con-low}$(a)$ and \ref{fig:con-low}$(d)$ are too large for these structures to be observed in practical turbomachinery applications because of the limited length of turbine blades. As shown by the abscissas at the top of figure \ref{fig:curvature-effect}$(a)$, locations beyond $\bar x=1$ considerably exceed the length of a turbine blade, estimated to be $x_s=1.65$ by using the flow parameters in the experiments of \citet{arts1990aero}.

Figure \ref{fig:con-high} shows that, for the high-intensity FVD level $Tu=6\%,$ the boundary-layer disturbances do not exhibit the typical mushroom shape for $\mathcal{G}=35.2$ and instead resemble the streaks evolving over a flat plate. This occurrence is due to the destabilising effect of the concave wall not being sufficiently intense to alter the character of the disturbances when the FVD level is large. Figures \ref{fig:con-high}$(c,f)$ show that the stabilising effect of the convex wall is also insignificant in the presence of high-intensity FVD as the nonlinear disturbances over convex walls also resemble streaks over a flat plate. This dynamics is in stark contrast with the quiet environment observed in figure \ref{fig:con-low}(c,f) for the convex-wall case at the lower FVD level $Tu=1\%$.

\subsection{Wall-shear stress and wall-heat transfer}
\label{sec:cf}
Motivated by the dominance of the velocity and temperature modes (0,0) observed in figure \ref{fig:fourier-modes}, we study the streamwise evolution of the skin-friction coefficient and Stanton number, defined as \citep{anderson2000hypersonic}
\begin{eqnarray}
\mathcal{C}_f
=\left.\dfrac{2\mu_w}{R_\Lambda}
{\dfrac{\p \left(U+r_t \hat u_{0,0}\right)}{ \p y}} \right|_{y=0},
\\
\mathcal{S}_t
=
\left.
\dfrac{\upkappa_w}{(T_{ad}-T_w) R_\Lambda \ Pr}{\dfrac{\p \left(T+r_t \hat \tau_{0,0}\right)}{ \p y}} \right|_{y=0},
\end{eqnarray}
where $\mu_w$ and $\upkappa_w$ are constant because the wall is isothermal.

\begin{figure}
\centering
\includegraphics[width=0.7\textwidth]{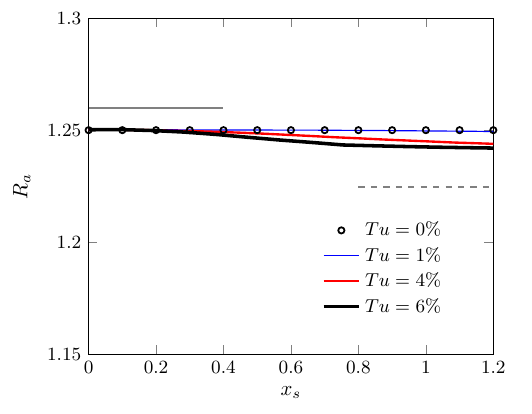}
\captionsetup{justification=raggedright}
\caption{Reynolds analogy factor along the streamwise direction for different FVD levels at $\mathcal{M}_\infty=0.69.$ The solid grey line indicates the Reynolds analogy factor for the incompressible laminar flow and the dashed line indicates the experimental measurement of an incompressible turbulent boundary layer by \citet{bons2005critical}.}
\label{fig:ra-tu}
\end{figure}

Another quantity of interest is the Reynolds analogy factor, $R_a=2\mathcal{S}_t/\mathcal{C}_f$ \citep{roy2006review}, shown in figure \ref{fig:ra-tu}. It can be utilised to obtain either $\mathcal{C}_f$ or $\mathcal{S}_t$ when the other quantity is known. \citet{bons2005critical} showed that the Reynolds analogy factor depends on the pressure gradient, but it is almost constant for a boundary layer without a pressure gradient. The solid grey line in figure \ref{fig:ra-tu} denotes the so-called Chilton–Colburn relation for incompressible laminar boundary layers, namely, $R_a=Pr^{-2/3}$ \citep{chilton1934mass}, based on experimental data. The Chilton–Colburn value is slightly higher than $R_a=1.25$, obtained using the Blasius boundary-layer theory. The dashed grey line denotes the value for turbulent boundary layers, reported by \citet{bons2005critical}. Figure \ref{fig:ra-tu} shows that the Reynolds analogy factor for nonlinear G\"ortler vortices slightly decreases downstream and lies between the laminar and turbulent values. This result is expected since the G\"ortler vortices develop in a transitional boundary layer. Our computations also show that, as the FVD level increases, the Reynolds analogy factor decreases. This behaviour is opposite to that found by \citet{bons2005critical} for turbulent boundary layers.

\begin{figure}
\centering
   \subfigure{
      \put(68,146){Present calculations}
      \put(0,146){$(a)$}
      \includegraphics[width=0.48\textwidth]{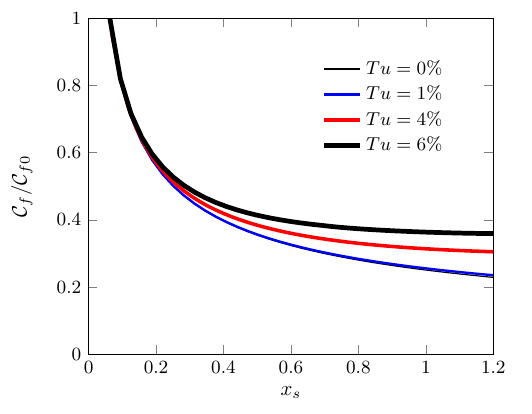}
      \label{fig:skin-friction-a}
  }
  \subfigure{
      \put(48,146){Measurements \& Simulations}
      \put(0,146){$(b)$}
      \includegraphics[width=0.48\textwidth]{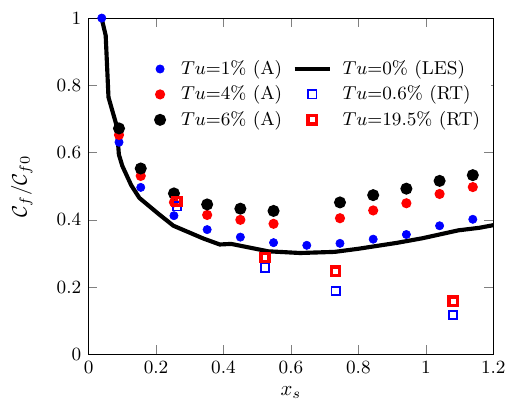}
      \label{fig:skin-friction-b}
   }
\captionsetup{justification=raggedright}
\caption{Comparison of ($a$) the computed skin-friction coefficients with ($b$) the experimental data of \citet{arts1990aero} (A) and \citet{radomsky2002detailed} (RT). The coefficients are normalised by the value $\mathcal{C}_{f0}$ at ${x}_s=0.06.$ The line in ($b$) shows the skin-friction coefficient computed by large eddy simulations (LES) without inflow disturbances \citep{bhaskaran2010large}.}
\label{fig:skin-friction}
\end{figure}

Figures \ref{fig:skin-friction}$(a,b)$ show the comparison between our computed skin-friction coefficients and other experimental and numerical data. In figure \ref{fig:skin-friction}$(a)$, the skin-friction coefficient is largely unaffected by the FVD level for $Tu$$\leq$1\% and it increases with $Tu$ for $Tu$$>$1\%. 
These results are consistent with the experimental data of \citet{radomsky2002detailed}. As evidenced in figure 8a of \citet{radomsky2002detailed}, their measured skin friction on the pressure side of a turbine blade for $Tu=0.6\%$ is almost the same as that of the laminar flow. Our figure \ref{fig:skin-friction}$(b)$ shows that their skin-friction coefficient is enhanced by an increase of FVD level. The decrease of skin-friction coefficient with $x_s$ is also in agreement with our result in figure \ref{fig:skin-friction}$(a)$ as the pressure gradient is not included in our calculations and it is very small in \citet{radomsky2002detailed}. The main difference is that our skin-friction coefficient becomes almost independent of the streamwise location for $Tu=6\%$, while their skin-friction coefficient keeps decreasing at all FVD levels, for $Tu$ as large as 19.5\%.

Figure \ref{fig:skin-friction}$(b)$ also shows the experimental data of \citet{arts1990aero}. As the wall-shear stress was not measured by \citet{arts1990aero}, we have used their wall-heat transfer data and computed the skin-friction coefficients via our Reynolds analogy factors. Considering that the Reynolds analogy is not strictly valid in pressure-gradient and transitional flows, our estimate of the skin-friction coefficient can only be regarded as qualitative. Their skin-friction coefficients are enhanced as the FVD level increases, consistently with our results, and grow downstream following the initial decay. This result is markedly different from the decaying trends obtained in our computations and reported by \citet{radomsky2002detailed}. A reason behind this discrepancy is the difference in geometry of the turbine blades, which leads to different pressure gradients. In the experiments of \citet{radomsky2002detailed}, the pressure gradient is significantly lower than that of \citet{arts1990aero}, while in our computations the pressure gradient is absent. The direct numerical simulations conducted by \citet{zhao2020bypass} led to skin-friction coefficients that were independent of the FVD level (refer to their figure 7), a result that remains unexplained.

\begin{figure}
\centering
\subfigure{
    \put(68,146){Present calculations}
    \put(0,146){$(a)$}
\includegraphics[width=0.48\textwidth]{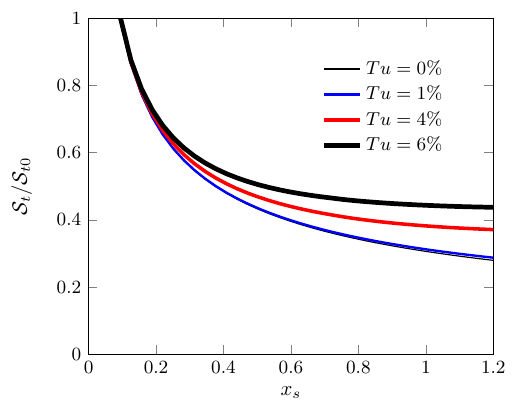}
}
\subfigure{
   \put(48,146){Measurements \& Simulations}
   \put(0,146){$(b)$}
\includegraphics[width=0.48\textwidth]{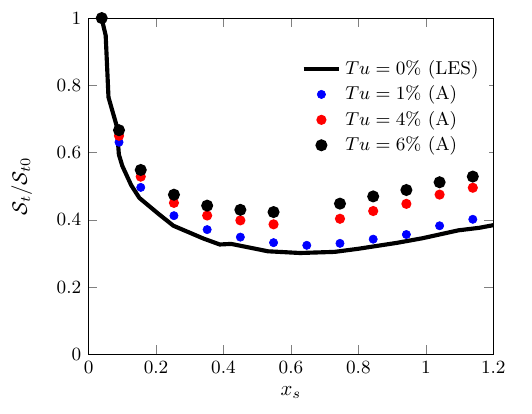}
}
\captionsetup{justification=raggedright}
\caption{Comparison of ($a$) the computed Stanton numbers with ($b$) the experimental data of \citet{arts1990aero} (A). The Stanton numbers are normalised by the value $\mathcal{S}_{t0}$ at ${x}_s=0.06.$ The line in ($b$) shows the Stanton number computed by large eddy simulations (LES) without inflow disturbances \citep{bhaskaran2010large}.}
\label{fig:heat-flux}
\end{figure}

The wall-heat flux over turbine-blade surfaces is also of interest since experimental measurements have shown its significant enhancement over pressure surfaces \citep{arts1990aero,butler2001effect}. As with the skin-friction coefficient, our computed Stanton numbers are unaffected by the change of FVD level up to $Tu=1\%$ and are enhanced by the FVD level for $Tu>1\%$, as reported in figure \ref{fig:heat-flux}$(a)$. As shown in figure \ref{fig:heat-flux}$(b)$, the large eddy simulations of \citet{bhaskaran2010large} resulted in an intensified laminar wall-heat transfer following the initial decay, a phenomenon not observed in our computations. As with the skin-friction coefficient, this discrepancy arises from the absence of streamwise pressure gradient in our case. Figure \ref{fig:heat-flux}$(b)$ also depicts the experimental data of \citet{arts1990aero}. Their wall-heat transfer for $Tu=1\%$ is only slightly larger than the laminar value and increases with the FVD level $Tu>1\%$. Both results agree with our computations and with the response of the skin friction to a change of FVD level. Ours is the first numerical verification of the effect of FVD level in the experiments of \citet{arts1990aero}, although the effect of streamwise pressure gradient needs to be further investigated.

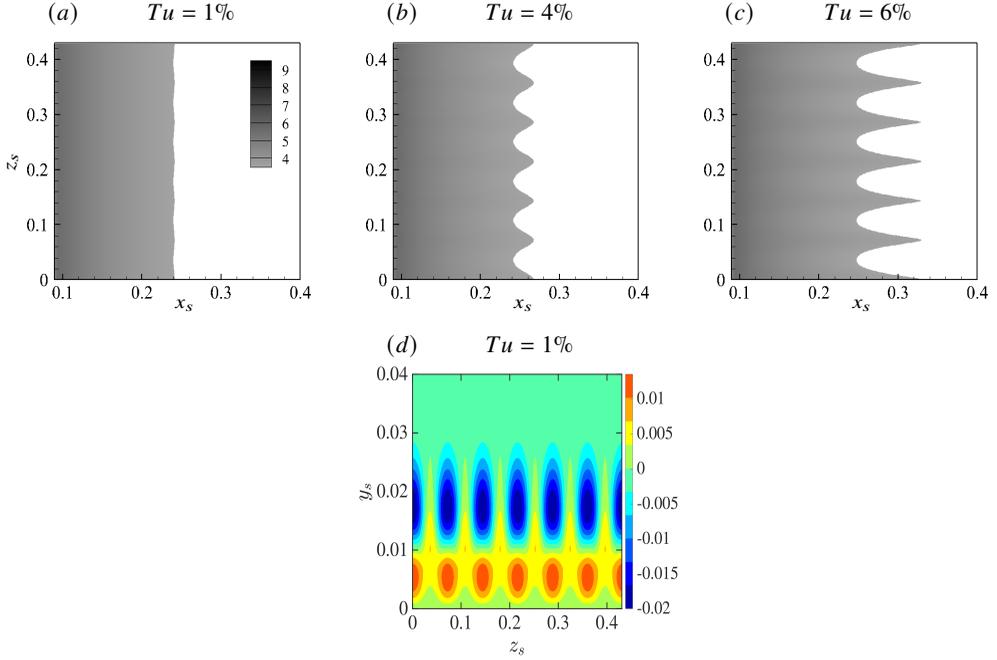
\begin{figure}
\centering
\subfigure{
\includegraphics[width=0.32\textwidth,height=0.3\textwidth]{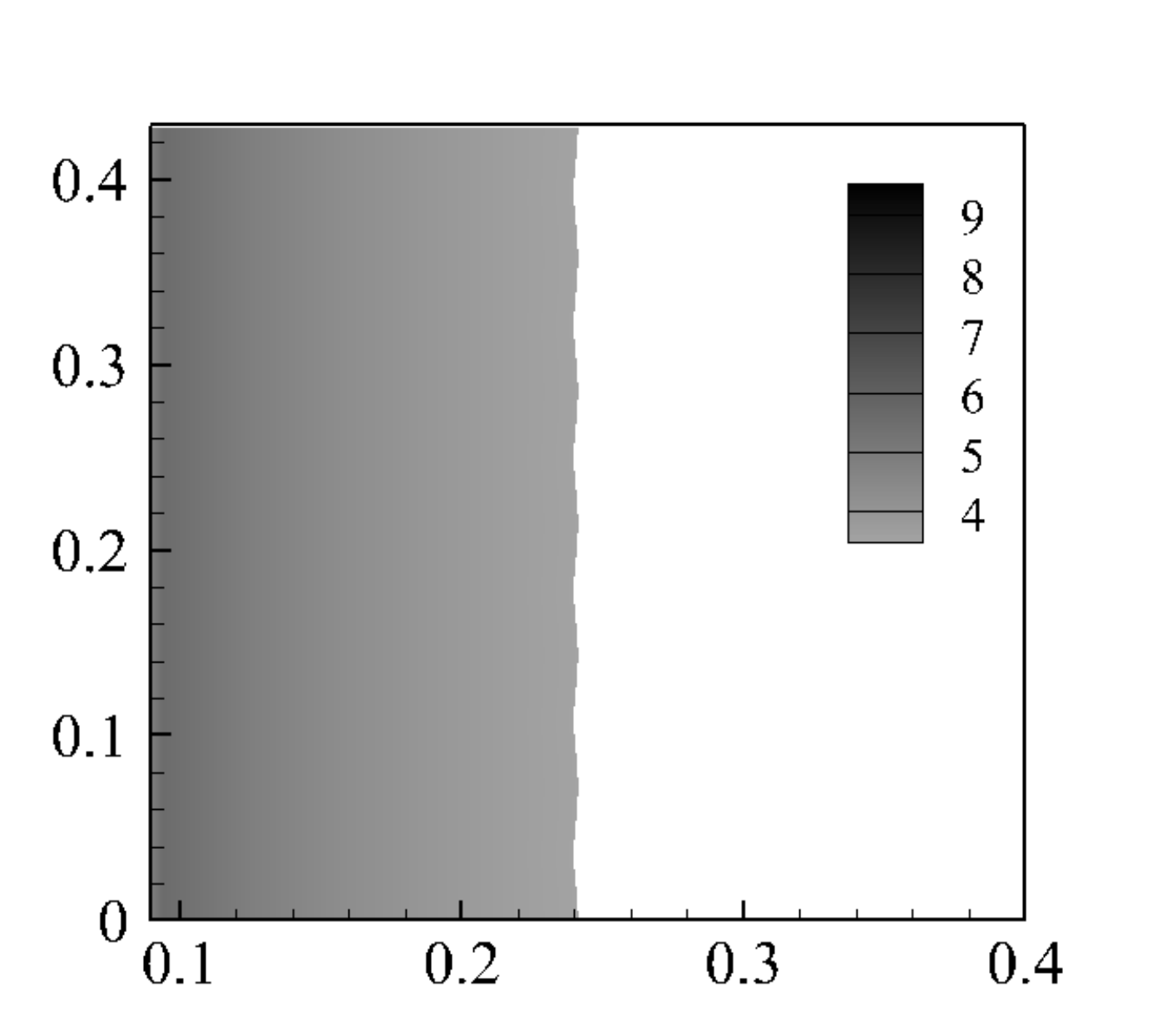}
\put(-110,110){$(a)$}
\put(-62,2){\scriptsize $x_s$}
\put(-72,110){$Tu=1\%$}
\put(-125,56){\scriptsize \rotatebox[origin=c]{90}{$z_s$}}
}
\subfigure{
\includegraphics[width=0.32\textwidth,height=0.3\textwidth]{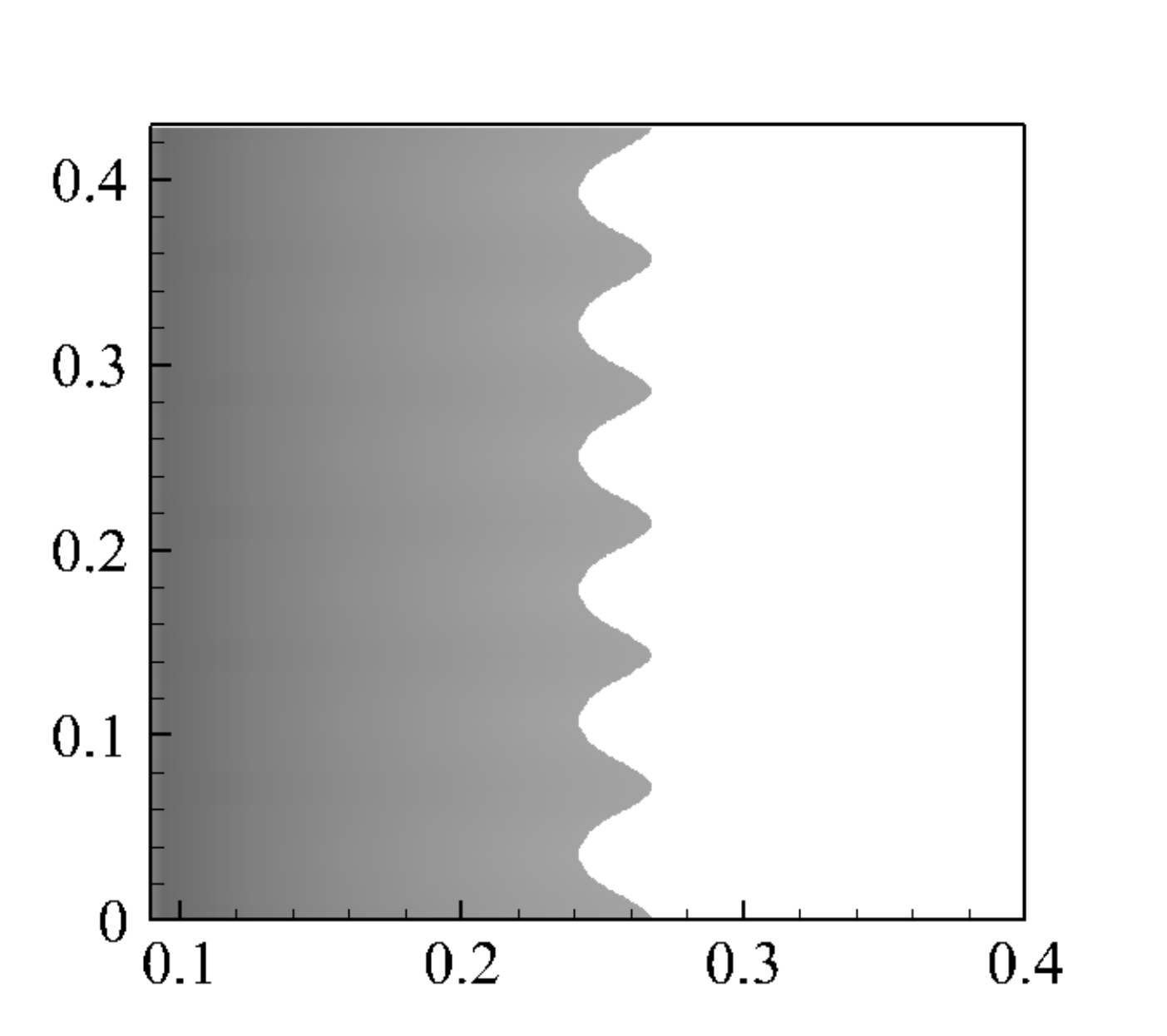}
\put(-110,110){$(b)$}
\put(-62,2){\scriptsize $x_s$}
\put(-72,110){$Tu=4\%$}
}
\vspace{-1mm}
\subfigure{
\includegraphics[width=0.32\textwidth,height=0.3\textwidth]{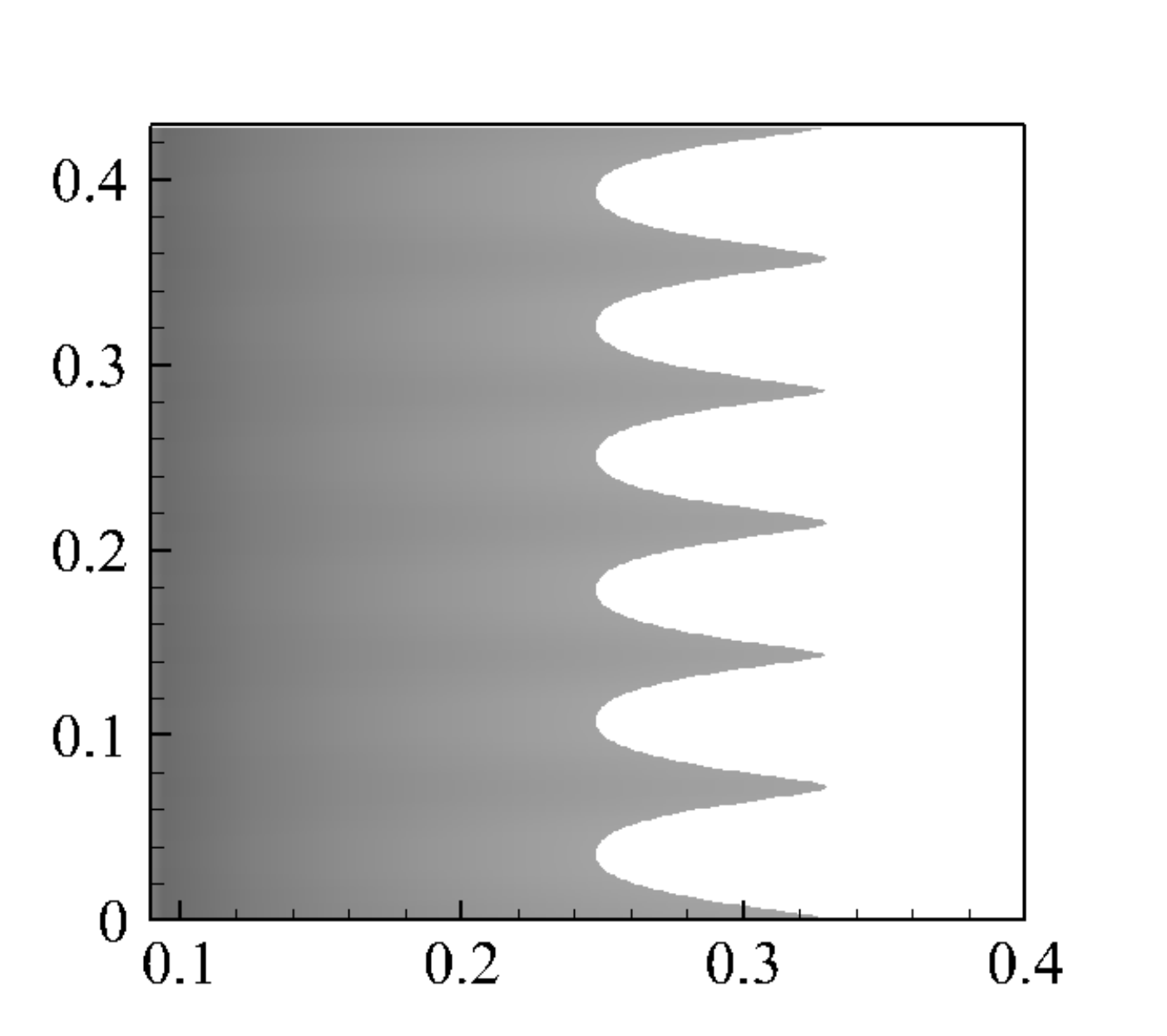}
\put(-110,110){$(c)$}
\put(-62,2){\scriptsize $x_s$}
\put(-72,110){$Tu=6\%$}
}
\subfigure{
\includegraphics[width=0.32\textwidth,height=0.3\textwidth]{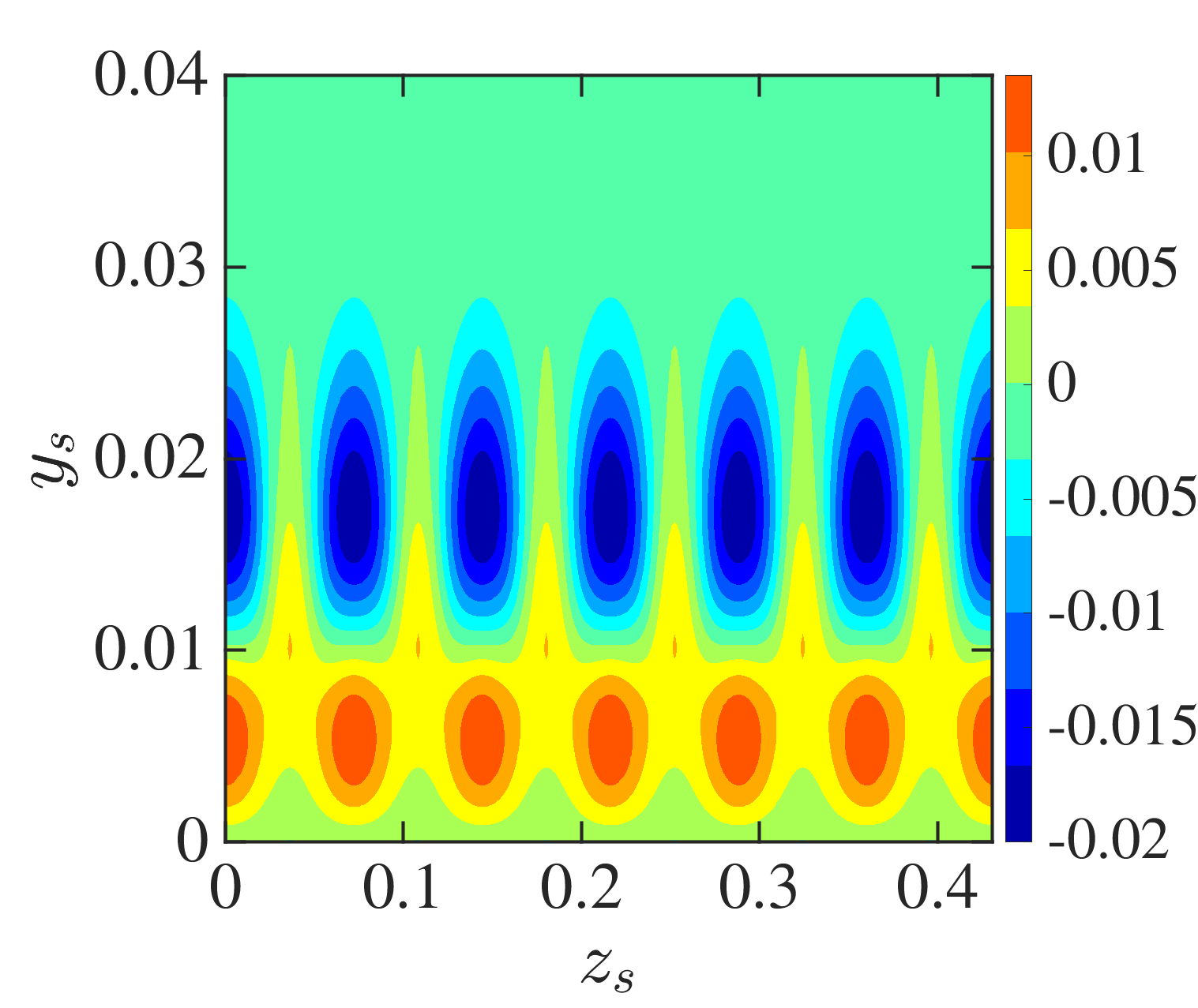}
\put(-110,115){$(d)$}
\put(-72,115){$Tu=1\%$}
}
\captionsetup{justification=raggedright}
\caption{($a-c$) Time-averaged wall-shear stress $\mathcal{F}(x_s,z_s)$, defined in equation \eqref{eq:skin-friction}, for different FVD levels. Panel ($d$) shows the contour of the timed-averaged streamwise velocity streaks, given by mode (0,2). The wall-normal coordinate is $y_s=y^*/C_{ax}^*$. The G\"ortler number is $\mathcal{G}=$35.2.}
\label{fig:skin-friction-3d}
\end{figure}

Figures \ref{fig:skin-friction-3d}($a-c$) show the time-averaged wall-shear stress
\begin{equation}
\label{eq:skin-friction}
\mathcal{F}(x_s,z_s)
=
\mu_w \pipe{\frac{\p U}{\p y}}_{y=0}
+
\mu_w r_t \sum_{n=-\infty}^{\infty} \pipe{\frac{\p \hat u_{0,n}}{\p y}}_{y=0} {\rm e}^{{\rm i} n k_z z},
\end{equation}
where $z_s=z^*/C_{ax}^*$. As the leading edge is approached, $\mathcal{F} \sim \mu_w \left(F''(0)/T_w\right)\sqrt{R_\Lambda/(2x)}=1.70/\sqrt{x_s}.$ The region close to the leading edge experiences an intense $\mathcal{F}$ that is almost uniform along the spanwise direction. Further downstream, a distinct streaky structure emerges, characterised by alternating streamwise-elongated low-$\mathcal{F}$ and high-$\mathcal{F}$ regions. These patterns become longer as the FVD level increases from $Tu=1\%$ to $Tu=6\%$. They are induced by the steady mode (0,2), as shown in figure \ref{fig:skin-friction-3d}($d$).

\begin{figure}
\centering
\subfigure{
\includegraphics[width=0.32\textwidth,height=0.3\textwidth]{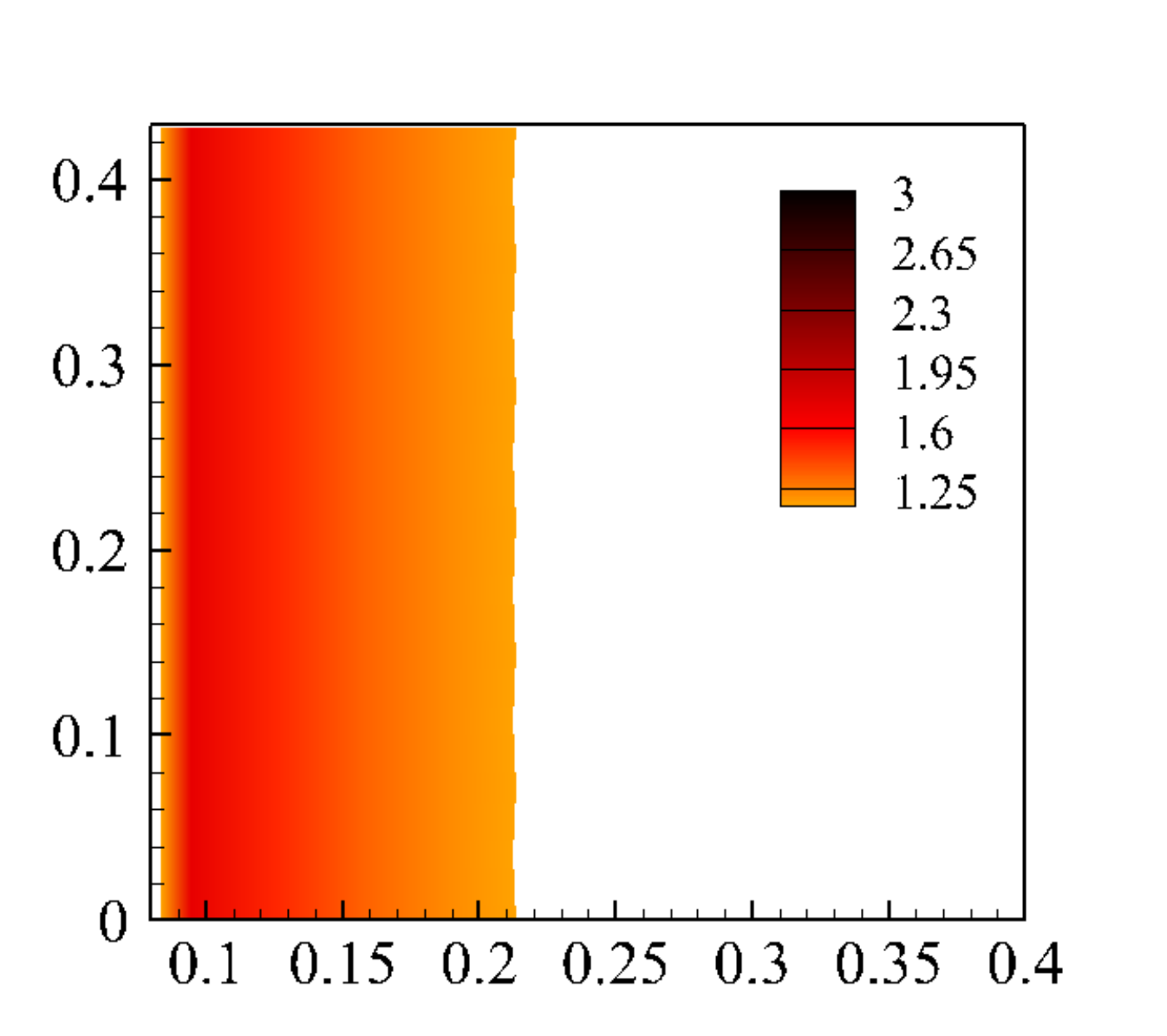}
\put(-110,110){$(a)$}
\put(-62,-1){\scriptsize $x_s$}
\put(-72,110){$Tu=1\%$}
\put(-125,56){\scriptsize \rotatebox[origin=c]{90}{$z_s$}}
}
\subfigure{
\includegraphics[width=0.32\textwidth,height=0.3\textwidth]{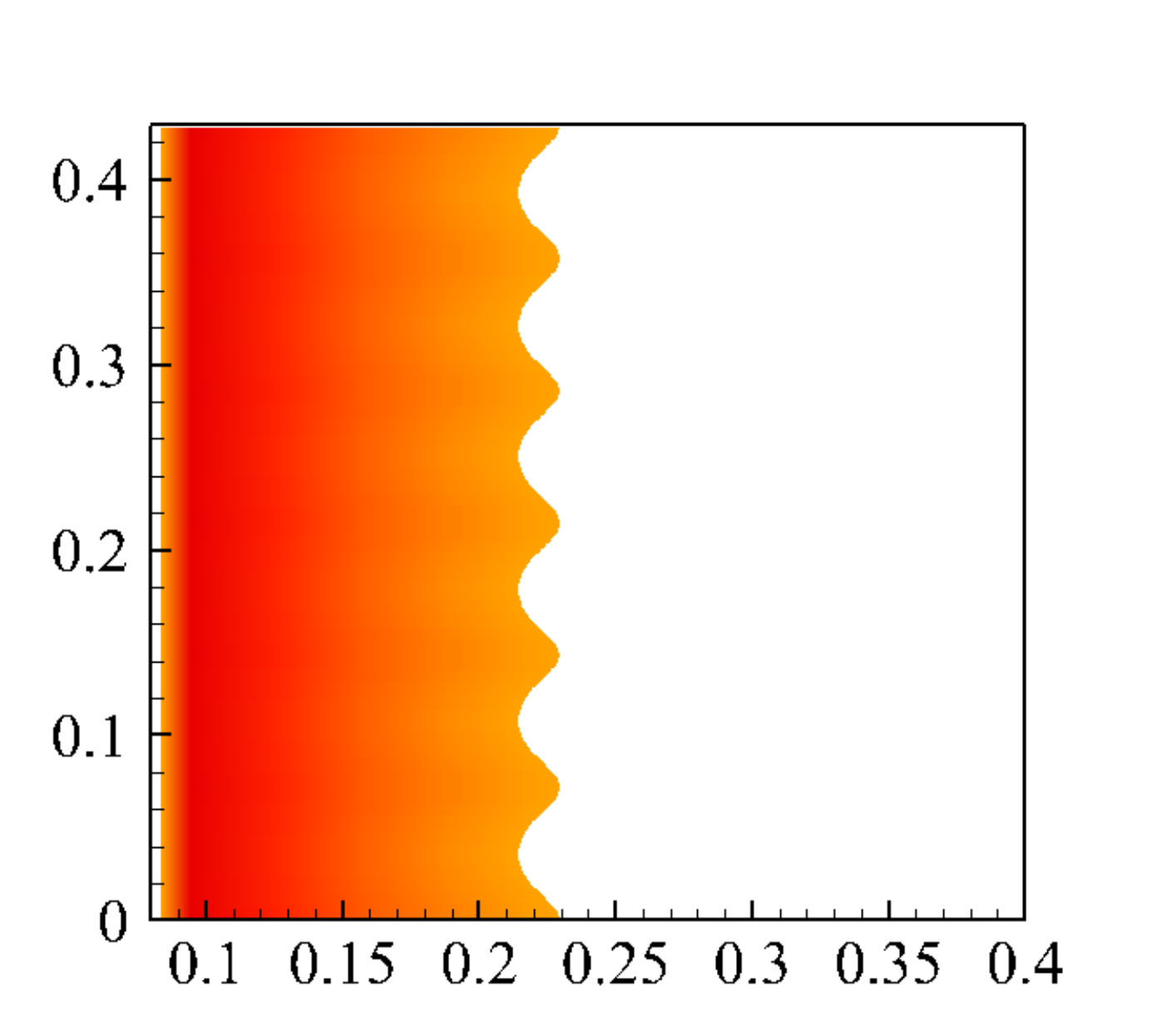}
\put(-110,110){$(b)$}
\put(-62,-1){\scriptsize $x_s$}
\put(-72,110){$Tu=4\%$}
}
\vspace{-1mm}
\subfigure{
\includegraphics[width=0.32\textwidth,height=0.3\textwidth]{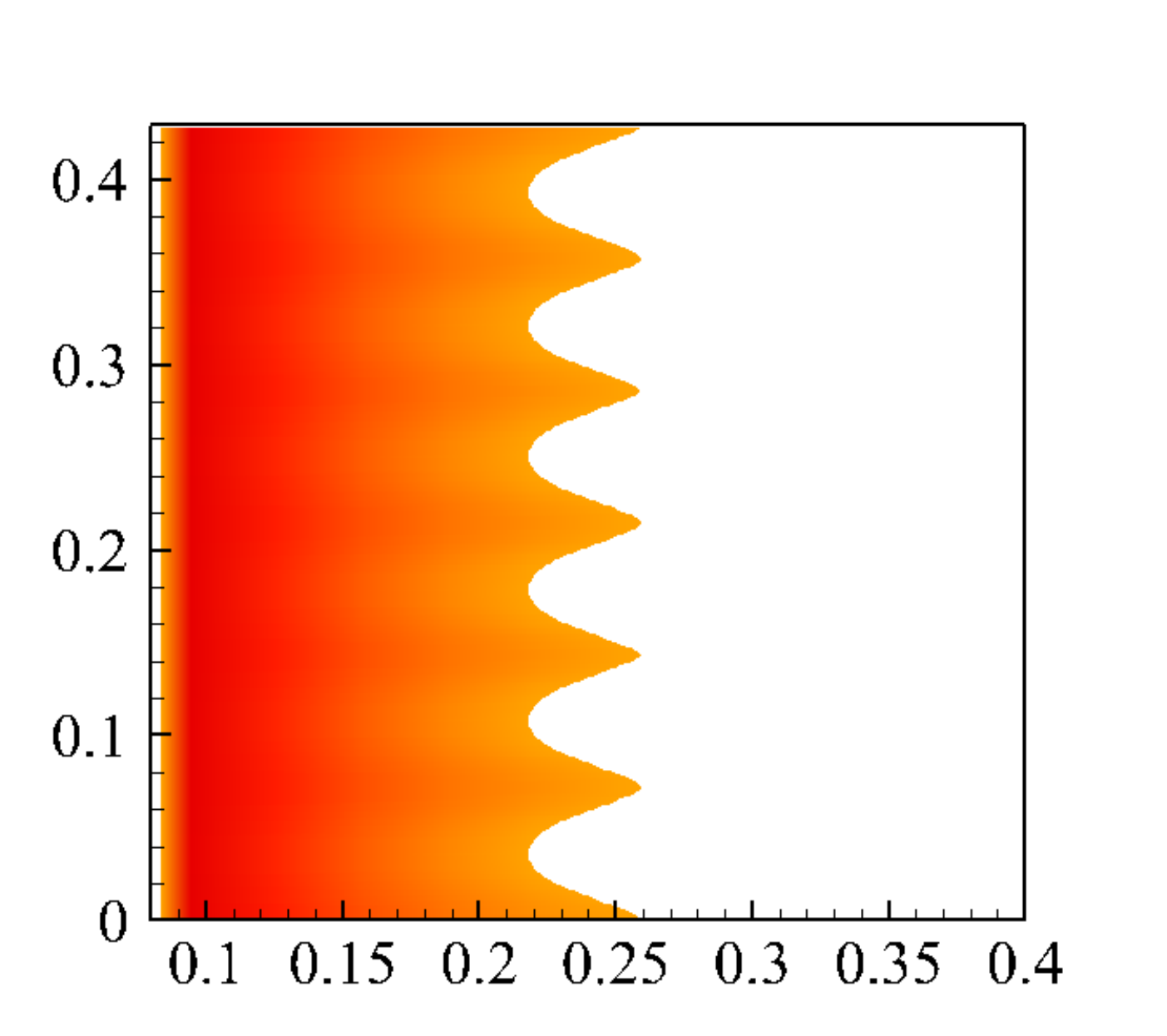}
\put(-110,110){$(c)$}
\put(-62,-1){\scriptsize $x_s$}
\put(-72,110){$Tu=6\%$}
}
\subfigure{
\includegraphics[width=0.32\textwidth,height=0.3\textwidth]{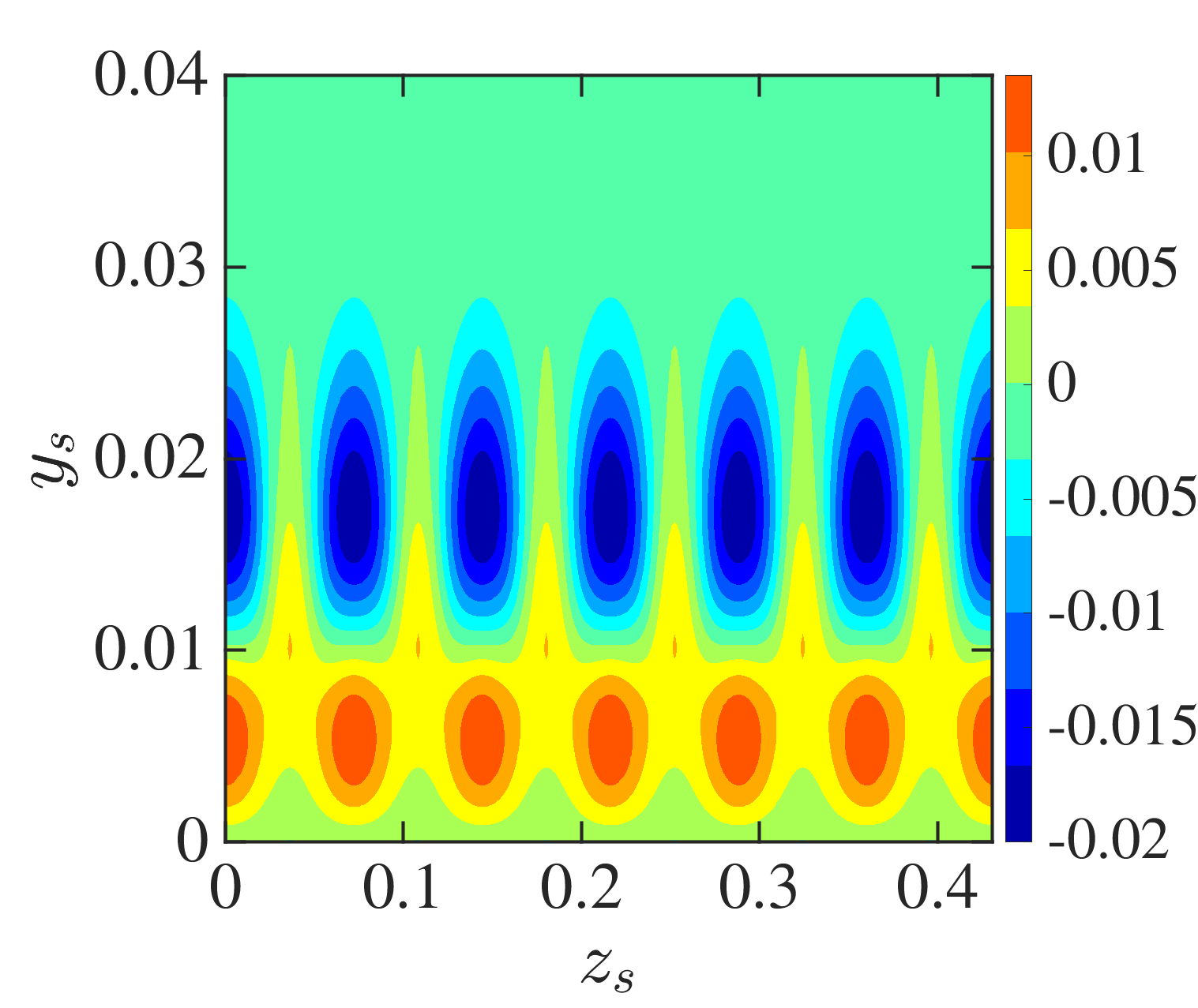}
\put(-110,115){$(d)$}
\put(-72,115){$Tu=1\%$}
}
\subfigure{
\includegraphics[width=0.32\textwidth,height=0.24\textwidth,trim=-6cm -6cm -4cm 5.5cm]{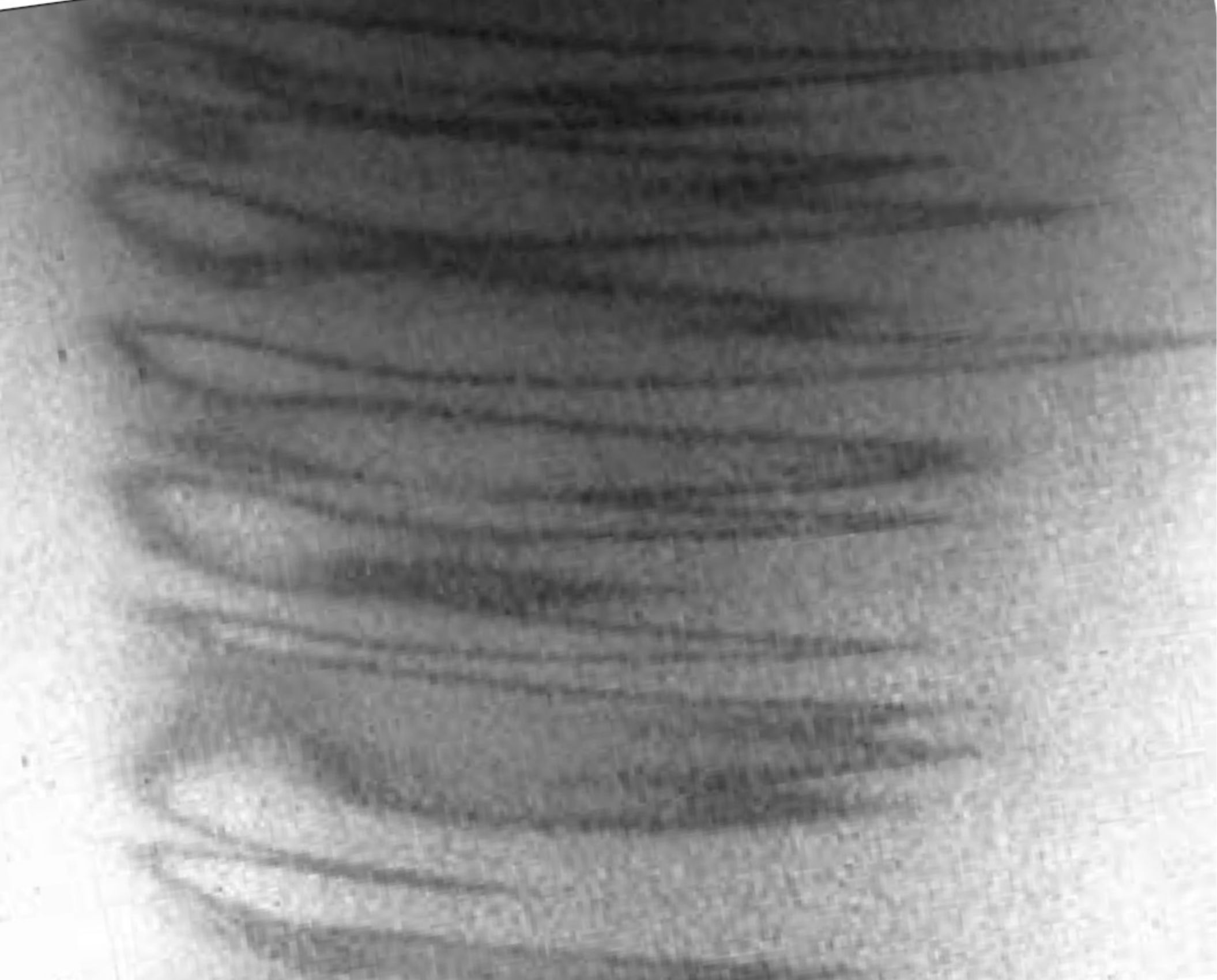}
\put(-110,115){$(e)$}
}
\captionsetup{justification=raggedright}
\vspace{-4mm}
\caption{($a-c$) Absolute value of the time-averaged wall-heat transfer, $|\mathcal{Q}(x_s,z_s)|$, defined in equation \eqref{eq:hot-fingers}, for different FVD levels. Panel $(d)$ is a contour of the timed-averaged temperature streaks, given by mode (0,2). Panel ($e$) shows the experimental measurements of \citet{butler2001effect}. The G\"ortler number is $\mathcal{G}=$35.2.}
\label{fig:skin-heat-3d}
\end{figure}

Figures \ref{fig:skin-heat-3d}($a-c$) show the time-averaged wall-heat transfer
\begin{equation}
\label{eq:hot-fingers}
\mathcal{Q}(x_s,z_s)
=
-\upkappa_w\pipe{\frac{\p T}{\p y}}_{y=0}
-\upkappa_w r_t \sum_{n=-\infty}^{\infty} \pipe{\frac{\p \hat \tau_{0,n}}{\p y}}_{y=0} {\rm e}^{{\rm i} n k_z z}.
\end{equation}
As the leading edge is approached, $\mathcal{Q} \sim -\upkappa_w \left(T'(0)/T_w\right)\sqrt{R_\Lambda/(2x)}=-0.41/\sqrt{x_s}.$ The spanwise streaky pattern observed in figure \ref{fig:skin-friction-3d}$(a-c)$ for the wall-shear stress is also detected for the wall-heat flux $\mathcal{Q}$, although $\mathcal{Q}$ is less affected by the FVD level than $\mathcal{F}$. Similarly to $\mathcal{F}$, the wall-heat flux modulation is induced by the steady mode (0,2), as shown in figure \ref{fig:skin-heat-3d}$(d)$. Our calculations qualitatively reproduce the experimental findings by \citet{butler2001effect}, shown in figure \ref{fig:skin-heat-3d}$(e)$ and also discussed in \citet{baughn1995experimental}. These streaky thermal patterns were obtained by liquid crystals on the pressure side of a turbine blade and have been termed hot fingers. Similarly to our numerical results, the picture in figure \ref{fig:skin-heat-3d}$(e)$ shows that the high-$\mathcal{Q}$ regions are elongated in the streamwise direction. However, the hot fingers in the experiments display a thin shape upstream before broadening downstream, a feature not observed in our numerical results. This difference could be ascribed to the variation of the streamwise pressure gradient along the blade and to the full spectrum of free-stream turbulence in the experiments, effects that are not included in our computations. \citet{butler2001effect} and \citet{baughn1995experimental} realised the importance of the FVD intensity on the formation of these patterns, although the occurrence of G\"ortler vortices was not confirmed.

High-frequency secondary-instability disturbances may influence the trailing edge of the hot fingers, potentially inducing small serrated structures as those depicted in figure \ref{fig:skin-heat-3d}($e$). These smaller structures are, however, less significant than the low-frequency components of the streaks in the formation of the hot fingers and are not computed herein because of our low-frequency assumption. They are discussed in \citet{huang2021inner} and \citet{feng2024investigation}.

\begin{figure}
\centering
\subfigure{
\includegraphics[width=0.32\textwidth,height=0.3\textwidth]{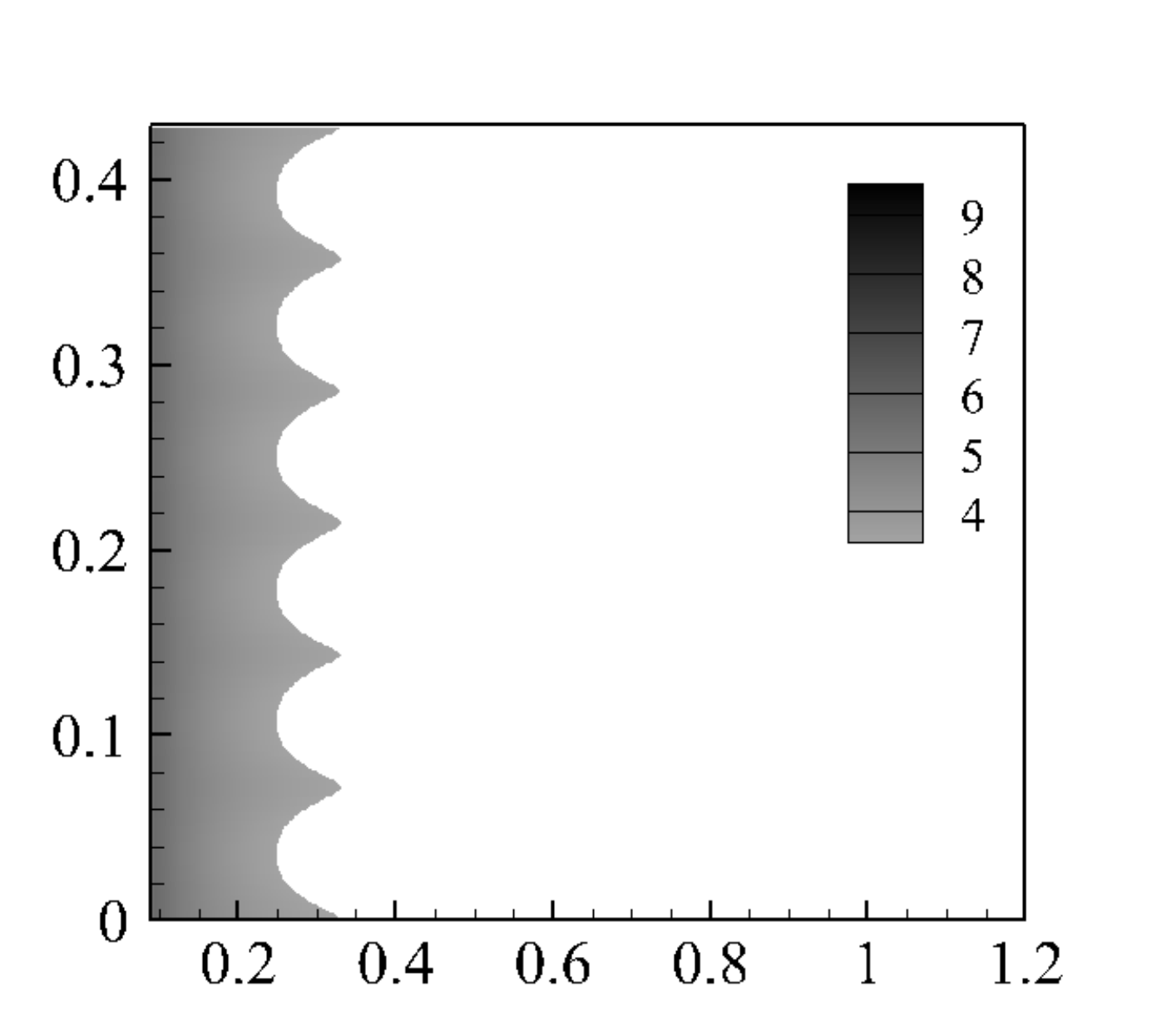}
\put(-120,110){$(a)$}
\put(-62,0){\scriptsize $x_s$}
\put(-76,110){$\mathcal{M}_\infty=0$}
\put(-125,58){\scriptsize \rotatebox[origin=c]{90}{$z_s$}}
}
\subfigure{
\includegraphics[width=0.32\textwidth,height=0.3\textwidth]{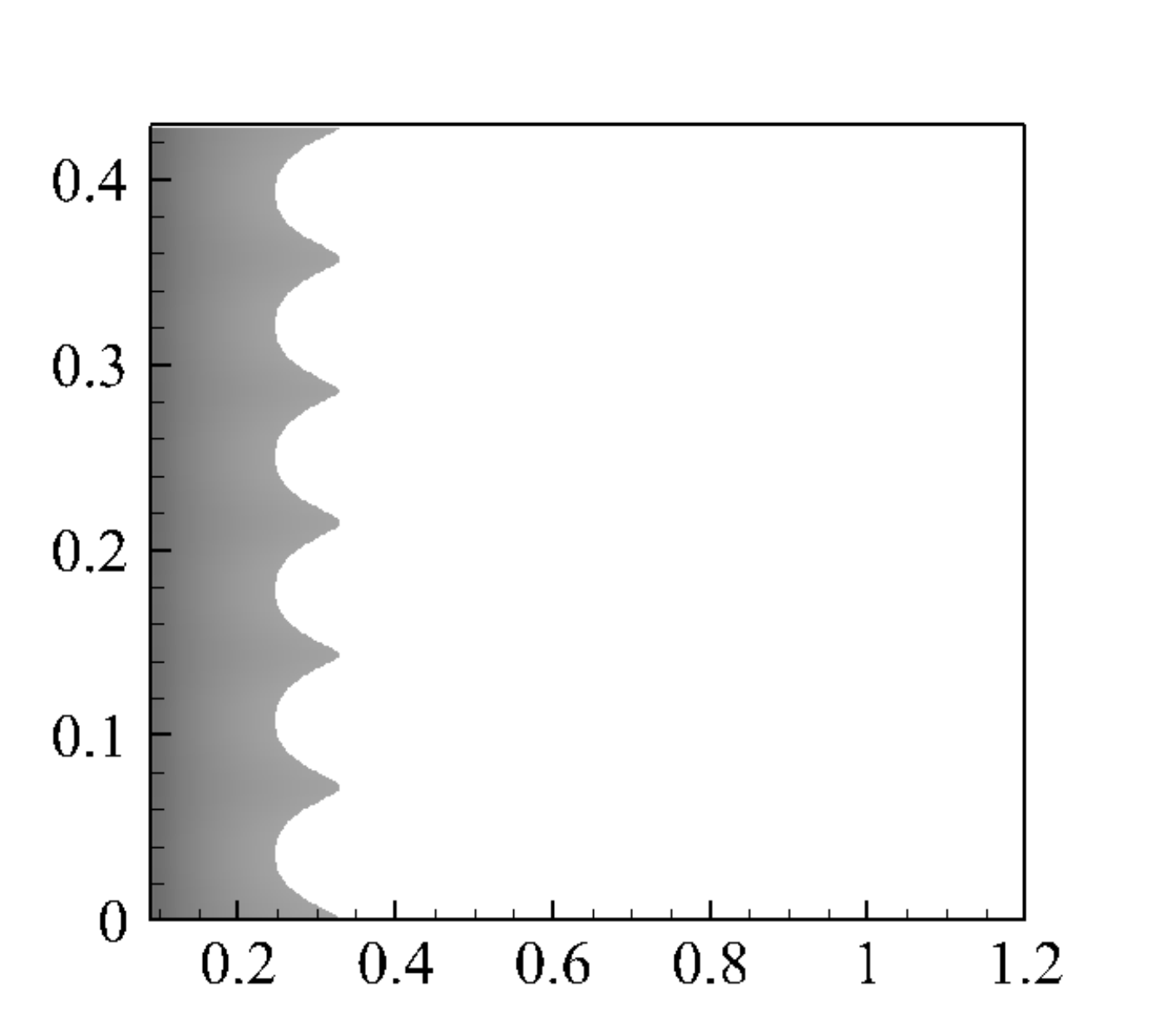}
\put(-120,110){$(b)$}
\put(-62,0){\scriptsize $x_s$}
\put(-80,110){$\mathcal{M}_\infty=0.69$}
\put(-125,58){\scriptsize \rotatebox[origin=c]{90}{$z_s$}}
}
\vspace{0mm}
\subfigure{
\includegraphics[width=0.32\textwidth,height=0.3\textwidth]{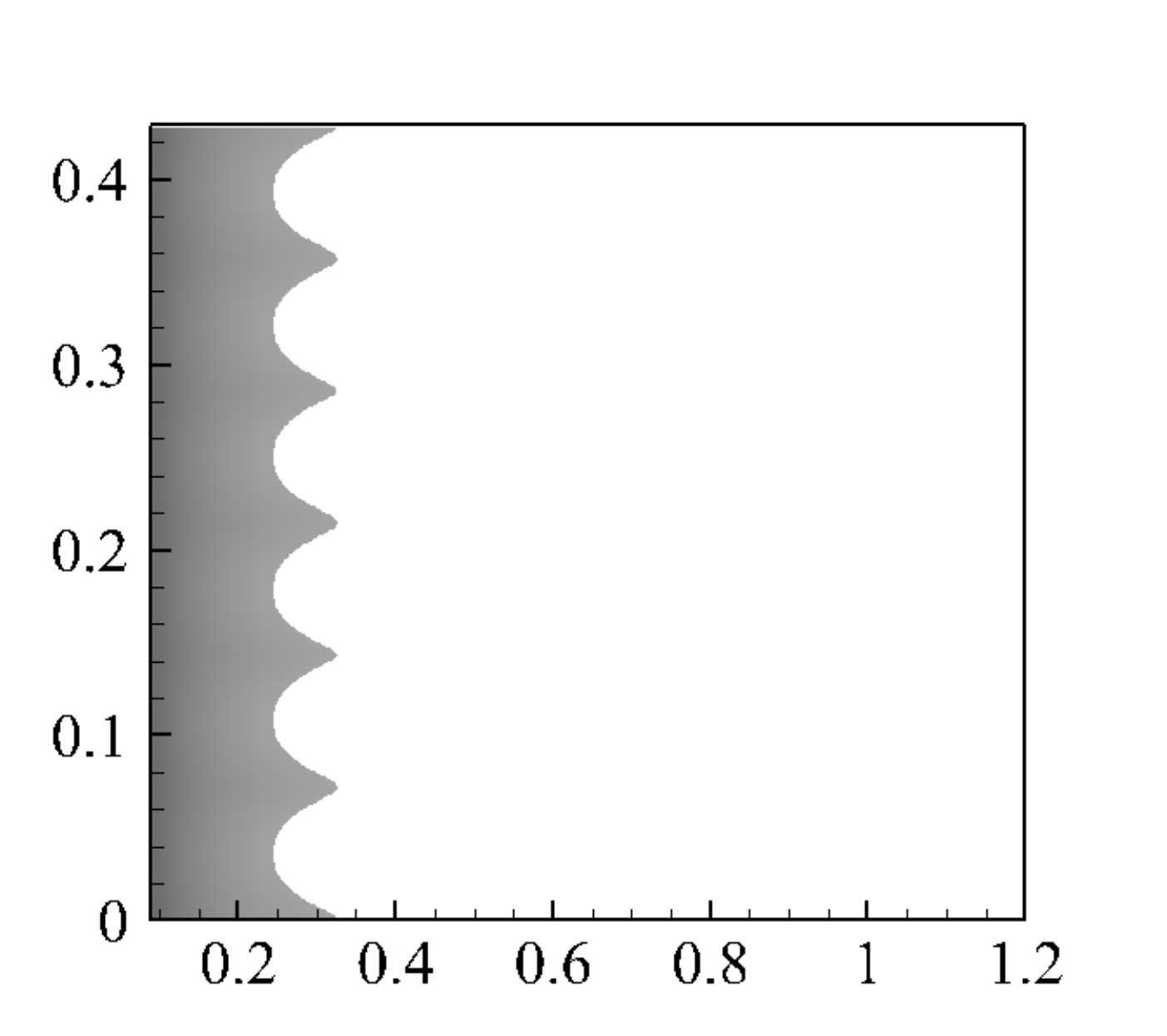}
\put(-120,110){$(c)$}
\put(-62,0){\scriptsize $x_s$}
\put(-82,110){$\mathcal{M}_\infty=1.1$}
\put(-125,58){\scriptsize \rotatebox[origin=c]{90}{$z_s$}}
}
\subfigure{
\includegraphics[width=0.32\textwidth,height=0.3\textwidth]{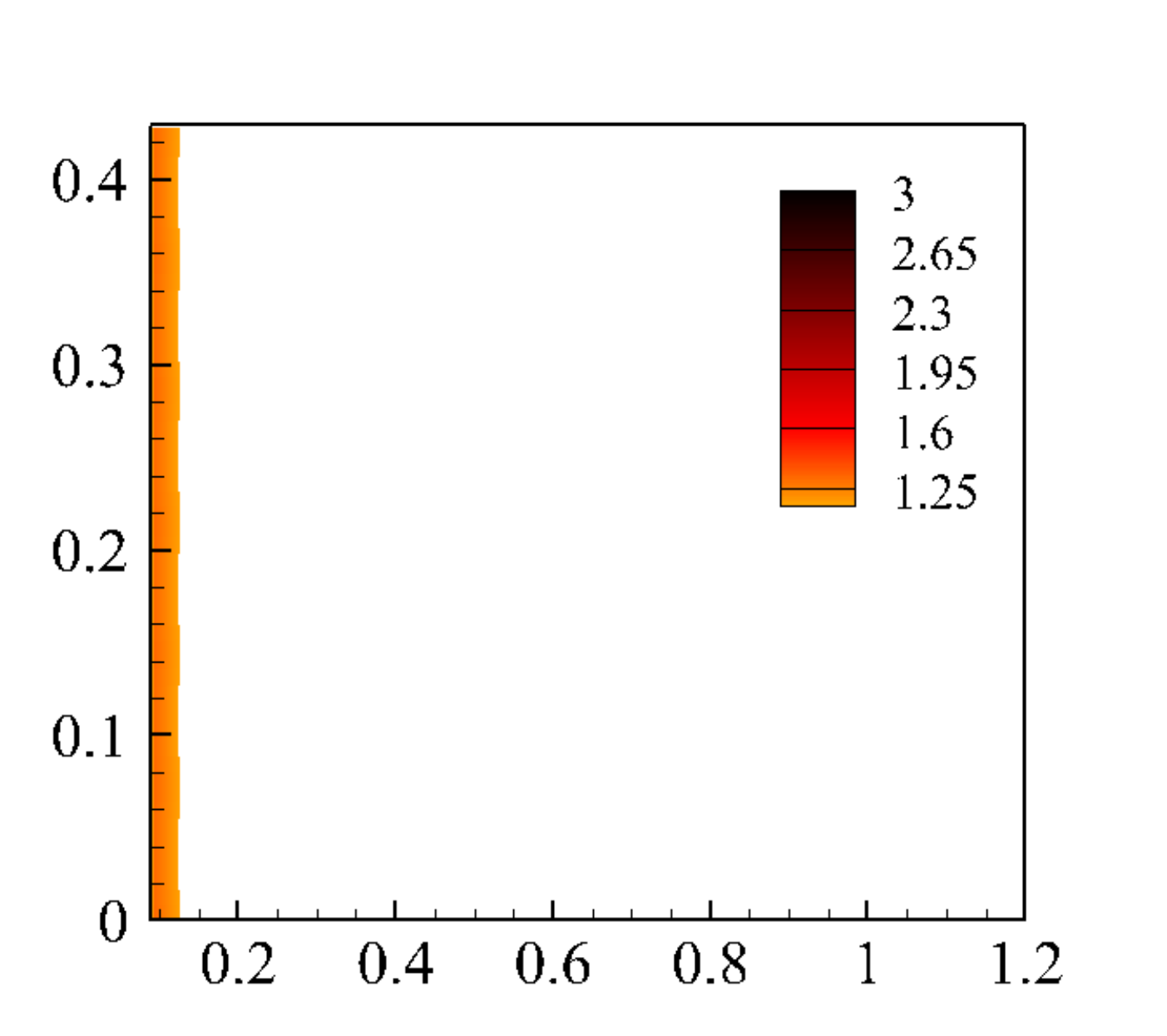}
\put(-120,110){$(d)$}
\put(-62,0){\scriptsize $x_s$}
\put(-76,110){$\mathcal{M}_\infty=0$}
\put(-125,58){\scriptsize \rotatebox[origin=c]{90}{$z_s$}}
}
\subfigure{
\includegraphics[width=0.32\textwidth,height=0.3\textwidth]{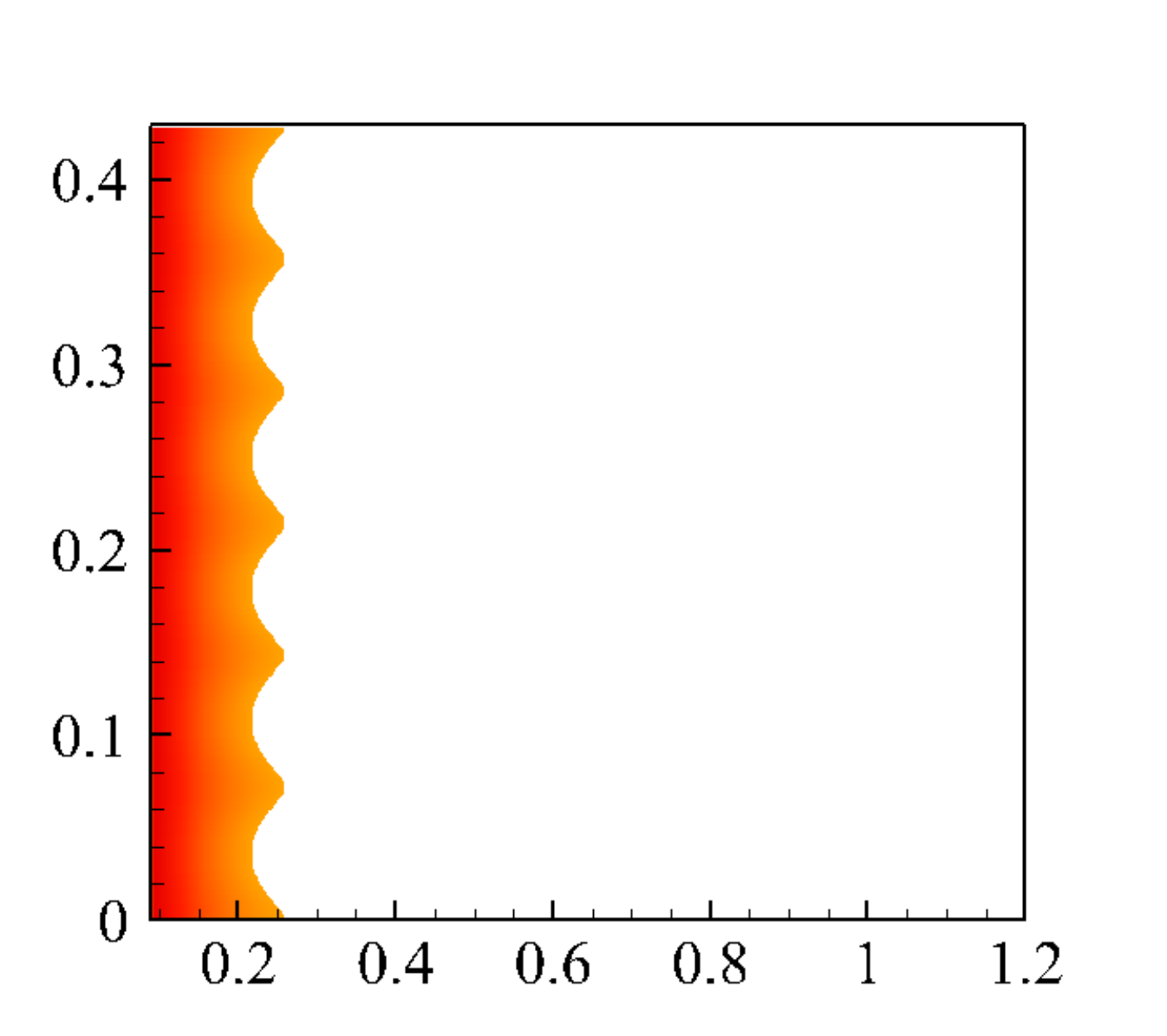}
\put(-120,110){$(e)$}
\put(-62,0){\scriptsize $x_s$}
\put(-80,110){$\mathcal{M}_\infty=0.69$}
\put(-125,58){\scriptsize \rotatebox[origin=c]{90}{$z_s$}}
}
\subfigure{
\includegraphics[width=0.32\textwidth,height=0.3\textwidth]{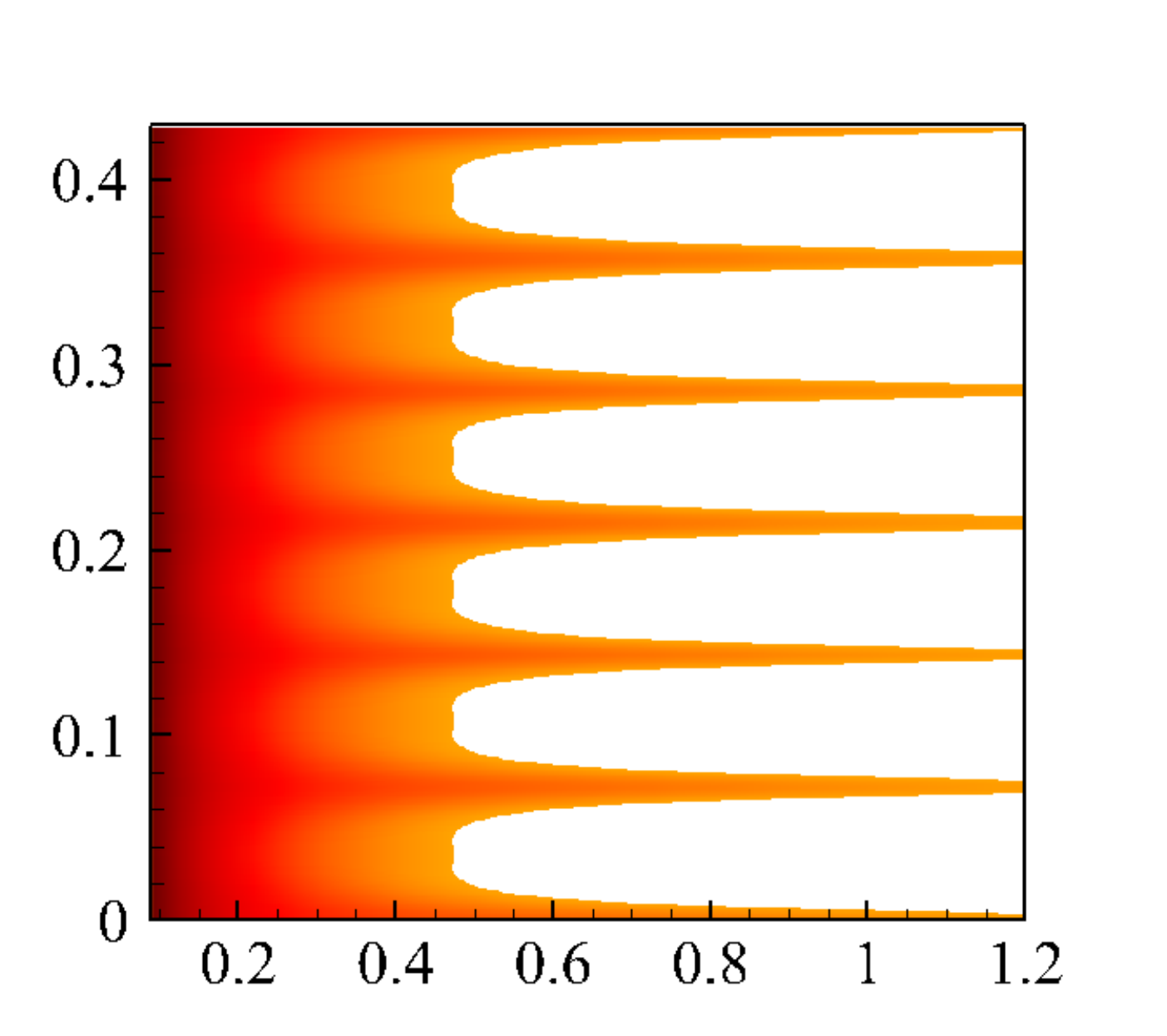}
\put(-120,110){$(f)$}
\put(-62,0){\scriptsize $x_s$}
\put(-82,110){$\mathcal{M}_\infty=1.1$}
\put(-125,58){\scriptsize \rotatebox[origin=c]{90}{$z_s$}}
}
\captionsetup{justification=raggedright}
\vspace{-4mm}
\caption{($a-c$) Time-averaged wall-shear stress, $\mathcal{F}(x_s,z_s)$, defined in equation \eqref{eq:skin-friction},  and ($d-f$) absolute value of the time-averaged wall-heat transfer, $|\mathcal{Q}(x_s,z_s)|$, defined in equation \eqref{eq:hot-fingers}. The numerical data are for ($a,d$) $\mathcal{M}_\infty=0$, ($b,e$) $\mathcal{M}_\infty=0.69$ and ($c,f$) $\mathcal{M}_\infty=1.1$.}
\label{fig:sf-ht-mach}
\end{figure}
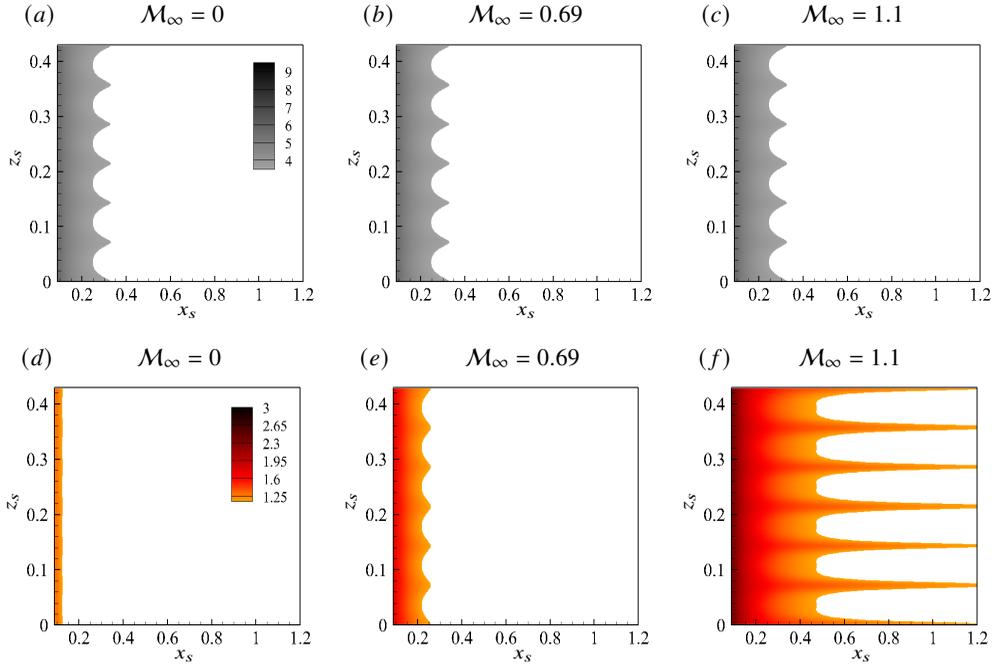

Figure \ref{fig:sf-ht-mach} illustrates the influence of Mach number on the enhancement of the spanwise modulated patterns of the wall-shear stress and the wall-heat flux. As the Mach number increases at a constant Reynolds number, the skin friction is not affected, while the wall-heat flux is significantly enhanced. We conclude that increasing the turbulence level enhances both the wall-shear stress and the wall-heat transfer, whereas increasing the Mach number only enhances the wall-heat transfer.

\subsection{Occurrence map for G\"ortler vortices and streaks}

\begin{figure}
\centering
\subfigure{
\includegraphics[width=0.7\textwidth]{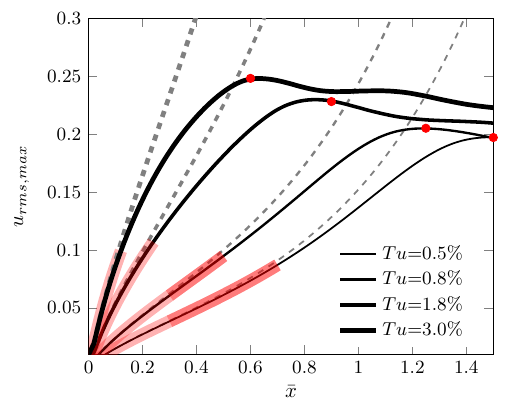}
\put(-258,197){$(a)$}
\vspace{-5mm}
}
\subfigure{
\includegraphics[width=0.32\textwidth,height=0.3\textwidth]{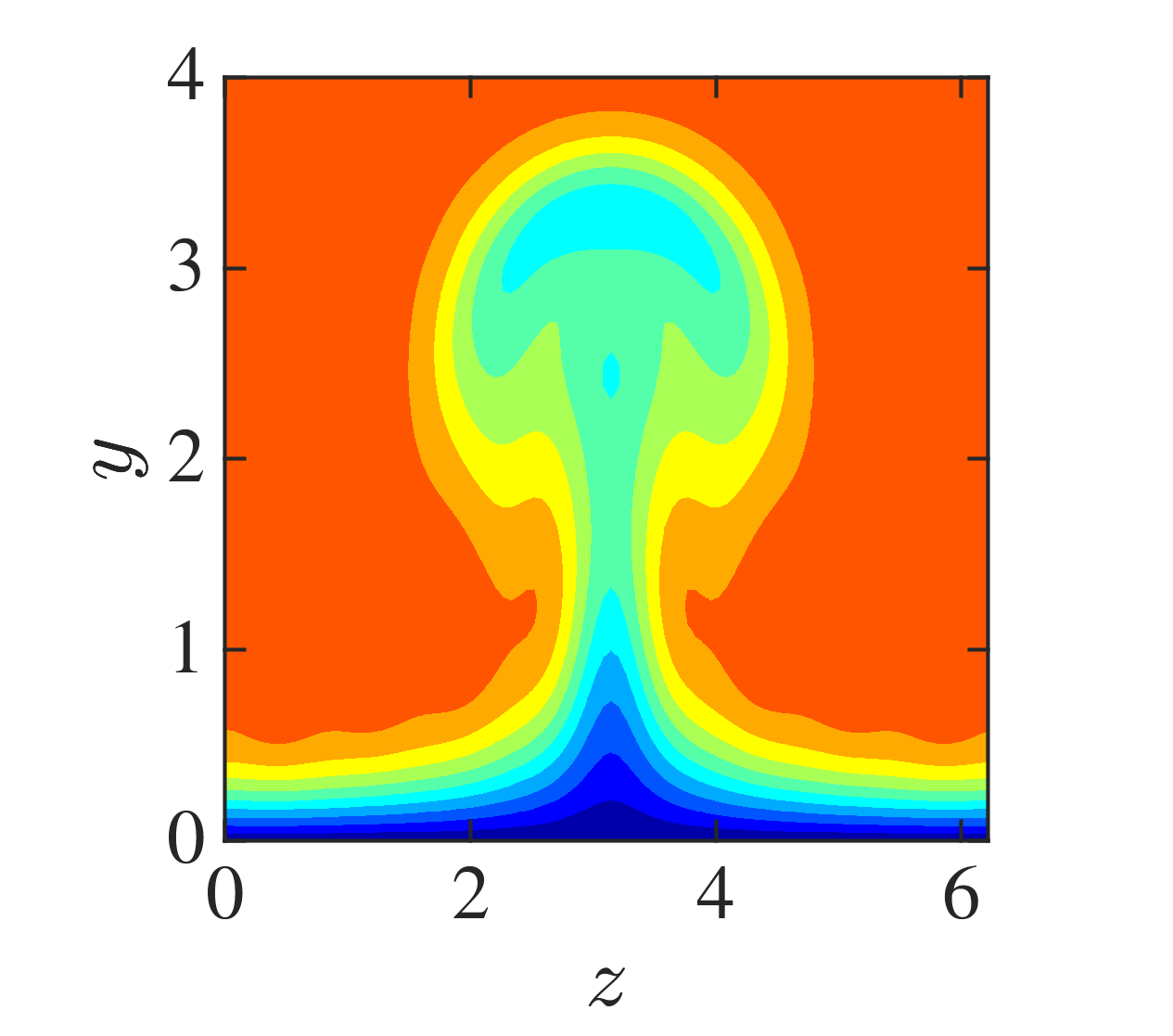}
\put(-110,114){$(b)$}
\put(-75,114){$Tu=0.5\%$}
}
\subfigure{
\includegraphics[width=0.32\textwidth,height=0.3\textwidth]{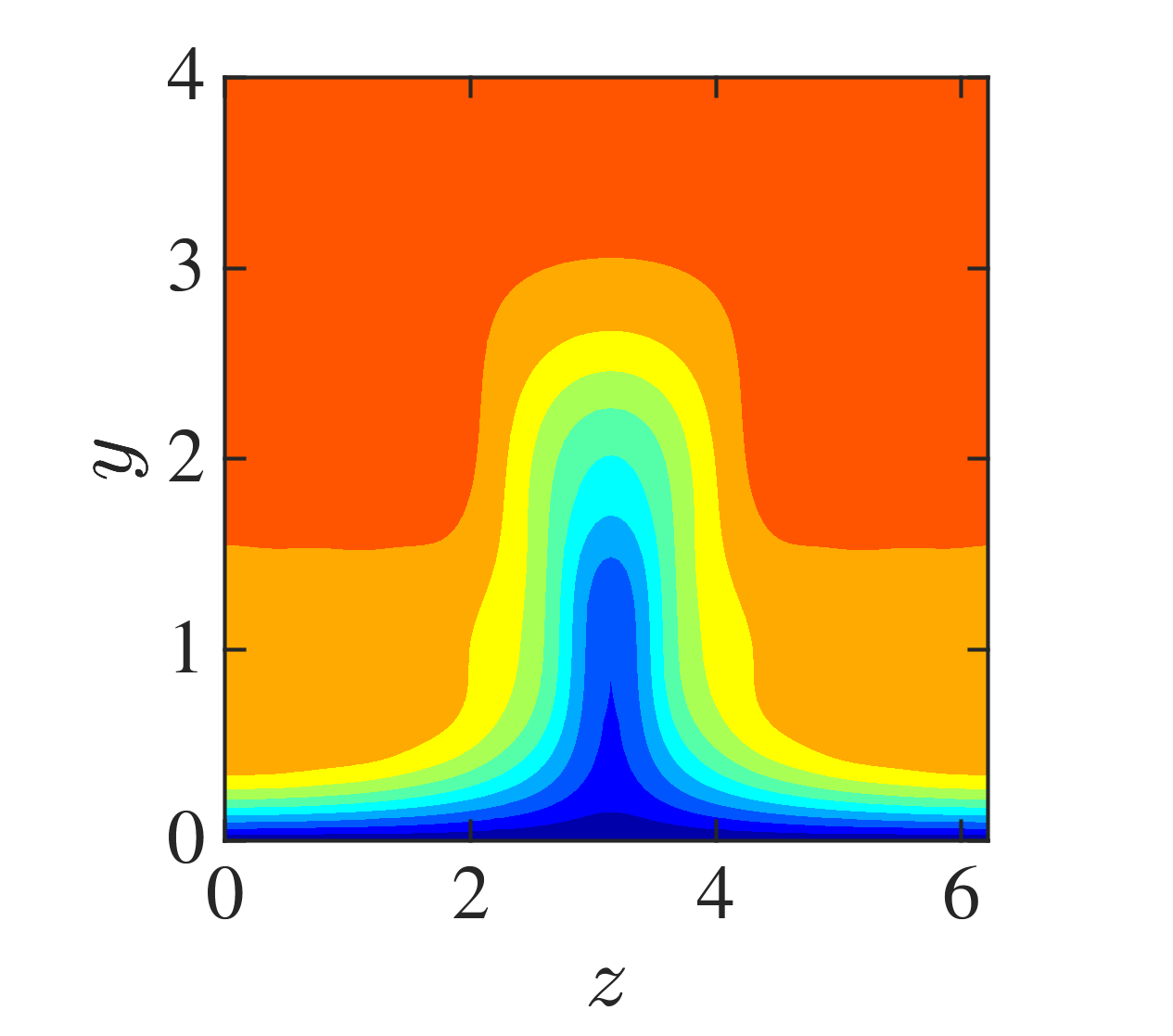}
\put(-110,114){$(c)$}
\put(-75,114){$Tu=1.8\%$}
}
\subfigure{
\includegraphics[width=0.32\textwidth,height=0.3\textwidth]{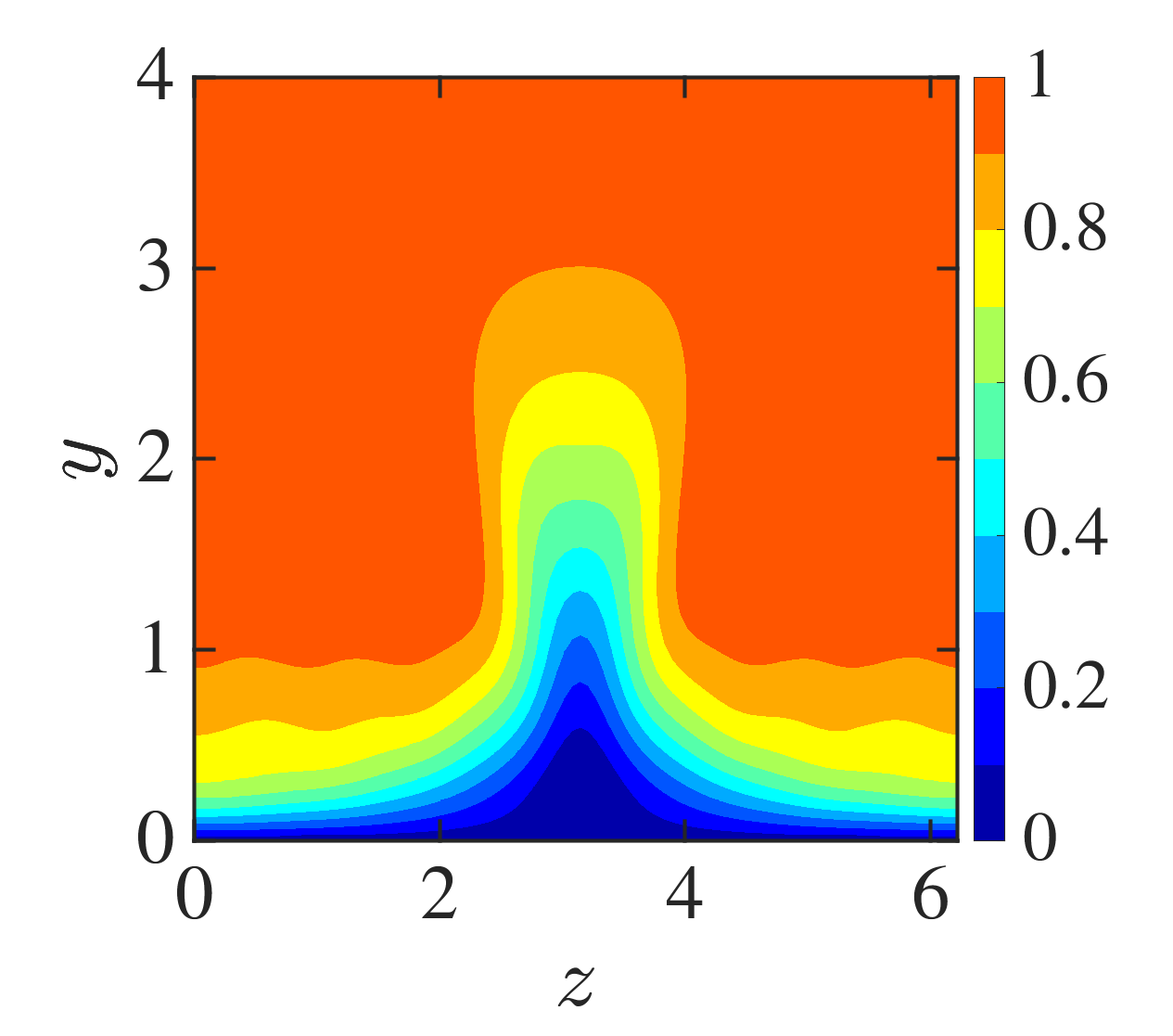}
\put(-110,114){$(d)$}
\put(-75,114){$Tu=3\%$}
}
\captionsetup{justification=raggedright}
\vspace{-3mm}
\caption{($a$) Growth of $u_{rms,max}$ for $\mathcal{G}=35.2$ and different FVD levels. The portions of the trends highlighted in red indicate where the linear and the nonlinear solutions overlap. The darker portions of the trends denote a $u_{rms,max}$ growth with positive concavity. The saturation points are marked by red circles. Panels ($b, c,d$) show contours of the instantaneous streamwise velocity $y-z$ plane at the saturation locations for different FVD levels.}
\label{fig:cri-tu}
\end{figure}

As discussed in \S\ref{sec:gortler}, both the FVD level and the wall curvature determine whether the boundary-layer disturbances evolve as G\"ortler vortices or streaks. It is thus useful to discern which type of disturbance occurs under which conditions. We utilise the top graph in figure \ref{fig:cri-tu} to this end. It shows the evolution of $u_{rms,max}$ for $\mathcal{G}=35.2$ and different FVD levels. Linear and nonlinear results are included. The portions of the lines highlighted in red indicate where the evolutions of the boundary-layer disturbances studied by the linear and nonlinear theory overlap. The light red portions of the $u_{rms,max}$ trends grow with a negative concavity, while the dark red portions grow with a positive concavity. The latter do not display a fully exponential growth because nonlinearity quickly sets in leading the disturbance flow to saturation. The dark red portion is clearly visible for $Tu=0.5\%$, becomes smaller as the FVD level increases to approximately $Tu=1.8\%$, and disappears for larger $Tu$ as the growth of $u_{rms,max}$ with positive concavity is fully bypassed.

\begin{figure}
\centering
\includegraphics[width=0.75\textwidth]{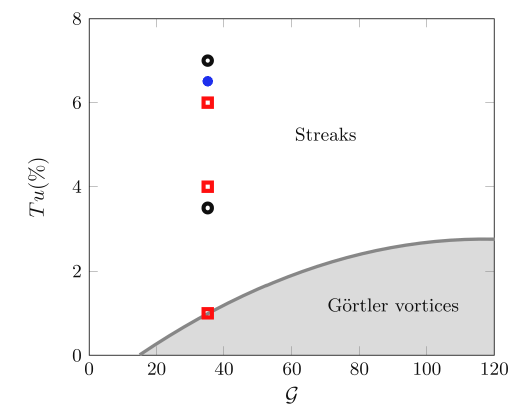}
\captionsetup{justification=raggedright}
\caption{Occurrence map of nonlinear streaks and G\"ortler vortices for $k_x=0.0073,$ $R_\Lambda=1124$ and $\mathcal{M}_\infty=0.69$. The symbols denote the data of \citet{zhao2020bypass} (solid circle), \citet{wheeler2016direct} (hollow circles) and \citet{arts1990aero} (hollow squares).}
\label{fig:tu-g}
\end{figure}

Nonlinear G\"ortler vortices are defined as boundary-layer disturbances that evolve through three stages from their inception near the leading edge, as shown in the top graph of figure \ref{fig:cri-tu}, i.e. a light-red growth (such as an algebraic-like $u_{rms,max}$ growth with negative concavity), a dark-red growth (a $u_{rms,max}$ growth with positive concavity) and a saturation stage, where nonlinearity is fully effective and the intensity of the vortices becomes almost independent of the streamwise position. When these flow conditions are met, cross-sectional contour plots of the saturated streamwise velocity and temperature feature the typical mushroom shape, shown in figure \ref{fig:cri-tu}$(a)$ for $Tu=0.5\%$. Streaks only exhibit a light-red algebraic-like growth of $u_{rms,max}$ instead of a dark-red growth with positive concavity and feature a bell shape instead of a mushroom shape, shown in figures \ref{fig:cri-tu}$(b,c)$. They saturate to a nearly constant amplitude, like the nonlinear G\"ortler vortices.

Using the observations of figure \ref{fig:cri-tu}, we have created the map shown in figure \ref{fig:tu-g}, which identifies the flow character as a function of $Tu$ and $\mathcal{G}$. The map represents subsonic nonlinearly saturated low-frequency disturbances in boundary layers over concave surfaces. The map is representative of flows over turbine blades with frequencies, Reynolds number and Mach number comparable to the reference values chosen herein. Convex-curvature effects are not included as our results indicate that the growth of disturbances is never enhanced with respect to the flat-wall case when the wall is convex.

In the linearised case, for which saturation does not occur, \citet{viaro2018neutral} distinguished G\"ortler vortices from streaks by applying a criterion solely based on the concavity of the amplitude of the streamwise velocity. This method is, however, inapplicable for nonlinear G\"ortler vortices because nonlinear disturbances saturate with a null or slightly negative growth rate. If the Viaro--Ricco criterion were applied to the saturated nonlinear disturbances, they would not be classified as G\"ortler vortices.

As represented in the map of figure \ref{fig:tu-g}, G\"ortler vortices appear when the FVD level is relatively low. The G\"ortler-vortex region expands as the G\"ortler number increases. Streaks are instead observed at larger $Tu$, i.e.  when the streamwise curvature is less influential, as discussed in \S\ref{sec:gortler}. As the G\"ortler number is increased beyond $\mathcal{G}=100$, the line that separates the two regions flattens to a FVD level slightly below $Tu=3\%$. This result indicates that, as the G\"ortler number increases, the wall curvature becomes less influential on whether the nonlinear disturbances evolve as streaks or G\"ortler vortices. Nonlinear streaks are likely to develop over turbine blades because free-stream disturbance environments characterised by $Tu>3\%$ certainly pertain to turbomachinery flows. Even if boundary layers over turbine blades were exposed to low FVD levels, i.e.  $Tu<3\%$,
streaks would still be more likely to occur than Görtler vortices. As discussed in \S\ref{sec:gortler}, the streamwise extent of turbine blades is indeed too limited for the disturbances to be influenced by the wall curvature and turn into G\"ortler vortices when $Tu$ is low, following the initial algebraic growth highlighted in light red in figure \ref{fig:cri-tu}. We also note that, while the nonlinear streaks evolving over concave surfaces saturate to a constant amplitude, the nonlinear streaks occurring over flat plates, also termed thermal Klebanoff modes \citep{marensi2017nonlinear}, typically decay after the initial algebraic growth. The line that distinguishes G\"ortler vortices from streaks in figure \ref{fig:tu-g} crosses the abscissa at a finite $\mathcal{G}$ value, i.e.  at any FVD level, small curvatures are not sufficient to trigger G\"ortler vortices because viscous dissipation overcomes the inviscid centrifugal imbalance \citep{wu2011excitation,viaro2018neutral}. Furthermore, although FVD are responsible for triggering G\"ortler vortices and streaks through receptivity, enhancing the FVD level always favours the formation of streaks over G\"ortler vortices.

In figure \ref{fig:tu-g}, experimental and direct numerical simulation data typical of flows over turbine blades and in subsonic wind tunnels are also shown. All those data are located in the `streaks' region, denoting the weak effect of the curvature in boundary layers over the pressure sides of turbine blades. The absence of G\"ortler vortices over the pressure side of turbine blades was also predicted by \citet{dhurovic2021free}, who utilised the criterion by \citet{Saric1994} based on the critical G\"ortler number. This approach, although successful in their case, is generally not applicable because it is based on three assumptions that are not often satisfied: the G\"ortler vortices are (i) fully developed along the streamwise direction, which may not be the case because of the limited extent of turbine blades, (ii) described by a linearised dynamics, which is unlikely to be the case for moderate and elevated FVD levels, typical of turbomachinery applications and (iii) unaffected by non-parallel effects, which instead play a leading role when $\mathcal{G}=\mathcal{O}(1)$ \citep{hall1983thelinear,wu2011excitation,xu_zhang_wu_2017,marensi2017growth}.

By unravelling the competition between the FVD level and the wall curvature, our occurrence map provides a theoretical explanation for the flow character and instabilities at play in boundary layers over concave walls in the presence of FVD. The map may be used to interpret experiments and simulations of subsonic turbomachinery flows.

\subsection{Secondary instability of G\"ortler vortices and streaks}
\label{sec:sec}
\begin{figure}
\centering
   \subfigure{
   \includegraphics[width=0.48\textwidth]{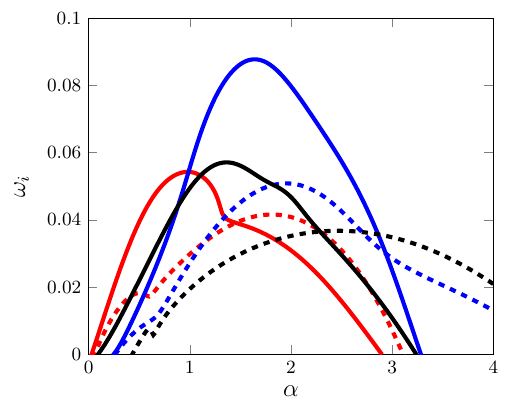}
    \put(-190,146){$(a)$}
    \label{fig:even-odd-a}
}
  \subfigure{
 \includegraphics[width=0.48\textwidth]{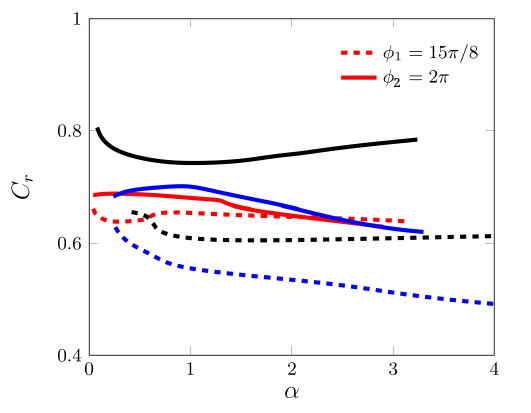}
   \put(-190,146){$(b)$}
   \label{fig:even-odd-b}
   }

    \subfigure{
   \includegraphics[width=0.48\textwidth]{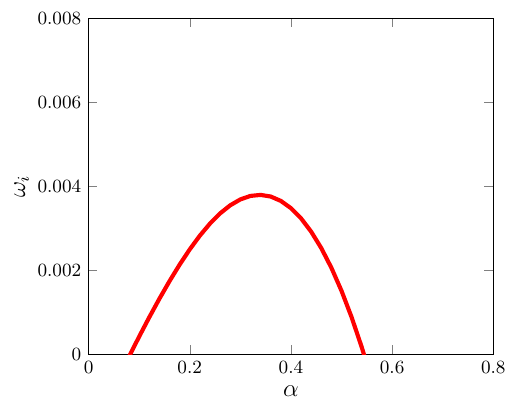}
    \put(-190,146){$(c)$}
    \label{fig:even-odd-c}
}
  \subfigure{
 \includegraphics[width=0.48\textwidth]{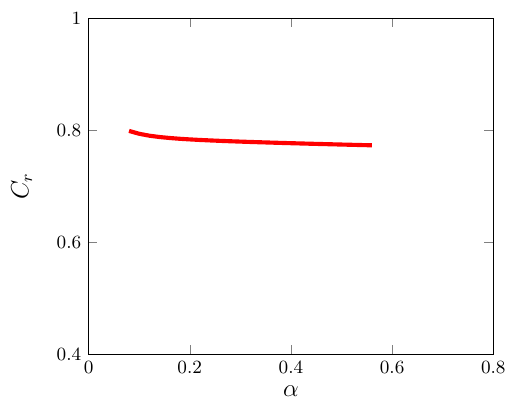}
   \put(-190,146){$(d)$}
   \label{fig:even-odd-d}
   }
\captionsetup{justification=raggedright}
\caption{($a,c$) Temporal growth rates and ($b,d$) phase speeds of the secondary-instability modes of G\"ortler vortices. Panels ($a,b$) are for the concave-wall case ($\mathcal{G}=35.2$) and panels ($c,d$) are for the flat-wall case ($\mathcal{G}=0$). The red lines represent odd mode I, the blue lines correspond to even mode I and the black lines indicate odd mode II. The parameters are $\bar x=1.5$, $Tu=1\%$ and $\mathcal{M}_\infty=0.69.$}
\label{fig:sec-g-rate}
\end{figure}

In this section, we present the results on the secondary instability of the G\"ortler vortices and streaks. We observe that the dominant fundamental modes are more unstable than the other modes; therefore, we only report the results for the fundamental modes. Figures \ref{fig:sec-g-rate}$(a,b)$ display the growth rate $\omega_i$ and the phase speed $c_r=\omega_r/\alpha$ of the secondary modes at $\bar x=1.5$ for $Tu=1\%,$ $\mathcal{M}_\infty=0.69$ and $\mathcal{G}=35.2$. For these conditions, the disturbances are nonlinear G\"ortler vortices reaching their maximum amplitude. The instability is analysed when the flow is unstable, i.e.  in two time windows within a period, from $3\pi/4$ to $\pi$ and from $7\pi/4$ to $2\pi$. Three dominant unstable modes are detected, one varicose mode (even mode I) and two sinuous modes (odd modes I and II), all of which were shown by \citet{ren2015secondary} and \citet{xu_zhang_wu_2017}. At each phase, the maximum growth rate is attained by the even mode I.

Figures \ref{fig:sec-g-rate}$(c,d)$ show the growth rate $\omega_i$ and the phase speed $c_r$ of the secondary modes at $\bar x=1.5$ for $Tu=1\%,$ $\mathcal{M}_\infty=0.69$ and $\mathcal{G}=0$. For these conditions, the disturbances are nonlinear streaks since the wall is flat. The growth rate of the odd mode I is relatively low, with a maximum value of about 0.004. Comparing the growth rates in the concave-wall case in figure \ref{fig:sec-g-rate}$(a)$ with the growth rate in the flat-wall case in figure \ref{fig:sec-g-rate}$(c)$  demonstrates that the curvature significantly increases the growth rate of this secondary-instability mode.

\begin{figure}
\centering
   \subfigure{
      \includegraphics[width=0.48\textwidth]{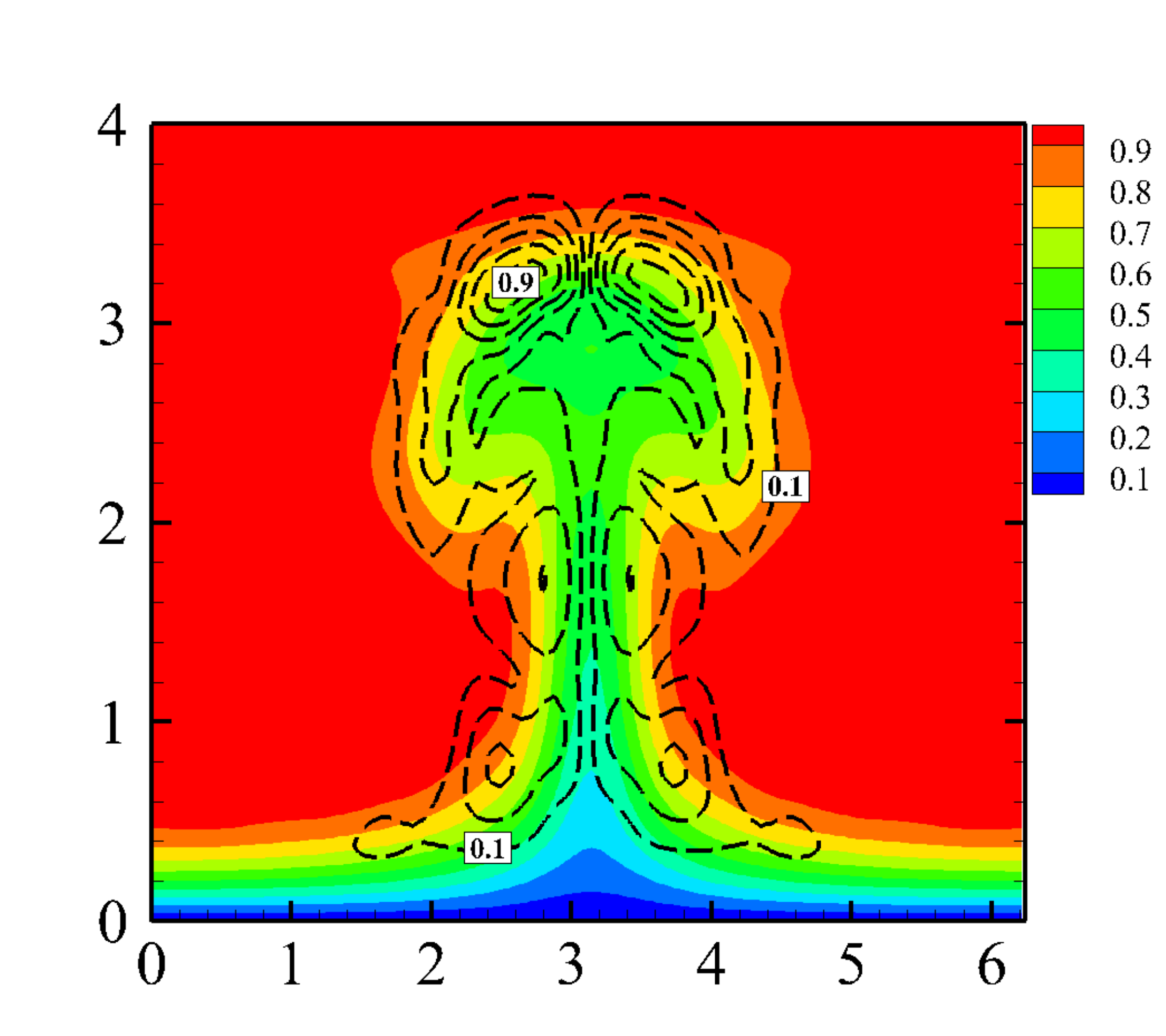}
      \put(-158,150){$(a)$}
      \put(-70,150){odd mode I}
      \label{fig:streak-a}
      \put(-96,0){$z$}
\put(-120,56){\rotatebox[origin=c]{90}{$y$}}
   }
   \subfigure{
      \includegraphics[width=0.48\textwidth]{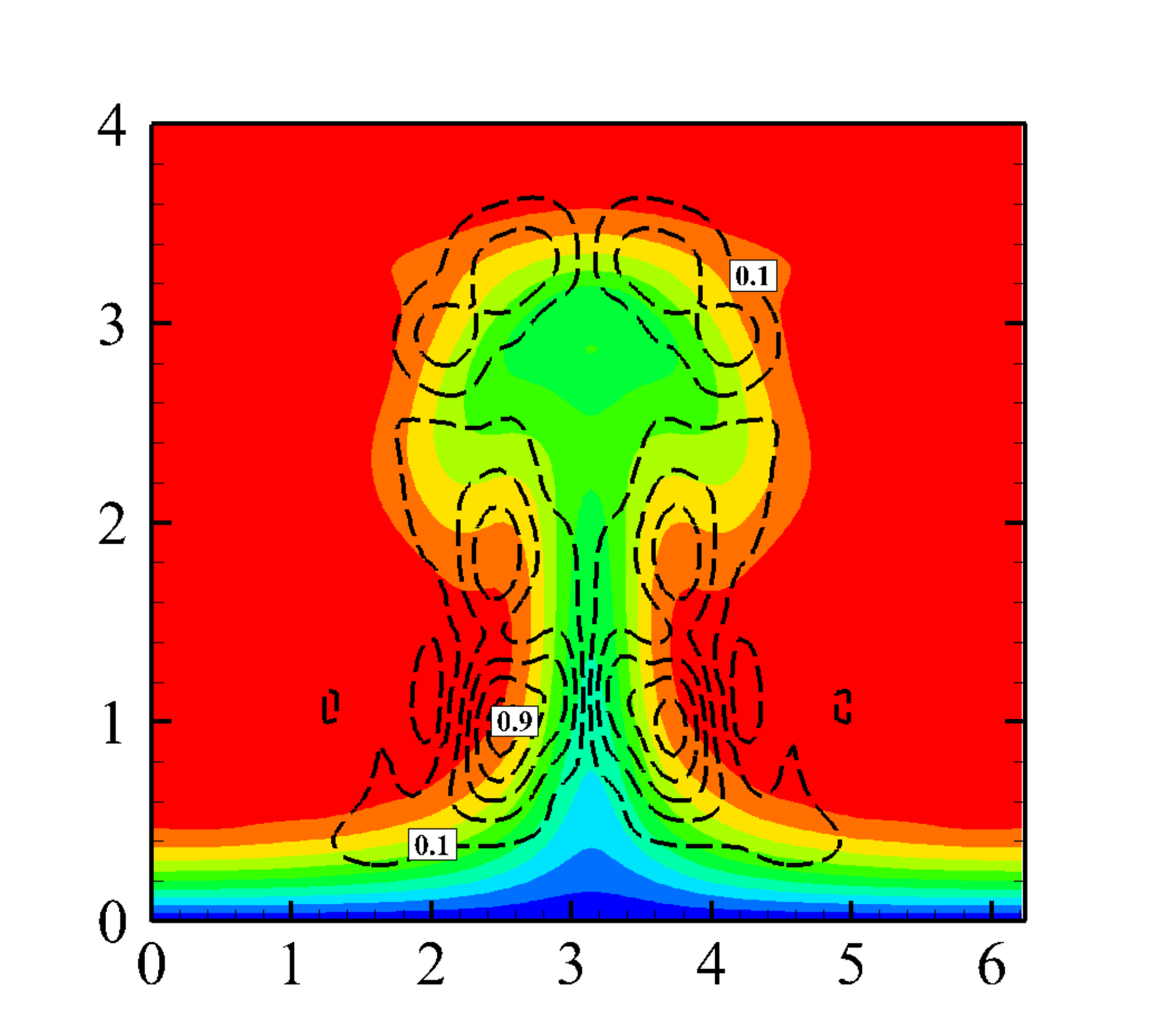}
      \put(-158,150){$(b)$}
      \put(-70,150){odd mode II}
      \put(-96,0){$z$}
      \put(-120,56){\rotatebox[origin=c]{90}{$y$}}
      \label{fig:streak-b}
   }
   \subfigure{
      \includegraphics[width=0.48\textwidth]{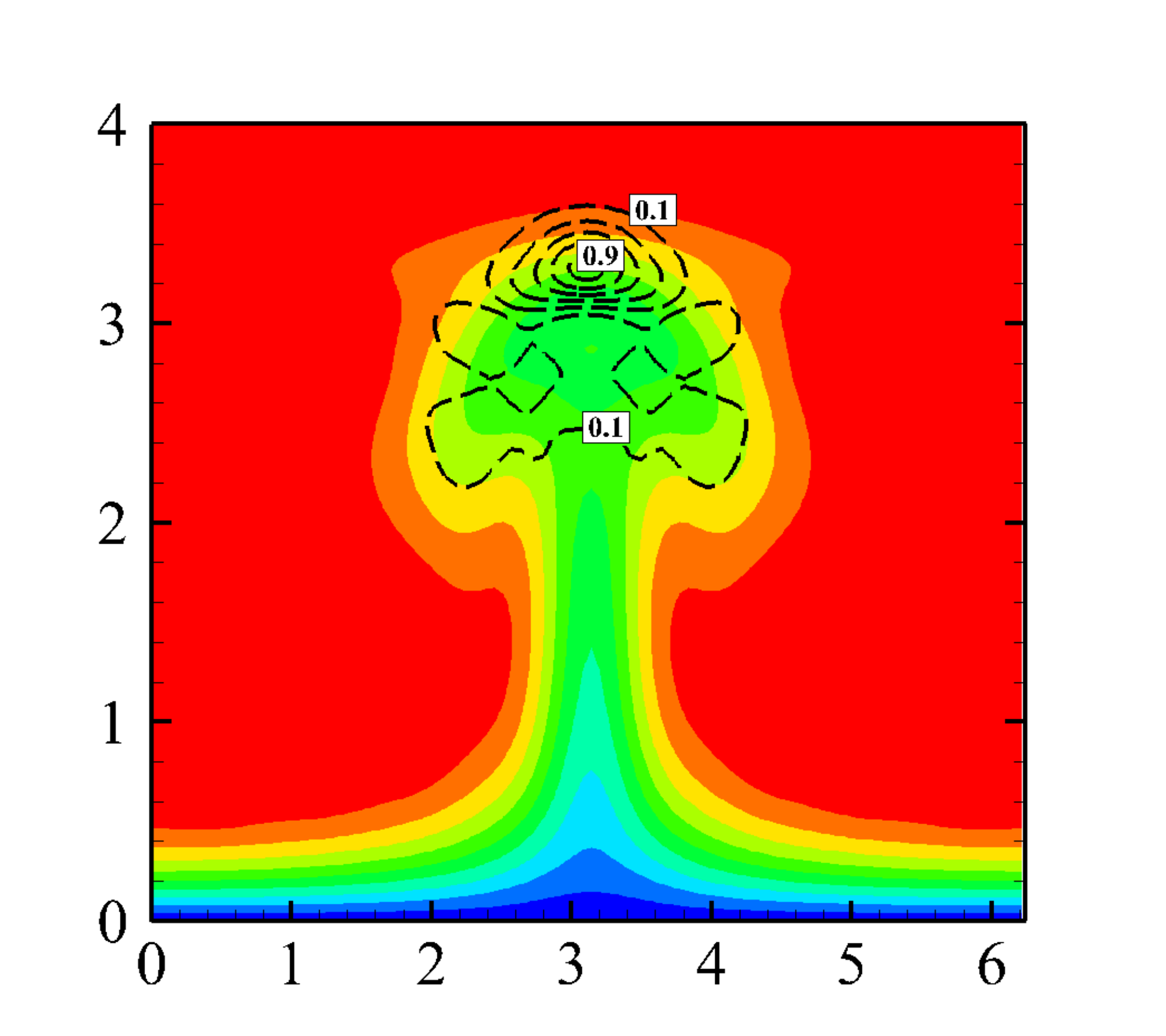}
      \put(-158,150){$(c)$}
      \put(-70,150){even mode I}
      \put(-96,0){$z$}
      \put(-120,56){\rotatebox[origin=c]{90}{$y$}}
      \label{fig:streak-c}
   }
   \subfigure{
      \includegraphics[width=0.48\textwidth]{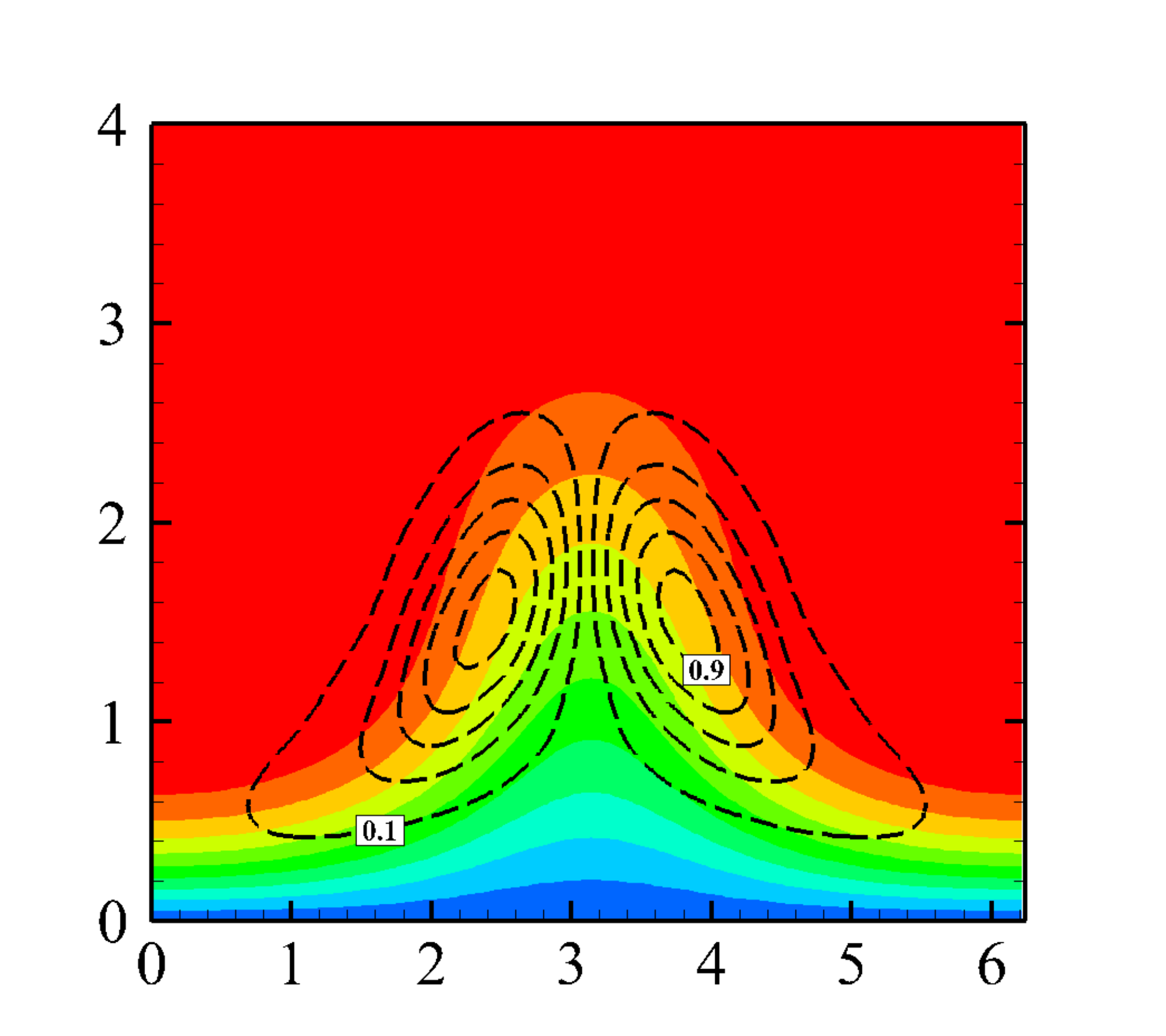}
      \put(-158,150){$(d)$}
      \put(-70,150){odd mode I}
      \put(-96,0){$z$}
      \put(-120,56){\rotatebox[origin=c]{90}{$y$}}
      \label{fig:streak-d}
   }
   \captionsetup{justification=raggedright}
   \caption{Eigenfunctions of secondary unstable modes, shown by contours of the streamwise velocity (absolute value, black lines). G\"ortler vortices ($\mathcal{G}=35.2$): ($a$) odd mode I; ($b$) odd mode II; ($c$) even mode I. Streaks ($\mathcal{G}=0$): ($d$) odd mode I. The coloured contours represent the streamwise velocity of the vortex base flow at $\bar x=1.5$. Five levels are specified, ranging from 0.1 to 0.9. The parameters are $Tu=1\%$ and $\mathcal{M}_\infty=0.69$.}
   \label{fig:sec-contours}
\end{figure}

Figures \ref{fig:sec-contours}$(a-c)$ shows the contours of the streamwise-velocity eigenfunctions of sinuous and varicose modes pertaining to nonlinear G\"ortler vortices for the same conditions of figures \ref{fig:sec-g-rate}$(a,b)$. The eigenfunctions of the unstable odd modes extend across the entire mushroom shape due to the highly distorted velocity profile, while the eigenfunctions of the even modes concentrate at the top of the mushroom shape. Figure \ref{fig:sec-contours}$(d)$ shows the eigenfunction of the odd mode I pertaining to the nonlinear streaks for the same conditions of figures \ref{fig:sec-g-rate}$(c,d)$.

Figure \ref{fig:sec-s-rate} presents the growth rate $\omega_i$ and phase speed $c_r$ of secondary modes growing on nonlinear streaks at $\bar x=0.36$ for $Tu=6\%$. Due to the high FVD level, the growth rate and phase speed are almost the same as those for G\"ortler vortices, as shown in figure \ref{fig:sec-g-rate}. Compared with the G\"ortler vortices, the time window of instability is shorter, although the dominant mode is still the odd mode I. A new even mode {(even mode II)} is detected for the nonlinear streaks, which has never been reported in the literature. Both its growth rate and phase speed are smaller than those of the odd mode I. This new mode is not the varicose mode reported in \cite{wu2003linear} as the new mode only appears for high-intensity FVD.

Figure \ref{fig:sec-contours-s} shows the contours of the streamwise-velocity eigenfunctions of the odd mode I and the even mode II for $Tu=6\%$. The structure of the odd mode I is similar to the odd mode I for the streaks shown in figure \ref{fig:sec-contours}$(d)$. The even mode II concentrates in the lower part of the streaks and it may thus promote transition to turbulence at the stem of nonlinear streaks.

Our analysis thus suggests that transition to turbulence over the pressure surface of turbine blades subject to high-intensity FVD is due to the breakdown of unsteady nonlinear streaks. We also conclude that transition to turbulence in subsonic wind tunnel can be caused by the breakdown of nonlinear G\"ortler vortices because of the low-intensity FVD environment and the long streamwise distance along which the vortices can develop.

\begin{figure}
\centering
   \subfigure{
      \includegraphics[width=0.48\textwidth]{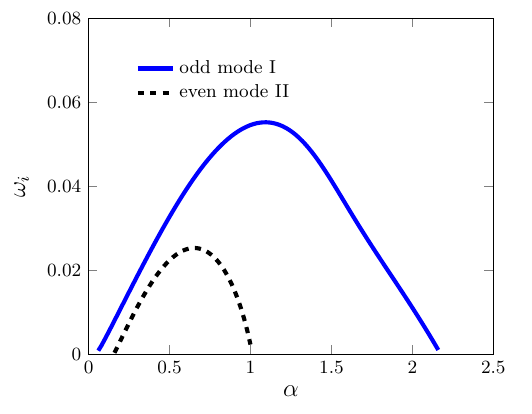}
      \put(-190,146){$(a)$}
   }
   \subfigure{
      \includegraphics[width=0.48\textwidth]{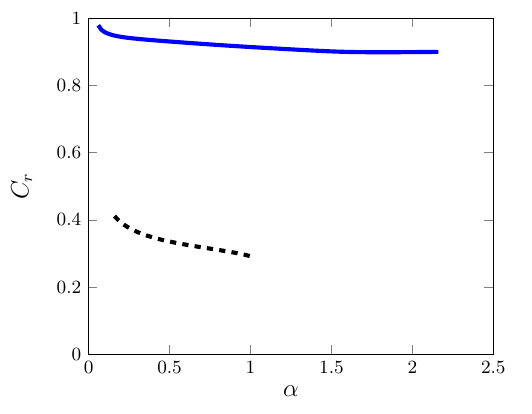}
      \put(-190,146){$(b)$}
   }
   \captionsetup{justification=raggedright}
   \caption{Characteristics of secondary instability of streaks: $(a)$ temporal growth rate and $(b)$ phase speed versus the streamwise wavenumber $\alpha.$ The parameters are $Tu=6\%,$ $\mathcal{G}=35.2$ and $\mathcal{M}_\infty=0.69$.}
   \label{fig:sec-s-rate}
\end{figure}
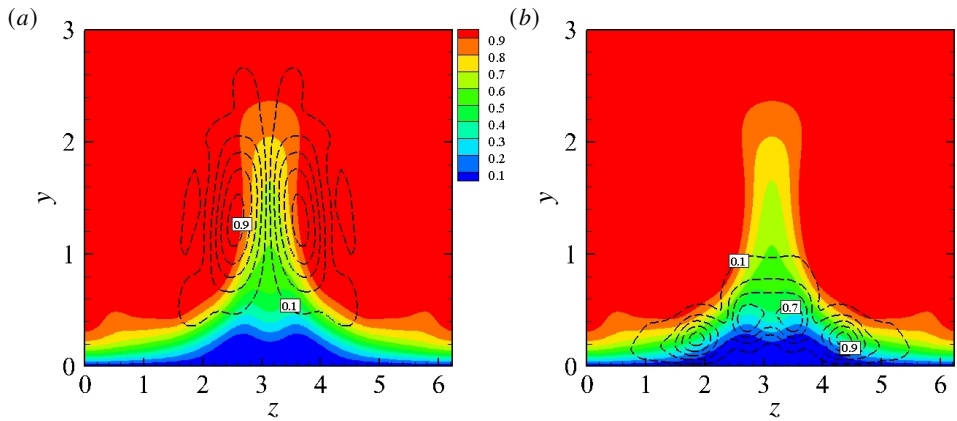
\begin{figure}
\centering
   \subfigure{
      \includegraphics[width=0.48\textwidth]{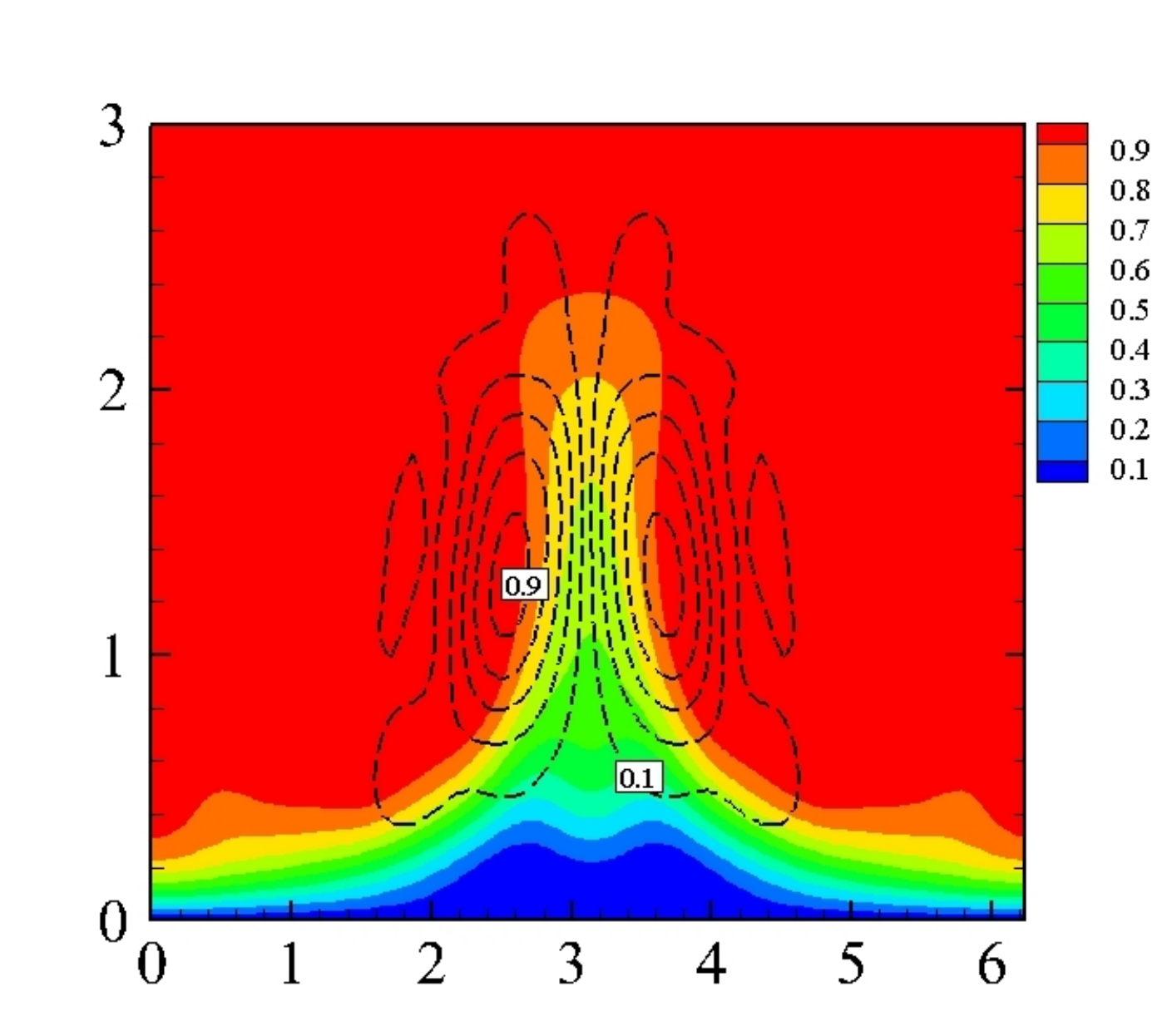}
      \put(-190,146){$(a)$}
      \put(-92,0){\normalsize $z$}
      \put(-178,80){\normalsize \rotatebox[origin=c]{90}{$y$}}
   }
   \subfigure{
      \includegraphics[width=0.48\textwidth]{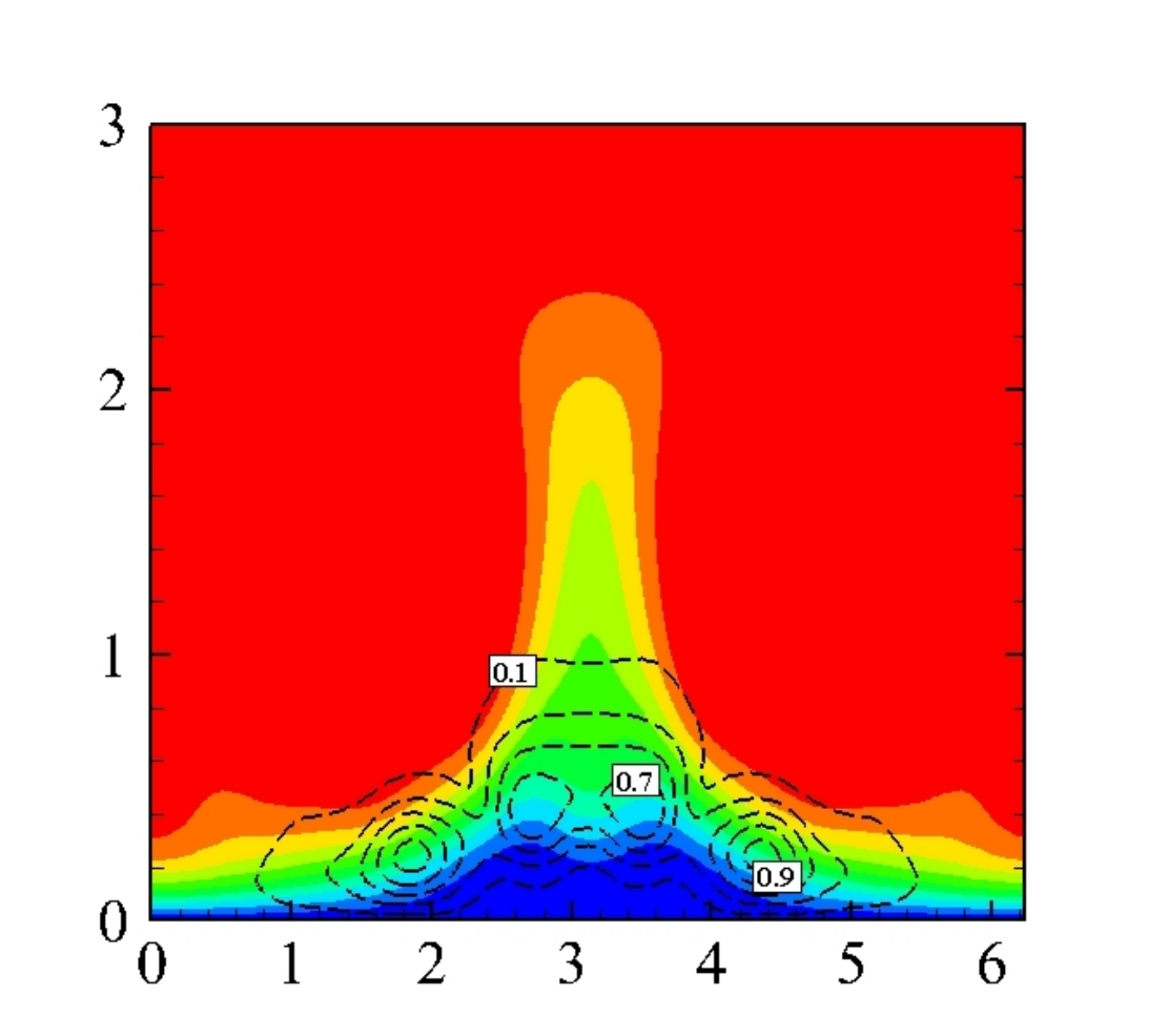}
      \put(-190,146){$(b)$}
            \put(-92,0){\normalsize $z$}
      \put(-178,80){\normalsize \rotatebox[origin=c]{90}{$y$}}
   }
   \captionsetup{justification=raggedright}
   \caption{Eigenfunctions (absolute value, black lines) of secondary unstable modes, shown by contours of the streamwise velocity. ($a$) odd mode I; ($b$) even mode II. The coloured contours represent the streamwise velocity of the vortex base flow at $\bar{x}=0.3$. Five levels are specified, ranging from 0.1 to 0.9. The parameters are $Tu=6\%,$ $\mathcal{G}=35.2$ and $\mathcal{M}_\infty=0.69.$}
   \label{fig:sec-contours-s}
\end{figure}

\clearpage
\section{Conclusions}
\label{sec:conclusions}

In this study, we have utilised receptivity theory to investigate the nonlinear response of compressible boundary layers over curved surfaces to unsteady free-stream vortical fluctuations of the convected-gust type. We have focused on low-frequency and long-wavelength disturbances because these disturbances penetrate the most into the core of a boundary layer, forming kinematic and thermal G\"ortler vortices or streaks. The free-stream disturbances are assumed to be strong enough to generate nonlinear interactions between velocity and temperature fluctuations, thus altering the original laminar base flow when the local boundary-layer thickness becomes comparable to the spanwise wavelength of the G\"ortler vortices or streaks. This boundary-layer response is governed by the compressible boundary-region equations, leading to a nonlinear initial-boundary-value problem that we have solved numerically. Our previous studies by \citet{xu_zhang_wu_2017}, \citet{marensi2017nonlinear} and \citet{viaro2019compressible} have been unified to account for compressibility, curvature and nonlinear effects simultaneously.

We have investigated transitional boundary layers for flow parameters pertaining to flows over pressure surfaces of turbine blades. Decreasing the frequency of the free-stream perturbations and increasing the wall concavity and the free-stream disturbance level energise the boundary-layer disturbances. The Mach number instead has no influence on the kinetic disturbances and has a slightly stabilising influence on the thermal disturbances in the subsonic conditions of interest. The disturbances are unsteady along an initial streamwise distance because the unsteadiness of the free-stream flow has a direct impact on the boundary layer. As the flow evolves, steady-flow distortions caused by nonlinearity become comparable to, and may even exceed, the unsteady components induced by the free-stream flow. Our numerical results have been compared with available experimental data for boundary-layer flows over curved pressure surfaces of turbine blades. The receptivity framework accurately predicts the streamwise-elongated spanwise patterns of enhanced skin friction and wall-heat transfer, often referred to as hot fingers.

We have also created a map that identifies the occurrence of saturated nonlinear G\"ortler vortices and streaks, for different G\"ortler numbers and free-stream disturbance levels. Nonlinear streaks are defined as disturbances that only grow algebraically and exhibit a bell-like shape. The streaks are more likely to occur at small G\"ortler numbers and at relatively high levels of ambient disturbances; for high G\"ortler numbers, a free-stream disturbance level slightly exceeding 3\% generates streaks only. Nonlinear G\"ortler vortices are instead defined as disturbances that display a growth with positive curvature following an initial algebraic growth and feature a mushroom-like shape. The G\"ortler vortices occur at low levels of free-stream disturbance and intensify as the G\"ortler number increases.

We have studied the secondary instability of the nonlinear boundary-layer disturbances to elucidate the subsequent stages of the transition process. Our numerical results indicate that the saturated disturbances are susceptible to exponentially growing high-frequency modes. Increasing the streamwise curvature promotes the growth of two odd modes and one even mode.
G\"ortler vortices and streaks excited by high-intensity free-stream disturbances are susceptible to a new even mode (even mode II), which has not been reported in earlier studies. This mode is important since it is located at the stem of the streaks and may thus initiate transition to turbulence there. The even mode II could potentially be more critical than the more unstable odd mode I because its concentration near the wall may cause the resulting transition to affect the skin friction and the wall-heat transfer immediately. In contrast, the odd mode I, located in the outer part of the boundary layer, will not substantially influence the skin friction and the wall-heat transfer until transition extends to the wall.

To conclude, the present study has provided a mathematical and numerical description of the generation, evolution and secondary instability of G\"ortler vortices and streaks in compressible boundary layers. The central result is that, thanks to the receptivity approach, the characteristics of the free-stream disturbance environment have been linked quantitatively to the transitional boundary layer. Our analysis can be readily extended to more realistic cases, including boundary layers exposed to broadband free-stream turbulence \citep{zhang-etal-2011b} or influenced by a streamwise pressure gradient \citep{xu2020gortler}.

An important avenue of future research is the study of amplified three-dimensional waves developing on the streaks, as recently observed in hypersonic boundary-layer flows by \cite{huang2021inner} and \cite{feng2024investigation}, and previously studied in incompressible boundary layers by \cite{lee2008transition}, \cite{jiang2020structure}, \cite{jiang2020experimental} and \cite{jiang2021metamorphosis}. As shown by \cite{huang2021inner} and \cite{feng2024investigation}, these three-dimensional waves feature overlapped temperature peaks and high-frequency modes, and play an important role in the breakdown to turbulence. In our future work, we plan to focus on the final stages of transition to turbulence and, therefore, it would be interesting to investigate how free-stream perturbations and wall curvature influence the formation of these three-dimensional waves.

For an accurate prediction of the transition location in boundary layers over turbine blades, the leading-edge bluntness should also be taken into account. Transition prediction methods would thus be possible for turbomachinery flows and other compressible flows of industrial interest.

\section*{Acknowledgments}
\indent
We thank the reviewers for their helpful comments that have helped improve the quality of the paper. DX would like to thank Professor Ming Dong, Dr Runjie Song and Dr Lei Zhao for the useful discussions. DR and PR wish to acknowledge the support of EPSRC (Grant No. EP/T01167X/1). PR has also been supported by the US Air Force through the AFOSR grant FA8655-21-1-7008 (International Program Office Dr Douglas Smith).

\vspace{0.25cm}
\section*{Declaration of Interests}
The authors report no conflict of interest.

\appendix
\noindent

\section{Initial conditions for the boundary-region equations}
\label{app:in}

The initial conditions are derived by first seeking a power series solution of the boundary-region equations for $\bar{x} \ll 1$ and $\eta=\mathcal{O}(1)$
\begin{equation*}
\{ \bar{u},\bar{v},\bar{w},\bar{\tau},\bar{p} \}=\sum_{j=0}^{\infty}(2\bar {x})^{j/2}\left\{ 2\bar{x}U_j(\eta), \sqrt{2\bar{x}}V_j(\eta), k_z^{-1}W_j(\eta), 2\bar{x}T_j(\eta), P_j(\eta)/\sqrt{2\bar{x}}\right\},
\end{equation*}
and by constructing a composite solution that is valid for all values of $\eta$. This procedure yields the initial conditions
\begin{flalign}
  \mbox{\footnotesize{$\x \rightarrow 0$}} ] \hspace{0.2cm}
  &\hat{u}_{1,\pm 1} \rightarrow
    q_{\pm}\left(2 \x U_0 + (2 \x)^{3/2} U_1\right), & \label{eq:BC-x0_u}\\[0.2cm]
    &\hat {v}_{1,\pm 1} \rightarrow
    q_{\pm}
    \left[V_0 + (2 \x)^{1/2} V_1
    - \left( V_c -\frac{1}{2} g_1 |\kappa_z|
      (2 \x)^{1/2} \right)
    e^{- |\kappa_z| (2 \x)^{1/2} \bar{\eta}}
    \right.
    \nonumber\\[0.2cm]
    &\hspace{0.7cm}
    \left.
    + \frac{{\rm i}}{(\kappa_y -
      {\rm i} |\kappa_z|)(2 \x)^{1/2}}
    \left( e^{{\rm i} \kappa_y (2 \x)^{1/2} \bar{\eta} -
        \left(\kappa_z^2 + \kappa_y^2\right) \x} -
      e^{- |\kappa_z| (2 \x)^{1/2} \bar{\eta}} \right) - \bar{v}_c\right],& \\[0.2cm]
    &\hat{w}_{1,\pm 1} \rightarrow
    \mp{\rm i} q_{\pm}\left[W_0 + (2 \x)^{1/2} W_1
    - V_c |\kappa_z| (2 \x)^{1/2}
    e^{- |\kappa_z| (2 \x)^{1/2} \bar{\eta}}
    \right.
    \nonumber\\[0.2cm]
    &\hspace{0.8cm}
    \left.
    +\frac{1}{\kappa_y -
      {\rm i} |\kappa_z|}
    \left( \kappa_y
      e^{{\rm i} \kappa_y (2 \x)^{1/2} \bar{\eta} -
        \left( \kappa_z^2 + \kappa_y^2 \right) \x} -
      {\rm i} |\kappa_z|
      e^{- |\kappa_z| (2 \x)^{1/2} \bar{\eta}}
    \right) - \bar{w}_c \right]
    ,& \\[0.2cm]
    &\hat{p}_{1,\pm 1} \rightarrow
    q_{\pm}\left[\frac{P_0}{(2 \x)^{1/2}} + P_1 + \left( g_1 -
      \frac{V_c}{|\kappa_z|
        (2 \x)^{1/2}} \right)
    e^{- |\kappa_z| (2 \x)^{1/2} \bar{\eta}} -
    \bar{p}_c\right],& \\[0.2cm]
    &\bar{\tau}_{1,\pm 1} \rightarrow
    q_{\pm}\left(2 \x T_0 + (2 \x)^{3/2} T_1\right), \label{eq:BC-x0_tau}
\end{flalign}
where $\bar{\eta} \equiv \eta - \beta_c$ and $\beta_c = \lim_{\eta \to \infty} (\eta -F)$. The common parts $\bar v_c,$ $\bar w_c$ and $\bar p_c,$ the constants $g_1$ and $V_c$ and the solutions $U_0$, $V_0$, $W_0$, $P_0$, $T_0$, $U_1$, $V_1$, $W_1$, $P_1$, $T_1$ are given in Appendix D of \citet{ricco2007thesis}. The term $q_{\pm}$ herein represents the amplitude of the induced disturbances. In the case of a pair of oblique modes, it is given by
\begin{equation*}
   q_{\pm}=\pm \dfrac{{\rm i} \kappa_z^2}{k_z}\left( \hat u_{z,\pm}^\infty \pm \dfrac{{\rm i}k_z}{\sqrt{k_x^2+k_z^2}}\hat u_{y,\pm}^\infty \right).
\end{equation*}

\bibliographystyle{jfm}
\bibliography{dx}

\end{document}